\documentclass[12pt]{article}
\usepackage{latexsym,cite,amssymb,amsthm,wasysym,textcomp,stmaryrd,times}
\input amssym.def
\input amssym.tex

\makeatletter

\@addtoreset{equation}{section} \makeatother

\newcommand{\q}{{\tilde{q}}}

\newcommand{\M}{{\cal M}}
\newcommand{\N}{{\cal N}}
\newcommand{\V}{{\cal V}}

\newcommand{\half}{{{\textstyle\frac{1}{2}}}}

\newcommand{\be}{\begin{equation} }
\newcommand{\ee}{\end{equation} }
\newcommand{\ba}{\begin{array}}
\newcommand{\ea}{\end{array}}

\newcommand{\dq}{\dot{q}}
\newcommand{\tq}{\dot{q}}
\newcommand{\D}{\displaystyle}
\newcommand{\T}{\textstyle}

\newcommand{\w}{\hat{w}}

\def\L{{\cal L}}
\def\O{{\cal O}}
\def\Q{{\cal Q}}

\def\Z{{\cal Z}}
\def\T{{\cal T}}
\def\G{{\cal G}}
\def\calD{{\cal D}}
\def\calS{{\cal S}}

\def\coeff{{+i}}
\def\w{{\omega}}

\def\F{{\cal F}}
\def\calB{{\cal B}_{\scriptscriptstyle {\rm F.P.}}}
\def\FP{{\scriptscriptstyle\rm{ F.P.}}}
\def\BRST{{\scriptscriptstyle {\rm BRST}}}
\def\Hodge{{\scriptscriptstyle {\rm Hodge}}}
\def\Ad{{\rm ad}}
\def\Haar{{H}}
\def\QB{{Q_{\BRST}}}
\def\Dirac{{\scriptscriptstyle\rm{Dirac}}}

\def\tr{{\rm tr}}

\def\I_M{{I_{\scriptscriptstyle M\times M}}}

\def\Deltap{\widehat{{\,\delta p}}{}}
\def\Deltaq{\widehat{{\,\delta q}}{}}
\def\DeltaK{\widehat{{\,\delta K}}{}}
\def\Deltatp{\widetilde{{\,\delta p}}{}}
\def\Deltatq{\widetilde{{\,\delta q}}{}}
\def\DeltatK{\widetilde{{\,\delta K}}{}}

\def\NG{{N_{\G}}}

\def\v{\widetilde{v}}
\def\free{{v}}

\theoremstyle{definition}
\newtheorem*{proposition}{Proposition}

\newlength{\blength}
\renewcommand{\proof}[1]{\vspace{-.05cm}
\begin{list}{\bf Proof:}
{\listparindent=\parindent\parsep=0pt \labelwidth=-0.5cm
\labelsep=\parindent \addtolength{\labelsep}{-\blength}
\addtolength{\labelsep}{1.5cm}
\itemindent=-\blength
\addtolength{\itemindent}{\parindent} \leftmargin=1.0cm}
\item
#1~$\qedsymbol$\end{list}
\vspace{.0cm}}

\begin{document}
\begin{titlepage}
\title{\vskip -60pt
{\small
\begin{flushright}
\end{flushright}}
\vskip 30pt
{\textbf{Symmetries and dynamics in constrained systems}}
\vskip
10pt
\author{Xavier Bekaert$^\ast$ ~~and~~ Jeong-Hyuck Park$^\dagger$}}
\date{}
\maketitle \vspace{-1.0cm}
\begin{center}
~~~\\
${}^{\ast}$Laboratoire de Math\'ematiques et Physique Th\'eorique\\
Unit\'e Mixte de Recherche $6083$ du CNRS, F\'ed\'eration Denis Poisson\\
Universit\'e Fran\c{c}ois Rabelais, Parc de Grandmont\\
37200 Tours, France\\
~{}\\
${}^{\dagger}$Department of Physics\\
Sogang University\\
Seoul 121-742, Korea\\
~{}\\
{\small{Electronic correspondence: {{{bekaert@univ-tours.fr}},~{{park@sogang.ac.kr}}}
{{{}}}}}\\
~~~\\
~~~\\
\end{center}
\begin{abstract}
\noindent We review in detail the Hamiltonian dynamics for constrained systems.
Emphasis is put on the \textit{total Hamiltonian} system rather than on the \textit{extended Hamiltonian} system. We provide a systematic analysis of (global and local) symmetries in total Hamiltonian systems.
In particular, in analogue to total Hamiltonians, we introduce the notion of \textit{total Noether charges}.
Grassmannian degrees of freedom are also addressed in details.
\end{abstract}
\today

\vspace{1cm}

\thispagestyle{empty}
\end{titlepage}
\newpage
\tableofcontents
\newpage

\section{Introduction}

Symmetries have always been a determinant guide for the understanding of Nature, because they seem to enable the simultaneous concretization of the two ideals underlying the scientific quest: simplicity and beauty.\footnote{One may recommend the collection of inspiring lectures given by Chandrasekhar or the celebrated book of Weyl on this topic \cite{Chandrasekhar}.} The symmetry principles have been essential for the development of modern physics, \textit{e.g.} in the birth of both relativity theories or in the building of the standard model. The importance of symmetries has been recognized since the very beginning of scientific inquiry but mankind waited until the twentieth century for a new paradigm to emerge: the gauge symmetry principle.\footnote{The many developments of this crucial chapter in the history of physics are very well summarized in the book \cite{Raifeartaigh}.} The aphorism ``symmetry dictates interaction'' can be considered as the cornerstone of modern theoretical physics. Both in classical general relativity and in quantum field theory, the gauge symmetries are the deep geometrical foundations of fundamental interactions. Indeed, gauge symmetries determine the terms which may appear in the action. Nevertheless, some qualifications need to be made because, unfortunately, the symmetries rarely fix uniquely the interactions although this dream underlies most unification models.
Even though, at first sight, gauge transformations could have been naively dismissed as auxiliary -if not irrelevant- tools since they are in some sense ``unphysical,'' they actually proved to be almost unavoidable! For instance, from a field theoretical point of view the light-cone formulation is perfectly consistent by itself, but it is extremely convenient to introduce spurious unphysical (in other words, ``gauge'') degrees of freedom in order to write down Lagrangians for massless particles which are manifestly local and covariant under Lorentz transformations. Another example is general relativity where the decisive role played by the requirement of covariance under the diffeomorphisms does not need to be stressed, even though a superficial glance at this issue would dismiss this requirement as irrelevant since any theory can be formulated independently of the coordinate system by introducing an affine connection.\footnote{A student, scared by some of the conceptual subtleties arising from the gauge symmetry principle, could find some recomfort in the following surprising anecdote: during his quest for a reconciliation between gravity and relativity, Einstein himself initially argued that the equations of motion for the metric must \textit{not} be diffeomorphism covariant! The ``physical'' bases of this wrong initial requirement were related to the subtle issues mentioned above.}

Like for every deep and fundamental concept in physics, gauge symmetries exhibit many faces and can be approached in different ways.
The investigations of Dirac on the Hamiltonian formulation of gravity opened a new door for entering into the world of gauge theories. As explained in his seminal works \cite{Dirac:1950pj}, the presence of gauge symmetries in the Lagrangian framework implies, from a Hamiltonian point of view, the existence of ``constraints'' on the phase space variables. Conversely, the study of constraints in the Hamiltonian framework may serve as a path leading towards some understanding of gauge symmetries.
The present lecture notes are intended to be a self-contained introduction to the Hamiltonian formulation of systems with constraints.
Since the seminal investigations of Dirac, the development of this topic has been so dramatic that we would not pretend to be complete.
At most, we do hope that these notes might be useful to newcomers searching for a pedestrian and concrete approach on the interplay between Lagrangian \textit{vs} Hamiltonian systems from the point of view of gauge symmetries \textit{vs} constraints.
The main particularity of the present notes is that the (rigid and gauge) symmetries and their associated Noether charges are discussed with many details. For instance, the various possible definitions according to the choice of formalisms (Lagrangian, total or extended Hamiltonian) are introduced and compared with each other. Their explicit relationship is provided, due to its importance for applications. Another original feature is that the fermionic case is included in the presentation from the very beginning. This case is so relevant in physics that we found better to discuss the general case immediately in order to allow a uniform treatment of all physical cases, rather than devote later a specific section to this `particular' case.
We also insert all the details and proofs of the properties of supermatrices that are used in this text. More generally, pretty much all results presented here are given with their proofs, in order to be entirely self-contained. Nevertheless, we have not aimed at complete mathematical rigour (in the sense that all symbols written in the text are supposed to exist under suitable regularity conditions and that the formal manipulations they are subject to, are allowed). The major emphasis is on the classical level, though we provide some flavour of the quantization process at the end of these notes, for both first and second class constraints.

In contradistinction to most of the fundamental textbooks on the subject, such as \cite{Dirac,Hanson,Tyutin:1990,Henneaux:1992ig}, we focus on the total Hamiltonian instead of the extended Hamiltonian.\footnote{Of course, these textbooks do include very detailed discussions on the total Hamiltonian formalism, we only mean that it is not their chief emphasis and guideline. Of course, the formalism of constraints is discussed in many other textbooks, \textit{e.g.} \cite{Blago,Sardanachvily}.} On the one hand, the main advantage of this choice is that the dynamics determined by the former is always equivalent to the Lagrangian dynamics. On the other hand, its drawback is precisely that the primary constraints play a privileged role, while such a distinction is not relevant from a purely Hamiltonian perspective. Of course, the evolution of the physical quantities (\textit{i.e.} the observables) through the dynamics of either the Lagrangian, the total Hamiltonian or the extended Hamiltonian always agrees.
Therefore, the preference between total and extended Hamiltonian is somehow `philosophical', in the sense that it reflects the opinion whether, respectively, the Lagrangian formulation is more fundamental than the Hamiltonian one, or the contrary. Mathematically, one may argue that none of these opinions is more valid than the others because some Lagrangian systems do not allow an Hamiltonian formulation, and conversely. (Both types of examples are reviewed in this text.) Physically, the quantization process seems to rely heavily on the Hamiltonian formulation, even if Feynmann's path integral could plead in favour of the Lagrangian as well. Still, a rigorous definition of the path integral measure, \textit{etc}, seems to be more natural in terms of the phase space. Although the total Hamiltonian formulation is emphasized here because our perspective is on the relationship with the Lagrangian formulation, we will adopt an `oecumenic' attitude by discussing both approaches.\\

The plan of these notes is as follows:
The Lagrangian dynamics is reviewed in Section \ref{Lagrangiand} under complete generality, scrutinizing on various aspects of the symmetries of the action principle which are not always addressed in details (higher derivatives, finite \textit{versus} infinitesimal transformation, invertibility, \textit{etc}).
The canonical Hamiltonian formalism for a dynamical system with constraints is reviewed in Section \ref{HamLag} and the conditions of the equivalence between the Lagrangian and the Hamiltonian formalism are mentioned.
In the section \ref{TotaL}, the total and extended Hamiltonian dynamics are introduced together with the distinct types of constraints: primary or secondary, first or second class. Although these distinctions become somewhat irrelevant at some deeper level from the Hamiltonian point of view, they are very important when addressing the quantization process or the concrete relation between the Lagrangian and the Hamiltonian formulations.
The symmetries of dynamical systems and their associated conserved charges are discussed thoroughly in Section \ref{SymmetrY} from several perspectives. The equivalences between the many approaches are not shown through general theorems but through direct computations in order to provide for the reader a concrete grasp of the formalism.
The Dirac quantization, suitable for second class constraints and reviewed in Section \ref{Diracqu},
has a rather straight interpretation: eliminate the spurious degrees of freedom by making use of the Dirac bracket. The main drawback of this method is that, in general, one is unable to compute the Dirac bracket explicitly.
While the BRST quantization, suitable for first class constraints and presented in Section \ref{BRSTquant}, is more subtle conceptually (because it is formulated as a cohomological problem) and technically (because many fields such as ghosts, \textit{etc}, have to be added) it possesses at least one great virtue: if the Hamiltonian constraints (or the Lagrangian gauge symmetries) have been entirely determined, then this method can always be settled concretely in order to write down the gauge-fixed path integral, even if the BRST cohomology group cannot be computed explicitly. At the end come some appendices: some proofs of various properties are placed in the appendix \ref{someProofs} in order to lighten the core of the text. A rigorous treatment of the fermionic variables is provided in Appendix \ref{subsectionGrassmann} through a review of the Grassmann algebras, while all the necessary definitions and basic properties of supermatrices are provided in Appendix \ref{supermatrix}. The proofs of some propositions on the canonical forms of supermatrices are presented in details in the appendix \ref{lemmasupermatrix}. Finally, the appendix \ref{parexample} is devoted to a very simple and illustrative example of the general discussion contained in the body of these notes. We advise the reader to progressively go through this example, while (s)he goes through the general material in the core of the text.\\

These notes are an expanded version of some lectures given by JHP at Sogang University during the years 2007 and 2008\,.

\newpage
\section{Lagrangian dynamics: symmetry and Grassmann variables}\label{Lagrangiand}

\subsection{Euler-Lagrange equations}

We first consider a generic Lagrangian  depending on $\N$ variables,\footnote{For some issues in the continuous limit $\cal{N}\rightarrow\infty$, see \textit{e.g.} \cite{Sardanachvily}.} $q^{A}$, $1\leq A\leq \N$, their time derivatives, and time is allowed to appear explicitly,
\be
\L(q_{n},t)\,,
\ee
where
\be
\ba{ll}
q^{A}_{n}=\D{\left(\frac{{\rm d}}{{\rm d}t}\right)^{\!n}q^{A}}\,,~~~~&~~~~n=0,1,2,3,\cdots\,.
\ea
\ee
The $q_{n}^{A}$'s form the coordinates of the so called ``jet space''. Note that some of the variables can be fermionic.\footnote{One of the only prerequisite of these lectures is that the reader is supposed to be familiar with graded algebras and related \textit{super} objects. A good self-contained introduction to supersymmetry is \cite{Buchbinder:1995uq}.}
In our conventions, unless explicitly mentioned, all derivatives act from the left to the right,
\be
\D{\frac{\partial F}{\partial q^{A}}=\frac{\overrightarrow{\partial} F}{\partial q^{A}}=
(-1)^{\#_{A}(\#_{F}+\#_{A})}\frac{\overleftarrow{\partial} F}{\partial q^{A}}\,,}
\label{LRD}
\ee
where $\#_{A}$ is the  ${\mathbb Z}_{2}$-grading of the $2\N$-dimensional tangent  space with coordinates $(q^{A}, \dq^{B})$,
\be
\#_{A}=\left\{\ba{ll} 0~&~~\mbox{for~bosonic~}~A\\
1~&~~\mbox{for~fermionic~}A\ea\right.\,.
\ee

From the variation of the action
\begin{equation}
S[\,q^A]=\int \L(q_{n},t)\,dt\,,
\end{equation}
and up to the boundary terms, one obtains
\be
\D{\int {\rm d}t~\delta\L(q_{n},t)=\int {\rm d}t~\delta q^{A}
\left[\sum_{n=0}^{\infty}\,\left(-\frac{{\rm d}}{{\rm d}t}\right)^{\!n}\frac{\partial\,}{\partial q^{A}_{n}}\right]\L(q_{m},t)\,,}
\label{deltaL}
\ee
so that the corresponding Euler-Lagrange equations
\be
\frac{\delta\L(q_{m},t)}{\delta q^{A}}\equiv 0\,, \label{EOML}
\ee
are given by acting on the Lagrangian with a linear differential operator called the Euler-Lagrange operator,\footnote{For the fermionic degrees of freedom, there  arises a subtle point which we discuss in Appendix \ref{subsectionGrassmann}.}
\be
\frac{\delta~~}{\delta q^{A}}:=\sum_{n=0}^{\infty}\,\left(-\frac{{\rm d}}{{\rm d}t}\right)^{\!n}\frac{\partial\,}{\partial q^{A}_{n}}\,.
\label{EL}
\ee
In the jet space, the submanifold defined by Eq.(\ref{EOML}) is called the ``stationary (hyper)surface''.
We have introduced the symbol $\equiv$ which will stand, from now on, for `equal on the stationary surface' or, equivalently, `equal modulo the Lagrangian equations of motion (\ref{EOML})'.
Remark that the Euler-Lagrange operator (\ref{EL}) is not a derivation, \textit{i.e.} it does not obey to the Leibniz rule.

It is useful to understand that in the jet space the total derivative ${\rm d}/{\rm d}t$ is defined as
\be
\D{\frac{{\rm d}}{{\rm d}t}\,=\,\frac{\partial}{\partial t}\,+\,\sum_{n=0}^{\infty}\,q_{n+1}^{A}\frac{\partial\,}{\partial q^{A}_{n}}\,,}
\ee
so that
\be
\ba{lll}
\D{\frac{\partial}{\partial q^{A}}\frac{{\rm d}}{{\rm d}t}=\frac{{\rm d}}{{\rm d}t}\frac{\partial}{\partial q^{A}}\,,}~~~~~&~~~~\D{\frac{\partial\,}{\partial q^{A}_{n}}\frac{{\rm d}}{{\rm d}t}=\frac{{\rm d}}{{\rm d}t}\frac{\partial\,}{\partial q^{A}_{n}}+\frac{\partial\,}{\partial q^{A}_{n-1}}\,,}~~&~~n\geq 1\,.
\ea
\label{dtq}
\ee
This implies that {\it{the Euler-Lagrange equations of a total derivative term are identically vanishing}},
\be
\D{\frac{\delta~~}{\delta q^{A}}
\left(\frac{{\rm d}K}{{\rm d}t}\right)=0\,.}
\label{totaltrivial}
\ee
~\\
The converse is also true, therefore
\be
\D{\frac{\delta F}{\delta q^{A}}
(q_{n},t)=0~~~\Longleftrightarrow~~~  F(q_{n},t)=\frac{{\rm d}K(q_{n},t)}{{\rm d}t}}\,\,.
\label{FdK}
\ee
We present a proof of these statements in Appendix \ref{someProofs}. Note that a generalization of Eq.(\ref{FdK}) holds for field theories ($\N=\infty$) too, in which case it is referred to as ``algebraic Poincar\'e lemma''.\footnote{For more details on this lemma, on jet space, \textit{etc}, see \textit{e.g.} the section 4 of \cite{Barnich:2000zw} and references therein.}\\

\subsection{Symmetries in the Lagrangian formalism}\label{subsecPRE}

In general, a symmetry of the action involves a certain change of variables,
\be
q^{A}~~\longrightarrow~~ {\q}^{A}(q_{n},t)\,,
\label{symt}
\ee
which may explicitly depend on the $q_{n}^{B}$'s and the time $t\,$.  It corresponds to a symmetry of the action if the Lagrangian is invariant under the transformation up to total derivative terms,
\be
\L(\q_{n},t)=\L(q_{n},t)+\D{\frac{{\rm d}}{{\rm d}t}K(q_{n},t)\,,}
\label{symmetry}
\ee
where
\be
{\q}^{A}_{n}=\D{\left(\frac{{\rm d}}{{\rm d}t}\right)^{\! n}{\q}^{A}=\left(\frac{\partial}{\partial t}+\sum_{m=0}^{\infty}\,q_{m+1}^{B}\frac{\partial\,}{\partial q^{B}_{m}}\right)^{\! n}{\q}^{A}(q_{l},t)\,.}
\ee
Surely this imposes  nontrivial conditions on the form of $\q(q_{n},t)$ in terms of the Lagrangian, $\L(q_{n},t)$.   \\

One useful identity for an arbitrary function $F$ is\footnote{See Eq.(\ref{usefulQqproof}) for a proof.}, for $m\geq 1$,
\be
\D{\sum_{n=0}^{\infty}\,\left(-\frac{{\rm d}}{{\rm d}t}\right)^{\! n}\left(
\frac{\partial \q_{m}^{B}}{\partial q^{A}_{n}}\,F\right)}=\D{\sum_{n=0}^{\infty}\,\left(-\frac{{\rm d}}{{\rm d}t}\right)^{\! n}\left[\frac{\partial \q_{0}^{B}}{\partial q^{A}_{n}}\left(-\frac{{\rm d}}{{\rm d}t}\right)^{\! m}F\right]\,,}
\label{usefulQq}
\ee
which gives the following algebraic identity for any change of variables,
\be
\D{\frac{\delta\L(\q_{l},t)}{\delta q^{A}}}=\D{\sum_{n=0}^{\infty}\left(-\frac{{\rm d}}{{\rm d}t}\right)^{\!n}\left(
\sum_{m=0}^{\infty}\,\frac{\partial\q_{m}^{B}}{\partial q_{n}^{A}}\,\frac{\partial\L(\q_{l},t)}{\partial \q_{m}^{B}}\right)}=
\D{\sum_{n=0}^{\infty}\,\left(-\frac{{\rm d}}{{\rm d}t}\right)^{\! n}\left(
\frac{\partial \q_{0}^{B}}{\partial q^{A}_{n}}\frac{\delta\L(\q_{l},t)}{\delta \q^{B}}\right)\,.}
\label{idEEp}
\ee
Infinitesimally,  $q_{0}^{A}~\rightarrow~q_{0}^{A}+\delta q_{0}^{A}(q_{n},t)$, Eq.(\ref{idEEp}) implies
\be
\D{\left[\frac{\delta~}{\delta q^{A}}\,,\,\delta\,\right]\L(q_{l},t)}
=
\D{\sum_{n=0}^{\infty}\,\left(-\frac{{\rm d}}{{\rm d}t}\right)^{\! n}\left(
\frac{\partial (\delta q_{0}^{B})}{\partial q^{A}_{n}}\frac{\delta\L(q_{l},t)}{\delta q^{B}}\right)\,,}
\ee
{since}
\begin{eqnarray}
&&\Big(\D{\frac{\delta\L(\q_{l},t)}{\delta q^{A}}}-\D{\frac{\delta\L(q_{l},t)}{\delta q^{A}}}\Big)
-\Big(\D{\frac{\delta\L(\q_{l},t)}{\delta \q^{A}}}-\D{\frac{\delta\L(q_{l},t)}{\delta q^{A}}}\Big)\nonumber\\
&&=\D{\sum_{n=0}^{\infty}\,\left(-\frac{{\rm d}}{{\rm d}t}\right)^{\! n}\left(
\frac{\partial \q_{0}^{B}}{\partial q^{A}_{n}}\frac{\delta\L(\q_{l},t)}{\delta \q^{B}}\right)\,\,-\,
\D{\frac{\delta\L(\q_{l},t)}{\delta \q^{A}}}\,.}
\end{eqnarray}
Eq.(\ref{idEEp}) will be used later in order to show that the Eler-Lagrange equations are preserved by symmetries of the action, a fact which is natural to expect but is non trivial to prove.

Now assuming the symmetry (\ref{symt}), acting with the Euler-Lagrange operator on both sides of (\ref{symmetry}), using (\ref{totaltrivial}) and (\ref{idEEp}), we get
\be
\D{\frac{\delta\L(q_{l},t)}{\delta q^{A}}=\frac{\delta\L(\q_{l},t)}{\delta q^{A}}=
\D{\sum_{n=0}^{\infty}\,\left(-\frac{{\rm d}}{{\rm d}t}\right)^{\! n}\left(
\frac{\partial \q_{0}^{B}}{\partial q^{A}_{n}}\frac{\delta\L(\q_{l},t)}{\delta \q^{B}}\right)\,.}}
\label{idEE}
\ee

Before we discuss the generic cases, we first consider the simple case where  $\q_{0}^{B}(q,t)$ depends on $q$ and $t$ only, being independent of  $q_{n}$ ($n\geq 1$) and  invertible \textit{i.e.}  $\det(\partial \q/\partial q){\neq 0}$. We have
\be
\D{\frac{\delta\L(\q_{l},t)}{\delta\q^{A}}=\frac{\partial q^{B}(\q,t)}{\partial \q^{A}}\, \frac{\delta\L(q_{l},t)}{\delta q^{B}}\,,}
\label{qQEOM}
\ee
using (\ref{idEE}) with $q$ and $\q$ exchanged.
Hence, if  $q(t)$ is a solution of the equations of motion, then so is $\q(q,t)$, where $\q(q,t)$ depends on $q$ and $t$ only, \textit{i.e.} not on the derivatives $q_{n}$ ($n\geq 1$).

Now, for the generic cases where $\q(q_{n},t)$ depends on the $q_{n}$'s ($n\geq 0$): \textit{if there exists an inverse map} $q(\q_{n},t)$ - most likely  depending on the infinite set of variables\footnote{For example, the usual translational symmetry reads
\[
\ba{lll}
\D{\q^{A}(t)=q^{A}(t+a)=\sum_{n\geq 0}\frac{~a^{n}}{n!}\,q^{A}_{n}(t)}~~&~~\Longleftrightarrow
~~&~~ \D{q^{A}(t)=\q^{A}(t-a)=\sum_{n\geq 0}\frac{(-a)^{n}}{n!}\,\q^{A}_{n}(t)\,.}\ea\]}
$q_{n}$ ($n= 0,1,2,\cdots$) - \textit{then} the (inverse relation of) Eq.(\ref{idEE}) indeed shows that
{\textit{if $q(t)$ is a solution of the equations of motion, then so is $\q(q_{n},t)$.}}  \\

\subsubsection{Invertible transformations}

The existence of an inverse map is always guaranteed when there exists a corresponding infinitesimal transformation,
\be
\ba{ll}
q^{A}~\rightarrow ~q^{A}+\delta q^{A}\,,~~~&~~~
\delta q^{A}=f^{A}(q_{n},t)\,.
\ea
\ee
Consequently, the infinitesimal transformation of the coordinates of the jet space are given by $q_{n}^{A}~\rightarrow ~q_{n}^{A}+\delta q_{n}^{A}\,$, where
\be
\delta q_{n}^{A}=\D{\left(\frac{{\rm{d}}\,}{{\rm{d}t}}\right)^{n}f^{A}(q_{m},t)=
\left(\frac{\partial}{\partial t}+\sum_{l=0}^{\infty}\,q_{l+1}^{B}\frac{\partial\,}{\partial q^{B}_{l}}\right)^{n}f^{A}(q_{m},t)=:
f_{n}^{A}(q_{m},t)}\,.
\ee
More explicitly, we define an exponential map with a real parameter $s$,
\be
\ba{ll}
\q^{A}(s,q_{n},t)=\D{\exp\left(s \sum_{l=0}^{\infty}\,f^{B}_{l}(q_{m},t)\frac{\partial~}{\partial q_{l}^{B}}\right)q^{A}}\,,~~~&~~~\q^{A}(0,q_{n},t)=q^{A}\,.
\ea
\ee
From (\ref{dtq}) we first note the commutativity  property,
\be
\D{\frac{{\rm{d}}\,}{{\rm{d}t}}\left(\sum_{l=0}^{\infty}\,f^{B}_{l}\frac{\partial~}{\partial q_{l}^{B}}\right)
=
\sum_{l=0}^{\infty}
\left(
f^{B}_{l+1}\frac{\partial~}{\partial q_{l}^{B}}+f^{B}_{l}
\frac{{\rm{d}}\,}{{\rm{d}t}}\frac{\partial~}{\partial q_{l}^{B}}
\right)
=
\left(
\sum_{l=0}^{\infty}\,f^{B}_{l}\frac{\partial~}{\partial q_{l}^{B}}\right)\frac{{\rm{d}}\,}{{\rm{d}t}}\,,}
\label{comm1}
\ee
and hence,
\be
\q^{A}_{n}(s,q_{m},t)=\D{\left(\frac{{\rm{d}}\,}{{\rm{d}t}}\right)^{n}\q^{A}(s,q_{m},t)=
\exp\left(s \sum_{l=0}^{\infty}\,f^{B}_{l}(q_{m},t)\frac{\partial~}{\partial q_{l}^{B}}\right)q_{n}^{A}}\,.
\label{comm2}
\ee
~\newline

The main claim is then,
\be
\D{\frac{{\rm{d}}\q^{A}_{n}\,}{{\rm{d}s}}=\sum_{l=0}^{\infty}\,f^{B}_{l}(q_{m},t)\frac{\partial\q^{A}_{n}}{\partial q_{l}^{B}}=f^{A}_{n}(\q_{m},t)}\,.
\label{mainc1}
\ee
From (\ref{comm2}), the first equality in (\ref{mainc1}) is obvious.  The derivation of the other relation is carried in Eq.(\ref{usefulQqproof2}) of the appendix.
The equation (\ref{mainc1}) implies  that the following differential operator is `$s$'-independent,
\be
\D\sum_{l=0}^{\infty}\,{f^{B}_{l}(q_{m},t)\frac{\partial~}{\partial q_{l}^{B}}=
\sum_{l=0}^{\infty}\,f^{B}_{l}(\q_{m},t)\frac{\partial~}{\partial \q_{l}^{B}}}\,.
\ee
Using the above identities, it is   straightforward to  obtain  the explicit inverse map,  $\q^{A}\rightarrow q^{A}$,
\be
\D{q^{A}(s,\q_{n},t)=\D{\exp\left(-s f^{B}_{l}(\q_{m},t)\frac{\partial~}{\partial \q_{l}^{B}}\right)\q^{A}}\,.}
\ee

\subsubsection{Local symmetries}\label{gsym}

In the case of a ``local symmetry", namely if the transformation involves some arbitrary time dependent functions $\alpha^{i}(t)$ as
\be
\ba{ccc}
q^{A}~&~\longrightarrow~&~ \q^{A}\Big(q_{n},t,\alpha(t)\Big)
\,,
\ea
\ee
it is possible to have `different' solutions by varying the arbitrary functions $\alpha_{i}(t)\,$, even though one starts from the same initial data $\q_n(t_{0})\,$. However, one may consider that the Lagrangian \textit{alone} dictates the whole dynamics of the given system (and nothing else) and, furthermore,  that the dynamics is deterministic (\textit{i.e.} there is a unique solution to the Cauchy problem). As long as one takes this viewpoint for granted, then one must regard different trajectories as the \textit{same physical state}.\\

Namely \textit{any local symmetry must be a `gauge' (i.e. unphysical) symmetry. In mathematical terms, a physical state is given by an equivalence class for which the local symmetry defines the equivalence relation.}\footnote{A complete and detailed treatment of the gauge invariance of an action can be found in the chapter 3 of \cite{Henneaux:1992ig}.} Obviously, the presence of the gauge symmetries complicates the correct counting\footnote{The Hamiltonian framework enables a precise and `algorithmic' computation of the number of degrees of freedom, which leads to a precise and completely general criterion for counting the physical degrees of freedom directly from the form of gauge transformations in the Lagrangian formalism itself \cite{Henneaux:1990au}. This rigorous treatment clarifies the origin of some maxims from the physicist folklore (such as ``gauge shoots twice'') and thereby provides a supplementary argument in favor of the fruitful interplay between Hamiltonian and Lagrangian formalisms.} of the number of physical degrees of freedom (especially if the gauge symmetries are not independent, \textit{etc}). An ``observable'' quantity is a function on the space of physical states, hence it must be gauge invariant. We will turn back to these issues in the Hamiltonian context.

\subsection{Second order field equations}

Henceforth, we focus  on   the standard Lagrangian, $L(q,\dq,t)$,  which depends on $q^{A}$, $\dq^{A}$, $t$ only, and derive some algebraic identities for the later use. \\

When the infinitesimal symmetry transformation, $q^{A}\rightarrow q^{A}+\delta q^{A}(q,\dq,t)$, also depends  on $q^{A}$, $\dq^{A}$, $t$ only,  we have\footnote{Actually, in the case of a regular second-order Lagrangian, this can be assumed without loss of generality, as explained in the exercise 3.8 of the book \cite{Henneaux:1992ig}.}
\be
\D{\delta\dot{q}^{A}=\frac{{\rm d}}{{\rm d}t}\delta q^{A}(q,\dq,t)=\ddot{q}^{B}\frac{\partial (\delta q^{A})}{\partial \dq^{B}}+\dq^{B}\frac{\partial (\delta q^{A})}{\partial q^{B}}+\frac{\partial (\delta q^{A})}{\partial t}\,.}
\label{deltadotq}
\ee
Thus, the function $\delta K(q,\dq,t)$ in
\be
\delta L=\frac{{\rm d}~}{{\rm d}t}\delta K
\label{deltaLK}
\ee
must depend on $q^{A}$, $\dq^{A}$, $t$ only too, as
\be
\ba{ll}
\D{\delta L}&=\D{\delta q^{A}\frac{\partial L(q,\dq,t)}{\partial q^{A}}+\delta\dq^{A}\frac{\partial L(q,\dq,t)}{\partial \dq^{A}}}\\
{}&{}\\
{}&=\D{\delta q^{A}\frac{\partial L(q,\dq,t)}{\partial q^{A}}+\left(\ddot{q}^{B}\frac{\partial (\delta q^{A})}{\partial \dq^{B}}+\dq^{B}\frac{\partial (\delta q^{A})}{\partial q^{B}}+\frac{\partial (\delta q^{A})}{\partial t}\right)\frac{\partial L(q,\dq,t)}{\partial \dq^{A}}}\\
{}&{}\\
{}&=\D{\frac{{\rm d}~}{{\rm d}t}\delta K}=\D{\ddot{q}^{A}\frac{\partial (\delta K)}{\partial \dq^{A}}+\dot{q}^{A}\frac{\partial (\delta K)}{\partial q^{A}}+\frac{\partial(\delta K)}{\partial t}}\,.
\ea
\ee
This implies
\be
\D{\frac{\partial (\delta q^{B})}{\partial \dq^{A}}\frac{\partial L(q,\dq,t)}{\partial \dq^{B}}=\frac{\partial (\delta K)}{\partial \dq^{A}}}\,,
\label{KL1}
\ee
and
\be
\D{\delta q^{B}\frac{\partial L}{\partial q^{B}}+\left(\dq^{C}\frac{\partial (\delta q^{B})}{\partial q^{C}}+\frac{\partial (\delta q^{B})}{\partial t}\right)\frac{\partial L}{\partial \dq^{B}}}=\D{\dot{q}^{B}\frac{\partial (\delta K)}{\partial q^{B}}+\frac{\partial (\delta K)}{\partial t}}\,.
\label{KLqdot}
\ee
By taking the partial derivative of the latter equation with respect to $\dot{q}^{A}\,$, making use firstly of Eq.(\ref{deltadotq}) and secondly of Eq.(\ref{KL1}), we get
\be
\ba{l}
\D{\frac{\partial(\delta q^{B})}{\partial \dq^{A}}\frac{\partial L}{\partial q^{B}}+
\delta q^{B}\frac{\partial^{2} L}{\partial q^{B}\partial\dq^{A}}+
\delta\dq^{B}\frac{\partial^{2}L}{\partial \dq^{B}\partial\dq^{A}}}
+\D{\left(\frac{\partial(\delta q^{B})}{\partial q^{A}}+\dq^{C}\frac{\partial^{2}(\delta q^{B})}{\partial q^{C}\partial\dq^{A}}+\frac{\partial^{2}(\delta q^{B})}{\partial\dq^{A}\partial t}\right)\frac{\partial L}{\partial\dq^{B}}}\\
{}\\
=\D{\frac{\partial(\delta K)}{\partial q^{A}}+\dq^{B}\frac{\partial^{2}(\delta K)}{\partial q^{B}\partial\dq^{A}}+\frac{\partial^{2}(\delta K)}{\partial \dq^{A}\partial t}+\ddot{q}{}^{\,C}\frac{\partial (\delta q^{B})}{\partial \dq^{C}}\frac{\partial^{2}L}{\partial \dq^{B}\partial\dq^{A}}}\\
{}\\
=\D{\frac{\partial(\delta K)}{\partial q^{A}}+\left(\dq^{B}\frac{\partial~}{\partial q^{B}}+\frac{\partial~}{\partial t}\right)\left(\frac{\partial (\delta q^{C})}{\partial \dq^{A}}\frac{\partial L}{\partial \dq^{C}}\right)+\ddot{q}{}^{\,C}\frac{\partial (\delta q^{B})}{\partial \dq^{C}}\frac{\partial^{2}L}{\partial \dq^{B}\partial\dq^{A}}\,.}
\label{usedscdl}\ea
\ee
From Eq.(\ref{KL1}) we {find that the coefficient of $\ddot{q}{}^{\,C}$ must vanish, which implies} an integrability condition:
\be
\ba{l}
\D{\frac{\partial(\delta q^{B})}{\partial\dq^{C}}\,
\frac{\partial^{2}L}{\partial\dq^{B}\partial\dq^{A}}-(-1)^{\#_{A}\#_{C}}
\frac{\partial(\delta q^{B})}{\partial\dq^{A}}\,
\frac{\partial^{2}L}{\partial\dq^{B}\partial\dq^{C}}}\\
{}\\
\D{=(-1)^{\#_{A}\#_{C}}
\frac{\partial~}{\partial\dq^{A}}\left(\frac{\partial(\delta q^{B})}{\partial\dq^{C}}\frac{\partial L}{\partial\dq^{B}}\right)-
\frac{\partial~}{\partial\dq^{C}}\left(\frac{\partial(\delta q^{B})}{\partial\dq^{A}}\frac{\partial L}{\partial\dq^{B}}\right)=0\,.}
\ea
\label{ddqL}
\ee
{The transformation of the `momenta' ${\partial L}/{\partial\dq^A}$ is given by}
\be
\ba{ll}
\D{\delta\left(\frac{\partial L(q,\dq,t)}{\partial\dq^{A}}\right)}
&\D{=\delta q^{B}\frac{\partial^{2} L}{\partial q^{B}\partial\dq^{A}}+
\delta\dq^{B}\frac{\partial^{2}L}{\partial \dq^{B}\partial\dq^{A}}}\\
{}&{}\\
{}&\D{=\frac{\partial(\delta K)}{\partial q^{A}}-\frac{\partial(\delta q^{B})}{\partial q^{A}}
\frac{\partial L}{\partial\dq^{B}}}+\D{\frac{\partial (\delta q^{B})}{\partial\dq^{A}}\left[
\frac{{\rm{d}\,}}{{\rm{d}t}}\left(\frac{\partial L}{\partial\dq^{B}}\right)-\frac{\partial L}{\partial q^{B}}\right]\,,}
\ea
\label{deltap0}
\ee
{where Eqs. (\ref{usedscdl}) and (\ref{ddqL}) have been used in the derivation at the second line.}
Finally, for the transformation of the Hamiltonian we get
\be
\D{\delta\!\left(\dq^{A}\frac{\partial L(q,\dq,t)}{\partial \dq^A}-L(q,\dq,t)\right)
=\dq^{A}\frac{\partial (\delta q^{B})}{\partial\dq^{A}}
\left[
\frac{{\rm{d}\,}}{{\rm{d}t}}\left(\frac{\partial L}{\partial\dq^{B}}\right)-\frac{\partial L}{\partial q^{B}}\right]+\frac{\partial(\delta q^{A})}{\partial t}\left(\frac{\partial L}{\partial\dq^{A}}\right)-\frac{\partial(\delta K)}{\partial t}}\,.
\label{Htransf}
\ee
These relations will be used later in Section \ref{preSTH} where we  analyze  the symmetries in the Hamiltonian formalism.\\

The corresponding Noether charge is defined by
\be
\D{Q=\delta q^{A}\frac{\partial L}{\partial \dq^{A}}-\delta K\,.}
\label{Noether}
\ee
Up to the Euler-Lagrange  equation,
\be
\D{\frac{\partial L}{\partial q^{A}}\equiv\frac{{\rm d}\,}{{\rm d}t}\left(\frac{\partial L}{\partial \dq^{A}}\right)\,,}
\label{ELeq}
\ee
the Noether charge is conserved,
\be
\D{\frac{{\rm d}Q}{{\rm d}t}=\delta q^{A}\left[\frac{{\rm d}\,}{{\rm d}t}\left(\frac{\partial L}{\partial \dq^{A}}\right)-\frac{\partial L}{\partial q^{A}}\right]\equiv 0\,,}
\label{Noetherconserved}
\ee
due to Eq.(\ref{deltaLK}).
Further discussions on symmetries in the Lagrangian dynamics are carried out in Section \ref{preSTH}.


\newpage
\section{From Lagrangian to Hamiltonian and vice versa}\label{HamLag}

\subsection{Canonical momenta}

Given a standard Lagrangian $L(q,\dq,t)\,$, depending on $\N$ bosonic or fermionic variables $q^{A}$ (with $1\leq A\leq \N$) and their first time derivatives (and, possibly, on time as as well), the equations of motion read
\be
\D{\frac{\partial L}{\partial q^{A}}\equiv\frac{{\rm d}\,}{{\rm d}t}\left(\frac{\partial L}{\partial \dq^{A}}\right)=\ddot{q}^{B}\frac{\partial^{2} L}{\partial \dq^{B}\partial \dq^{A}}+
\dot{q}^{B}\frac{\partial^{2} L}{\partial q^{B}\partial \dq^{A}}}+
{\frac{\partial^{2} L}{\partial t\partial \dq^{A}}}\,.
\label{EOM}
\ee
If the $\N\times \N$ supermatrix, ${\partial^{2} L}/\partial \dq^{B}\partial \dq^{A}$, is nondegenerate,  all the $\ddot{q}^{A}$'s are uniquely determined by $q$ and $\dq$. Namely all the variables are completely  determined by the initial data, and also all the  `velocities'  $\dq^{A}$  may be expressed in terms of  $q$ and the canonical momenta $p_A:=\partial L/\partial \dq^{A}\,$.

Henceforth, we focus on the degenerate case,
\be
\D{\mbox{sdet}\!\left(\frac{\partial^{2} L}{\partial \dq^{B}\partial \dq^{A}}\right)=0\,.}
\ee
In a large class of examples, it may still possible that all the variables are uniquely determined from the initial data through the equations of motion, \textit{e.g.} for a Lagrangian $L(q,\dq,t)=\omega_{AB}\dq^{A}q^{B}-V(q,t)$, linear in the `$\dq^{A}$', and where the constant graded symmetric matrix, $\omega_{[AB\}}=\omega_{AB}$, is nondegenerate. The (anti)symmetrization has weight one, \textit{i.e.} \be \omega_{[AB\}}:=\frac12\left(\omega_{AB}-(-)^{\#A\#B}\omega_{BA}\right)\,.\ee We will not attempt to analyze and classify all
cases here  in the Lagrangian formalism, but we will do so in the Hamiltonian formalism  later.

\subsection{Primary constraints}

Now, let us start from the expressions for the $\N$  momenta in terms of $q,\dq,t$,
\be
\ba{ll}
p_{A}=\D{\frac{\partial L(q,\dq,t)}{\partial \dq^{A}}}=f_{A}(q,\dq,t)~~~~~&~~~~~A=1,2,\cdots,\N\,,
\ea
\label{ep}
\ee
and try to invert the map in order to express the velocities $\dq^{A}$ in terms of  $q, t$ and the momenta $p$.

We first consider a bosonic system having bosonic variables only.
If one of the $\N$ momenta (\ref{ep}), say $p_a$, depends nontrivially on a certain velocity, say  $\dq^{\hat a}$, then this velocity can be expressed in terms of $p_a$ and the remaining velocities, collectively denoted by $\dq^{\hat m}$, as well as $q^A$ and $t\,$. Then, substituting the expression
\be
\dq^{\hat a}=h^{\hat a}(q^A\,,\,p_{a}\,,\,\dq^{\hat m}\,,\,t)\,,
\label{hhat}
\ee
into the other momenta than $p_a\,$, collectively denoted by $p_m\,$, we get
\be
p_m=g_m(q^A\,,\,p_{a}\,,\,\dq^{\hat m}\,,\,t)\,.
\ee
This procedure can be repeated until the expressions for the momenta $p_m$ do not depend on any of the velocities. Performing the procedure \textit{step by step} a finite number of times, we get finally
\be
\ba{ll}
\dq^{\hat a}=h^{\hat a}(q^A\,,\,p_{a}\,,\,\dq^{\hat m}\,,\,t)\,,~~~~&~~~~
 p_m=f_m(q^A\,,\,p_a\,,\,t)\,,
\ea
\label{primary1}
\ee
where the {sets of momenta and velocities split into two disjoint groups each,}
\be
\ba{lll}
p_{A}=(\,p_a,\,p_m\,)~~&:~~~\{a\}\cup\{m\}=\{1,2,\cdots,\N\}\,,~~&\{a\}
\cap\{m\}=\emptyset\,,\\
{}&{}&{}\\
\dot{q}^{A}=(\,\dot{q}^{\hat a},\,\dot{q}^{\hat m}\,)~~&:~~~
\{{\hat a}\}\cup\{\hat{m}\}=\{1,2,\cdots,\N\}\,,~~&\{\hat{a}\}
\cap\{\hat{m}\}=\emptyset\,,
\ea
\ee
and in particular our procedure defines a one-to-one correspondence of $\{a\}\leftrightarrow \{\hat{a}\}$.\footnote{They are distinguished because  $p_{a}$ is not necessarily  the conjugate momentum of $\dq^{\hat a}$.} {In words, on one side the velocities have hatted indices, on the other side the momenta have unhatted indices. For the velocities, the $\hat{m}$'s correspond to the velocities which remain independent while the $\hat{a}$'s correspond to the velocities which are determined in terms of the former  velocities and momenta. For the momenta, the situation is opposite: the $a$'s correspond  to the ones which are independent while the $m$'s correspond to the momenta which are expressed in terms of the latter momenta.}

Notice that $p_m(q,\dq)=f_m\Big(q^A,p_a(q,\dq),t\Big)$ are identities on the tangent space of coordinates $(q^A,\dq^B)$ and that there exists an  invertible map between the momenta $p_a$ and the velocities $\dq^{\hat a}$ (keeping $q^A$, $\dq^{\hat m}$ and $t$ fixed),
\be
\ba{lll}
\dq^{\hat a}=h^{\hat a}(q^A\,,\,p_{a}\,,\,\dq^{\hat m}\,,\,t)~~~
&\Longleftrightarrow&~~~p_a=f_a(q^B\,,\,\dq^{A}\,,\,t)\,.
\ea
\label{qdotp}
\ee
The latter means that
\be
\D{\det\!\left(\frac{\partial p_a(q,\dq,t)}{\partial \dq^{\hat b}}\right)\neq 0\,,}
\ee
thus the rank of $\partial p_{A}/\partial\dq^{B}$ is equal\footnote{{This fact is an alternative starting point for getting the constraints and the decomposition of the indices in disjoint set. We prefered to provide a concrete explanation of the result (\ref{primary1}) instead of a slightly more abstract one in terms of the rank of the (super)Jacobian \textit{via} the implicit function theorem.}} to the dimension of the set $\,\{a\}$. Also, one may say that there is a one-to-one map\footnote{See (\ref{11maps}) for a more general result.}
\be
\left(\ba{l}
q^{A}\\
{}\\
\dq^{B}\ea\right)~\longleftrightarrow~
\left(\ba{l}q^{A}\\
{}\\
p_a\\
{}\\
\dq^{\hat m}\ea\right)
\,.
\ee
\\

Now we return to the generic systems having both bosons and fermions. In contrast to the bosonic system, the above procedure which expresses the velocities in terms of the momenta may not work even if the momenta depend on velocities nontrivially, mainly due to the non-existence of an inverse for any fermionic variable. We consider an example,
\be
L\,=\,i\,\dot{\theta}\,\theta\,\dot{x}\,,
\ee
which gives $p_{\theta}=i\,\theta\dot{x}$ and $p_{x}=i\,\dot{\theta}\theta$. When $\theta$ is fermionic while $x$ is bosonic (which is  the case with the usual notations) none of the expressions can be inverted. Furthermore, the corresponding hamiltonian reads $H=i\,\dot{\theta}\theta\dot{x}=L$ which again cannot be reexpressed by the momenta and coordinates only.\footnote{One possible way to circumvent the obstacle is to employ explicitly the Grassmann algebra basis and  work strictly with the real number coefficients, namely  $\left[ q^{A}\right]_{\!J}$\,, $~\left[ p_{A}\right]_{\!J}$\,,  as discussed in Appendix \ref{subsectionGrassmann}. One can then apply the above procedure in the bosonic system without any problem until one gets a similar expression to (\ref{primary1}). However, the corresponding Hamiltonian dynamics for all the coefficients, especially the Poisson bracket, will have to be decomposed into a complicated expression, and this will not be done here. We will always assume that the expressions of the constraints do not need any explicit use of the Grassmann algebra basis.} If $\theta$ were bosonic, then $H=-ip_{\theta}p_{x}\theta^{-1}$.  In the present paper we do not consider this case. We always assume that when the the momenta depend on velocities nontrivially, one can always obtain the inverse function until we achieve the expression (\ref{primary1}).  Namely we will restrict\footnote{{If the constraints are not independent from each other, they are said to be ``reducible.'' This more general case is treated in the section 1.3.4 of \cite{Henneaux:1992ig}.}} our analysis to systems of bosons and fermions, where the $2\N$ coordinates $(p_{A},q^{B})$ of the phase space ($1\leq A,B\leq \N$), are subject to $\M$ functionnally independent constraints,
\be
\ba{ll}
\phi_{m}(p,q,t)=0\,,~~~~&~~~~1\leq m\leq \M\,.
\ea
\label{prconstrts}
\ee
The $\M$ primary constraints are independent in the sense that the following $\M$ vectors are linearly independent,
\be
\ba{ll}
\left.{\vec{\partial}}_{p}{}\phi_{m}\right|_{V}=\D{\left.\left(\frac{\partial \phi_{m}}{\partial p_{1}},\,\frac{\partial \phi_{m}}{\partial p_{2}},\,\cdots,\,\frac{\partial \phi_{m}}{\partial p_{\N}}\right)\right|_{V}}\,,~~~~&~~~~1\leq m\leq \M\,,
\ea
\ee
where $|_V$ means that the left-hand-side is evaluated on the hypersurface defined by the system (\ref{prconstrts}), after taking the partial derivatives.
The implicit function theorem actually ensures that it is possible to solve Eq.(\ref{prconstrts}) for $\M$ of the momentas, as in Eq.(\ref{primary1}).
Also, the $\M$ primary constraints naturally define  a $(2\N-\M)$-dimensional hypersurface $V$ in the phase space, called the ``primary constraint (hyper)surface'',
\be
V=\{(p,q)\,|\,\phi_{m}(p,q,t)=0,~~1\leq m\leq \M\}\,.
\label{hyperV}
\ee
We furthermore assume that at fixed time $t$ all the constraints can in principle be solved to express any point on $V$ by  $2\N-\M$ independent variables $x^{i}$ ($1\leq i\leq 2\N-\M$),
\be V=\left\{\,(p\,,\,q)=f(x,t)\,\right\}\,,
\label{onV}\ee
which provide a local coordinate chart on $V$ (the time dependence is due to the fact that the primary constraints may depend explicitly on time).
On the other hand, $\M$ constraints of all $\phi_{m}$'s can be taken as coordinates for the linearly independent directions to $V$ in the full phase space.  The entire $2\N$-dimensional phase space has then two sets of coordinate charts,
\be
\ba{lll}
{}&
\left(p_{A}\,,\,q^{B}\right)\,,~~&~~1\leq A,B\leq \N\\
{}&{}\\
\Longleftrightarrow &\,\left(x^{i},\,\phi_{m}\right)\,,~~& ~~1\leq i\leq 2\N-\M\,,~~ 1\leq m\leq \M\,.
\ea
\label{newcoV}
\ee
Note that there may exist some freedom in choosing different sets of the independent momenta $\{p_{a}\}$ from the $\M$ primary constraints.\footnote{For more details on the regularity conditions and the properties they imply, some of which are used here, the reader is referred to the subsection 1.1.2 of \cite{Henneaux:1992ig}. Notice also that more general regularity conditions (\textit{e.g.} where the momenta do not play a distinguished role, or where the constraints are not assumed to be independent) can be defined, as is done in the first chapter of \cite{Henneaux:1992ig}.}

For an arbitrary function $F(p,q,t)$ on the $2\N$-dimensional phase space, with the coordinate system $(x,\phi)$ in (\ref{newcoV}), we define $\widetilde{F}(x,\phi,t):=F(p,q,t)$ and a set of $\M$ functions, $F^{m}(p,q,t)$, by
\be
F(p,q,t)=\widetilde{F}(x,\phi,t)=\widetilde{F}(x,0,t)+\phi_{m}F^{m}(p,q,t)=
\left.F(p,q,t)\right|_V+{\phi}_{m}F^{m}(p,q,t)\,,
\label{Fm}
\ee
where $\left.F(p,q,t)\right|_V=\widetilde{F}(x,0,t)$. In other words, $|_V$ means that we substitute $(p,q)$ by its expression in terms of $(x,t)$ on the stationary surface $V\,$.\\

Finally note already that when we study the Hamiltonian dynamics, there can appear more constraints, namely the  ``secondary constraints''. In this case, all the constraints will define a smaller hypersurface, $\V\subset V$, in the phase space.

\subsection{Prior to the Hamiltonian formulation: change of variables}\label{priorto}

In this subsection,   instead of $(q^{A},\dq^{B})$  we regard $(q^{\hat a},q^{\hat m},p_{a},\dq^{\hat m})$  in Eqs.(\ref{primary1}) as the independent variables, and discuss briefly  the time evolution of them. The Lagrangian equations of motion are equivalent to
\begin{eqnarray}
\D{\frac{{\rm d} p_{a}}{{\rm d}t}=\left(\frac{\partial L(q,\dq,t)}{\partial q^{a}}\right)_{\dq^{\hat a}\,=\, h^{\hat{a}}(q^A,\,p_{b},\,\dq^{\hat m},\,t)}\,,}\label{dtpa}\\
{}\nonumber\\
\D{\frac{{\rm d} f_{m}(q^A,p_{a},t)}{{\rm d}t}=\left(\frac{\partial L(q,\dq,t)}{\partial q^{m}}\right)_{\dq^{\hat a}\,=\, h^{\hat{a}}(q^A,\,p_{b},\,\dq^{\hat m},\,t)}\,,}\label{dtpm}
\end{eqnarray}
provided
\be
\ba{ll}
\D{\frac{\,{\rm d} q^{\hat a}}{{\rm d}t}=h^{\hat a}(q,p_{a},\dq^{\hat m},t)\,,}~~~~~&~~~~~
\D{\frac{\,{\rm d} q^{\hat m}}{{\rm d}t}=\dq^{\hat m}\,.}
\ea
\label{Tr2}
\ee
Essentially these equations lead to a set of ${\cal M}$ algebraic relations on $(q^{A},p_{a},\dq^{\hat m})$ by substituting (\ref{dtpa}) and (\ref{Tr2}) into (\ref{dtpm}):\footnote{In terms of the Hamiltonian $H(q^A,p_{a},t)$ and the Poisson bracket $[\,\,,\,\,\}_{P.B.}$ defined later (in Eq.(\ref{Hamiltonian}) and (\ref{PB}) respectively),  the equation (\ref{algeid}) can be reexpressed in a compact  form:
\[
\D{\Big[H(q,p_{a},t)+\dq^{n}(p_{n}{-f_{n}})\,,\,p_{m}-f_{m}\Big\}_{P.B.}+\frac{\partial f_{m}}{\partial t}=0\,,}
\]
where   $f_{m}{=f_{m}}(q,p_{a},t)$ and the explicit velocities $\dq^{n}$ are taken to be constant with respect to the phase space derivatives of the Poisson bracket. {Anticipating a bit, one may realize that, in such a way, there are ${\cal M}$ \textit{linear}  equations (\ref{dtpmH}) for the ${\cal M}$ variables $\dq^{m}$ rather than the ${\cal M}$ algebraic \textit{nonlinear} equations (\ref{algeid}) for  $\dq^{\hat m}$.}}
\be
\D{h^{\hat a}\frac{\partial f_{m}}{\partial q^{\hat a}}+
\dq^{\hat m}\frac{\partial f_{m}}{\partial q^{\hat m}}
+\left(\frac{\partial L}{\partial q^{a}}\right)_{\!\!\dq^{\hat a}= h^{\hat{a}}\!\!}\frac{\partial f_{m}}{\partial p_a}+
\frac{\partial f_{m}}{\partial t}=\left(\frac{\partial L}{\partial q^{m}}\right)_{\!\!\dq^{\hat a}= h^{\hat{a}}}\,,}
\label{algeid}
\ee
where $m$ runs from 1 to ${\cal M}$.
{Now these ${\cal M}$ algebraic relations can be thought as constraints  for the ${\cal M}$ variables $\dq^{\hat m}$. Such constraints fix some of the $\dq^{\hat m}$'s but may leave others as completely free parameters.}  Once all the $\dq^{\hat m}$'s are determined as functions of other variables or as free parameters, the time evolution of the remaining (here, taken to be independent) variables $(p_{a}, q^{\hat a},q^{\hat m})$ follows from (\ref{dtpa}) and (\ref{Tr2}).
However, the constraints (\ref{algeid}) are in general nonlinear in $\dq^{\hat m}$ and so they are difficult to solve.
Below, we move to the Hamiltonian formalism where the independent variables are $(q^{A},p_{a},\dq^{m})$ rather than $(q^{A},p_{a},\dq^{\hat m})$. One advantage is that the corresponding constraints will be linear in $\dq^{m}$ so that we can do a more explicit analysis (see Subsection \ref{Solving}).\\

\subsection{From Lagrangian to Hamiltonian}\label{LtoHK}

Suppose that a given Lagrangian leads to $\M$ primary constraints, say (\ref{primary1}):
\be
\phi_m(q^A,p_B,t):=p_m-f_m(q^A,p_a,t)\,.\label{prctrs}
\ee
Replacing $\dq^{\hat a}$ by  $h^{\hat a}(q,p_{a},\dq^{\hat{m}},t)$ in (\ref{prctrs}) we take again $(q^A,p_a,\dq^{\hat m},t)$ as the independent variables. We write the ``canonical Hamiltonian'' as
\be
\ba{ll}
H(q^A,p_a,t)&:=\dq^{A}p_{A}-L(q,\dq,t)\\
{}&{}\\
{}&=h^{\hat a}(q^A,p_a,\dq^{\hat m},t)\,p_{\hat a}+
\dq^{\hat m}p_{\hat m}-
L\left(q^A,h^{\hat a}(q^A,p_a,\dq^{\hat m},t),\,\dq^{\hat m}\right)\\
{}&{}\\
{}&=\dq^{a}p_{a}+\dq^{m}f_{m}(q^A,p_{a},t)-
L\left(q^{A},h^{\hat a}(q^{A},p_{a},\dq^{\hat{m}},t),\,\dq^{\hat m}\right)\,,
\ea
\label{Hamiltonian}
\ee
where we made use of Eq.(\ref{primary1}). Since there may exist some freedom in choosing different sets of the independent $\N-\M$ momenta,
the Hamiltonian is not uniquely specified, in general, from a given Lagrangian, but depends on this choice. However, on $V$, these Hamiltonians are all equal.
\\

From the fact that
\be
\D{\frac{\partial H}{\partial \dq^{\hat m}}=\frac{\partial h^{\hat a}}{\partial \dq^{\hat m}}\,p_{\hat a}+p_{\hat m}-\frac{\partial h^{\hat a}}{\partial \dq^{\hat m}}\frac{\partial L}{\partial \dq^{\hat a}}-\frac{\partial L}{\partial \dq^{\hat m}}=0}\,,
\ee
one can see that the canonical Hamiltonian is indeed a function of $q^{A}$, $p_a$ and $t$ only, \textit{i.e.} it is independent of $\dq^{\hat{m}}\,$.  Further direct calculations can lead to
\be
\ba{l}
\D{\frac{\partial H(q,p_b,t)}{\partial p_a}=(-1)^{\#_a}\dq^a-
(-1)^{\#_a\#_{m}}\dq^{m}\frac{\partial\phi_{m}}{\partial p_a}\,,}\\
{}\\
\D{\frac{\partial H(q,p_b,t)}{\partial p_n}=(-1)^{\#_n}\dq^n-
(-1)^{\#_n\#_{m}}\dq^{m}\frac{\partial\phi_{m}}{\partial p_n}=0\,,}\\
{}\\
\D{\frac{\partial H(q,p_b,t)}{\partial q^{A}}=-\left(\frac{\partial L(q,\dq,t)}{\partial q^{A}}\right)_{\dq^{\hat a}{= h^{\hat a}}}-
(-1)^{\#_A\#_{m}}\dq^{m}\frac{\partial\phi_{m}}{\partial q^{A}}\,,}
\ea
\label{partialH}
\ee
where  all the velocities are to be understood as functions of $(q^{A},p_{a},\dq^{\hat m},t)$ by the substitution
$\dq^{\hat a}=h^{\hat a}(q,p_{a},\dq^{\hat{m}},t)$. {The first two equations are easily obtained by making use of the Legendre transform philosophy, \textit{i.e.} the canonical Hamiltonian does not really depend on the velocities. Concretely, it is enough to perform the partial differentiation only of the momenta in the term $\dq^A p_A$ in order to compute the right-hand-side of the first lines from (\ref{partialH}).} In a unified manner, any velocity can be expressed as a function of $(q,p_a,\dq^m,t)$
\be
\D{\tq^{A}(q,p_a,\dq^m,t):=(-1)^{\#_{A}}
\frac{\partial H(q,p_b,t)}{\partial p_{A}}+(-1)^{\#_{A}(1+\#_m)\,}\dq^{m}
\frac{\partial\phi_{m}}{\partial p_{A}}\,.}
\label{velocitydqm}
\ee
Now this formula suggests that we can take not only $(q,p_a,\dq^{\hat m},t)$ but, alternatively, $(q,p_a,\dq^m,t)$ as independent variables. {As follows from Subsection \ref{priorto},} the Hamiltonian dynamics is consistent with the Euler-Lagrangian equations only if
\begin{eqnarray}
&&\D{\frac{{\rm d}p_{a}}{{\rm d}t}=-\frac{\partial H(q,p_b,t)}{\partial q^{a}}+
(-1)^{\#_a\#_{n}}\dq^{n}\frac{\partial f_{n}}{\partial q^{a}}\,,}\label{dtpaH}\\
{}\nonumber\\
&&\D{\frac{{\rm d}f_{m}}{{\rm d}t}=-\frac{\partial H(q,p_b,t)}{\partial q^{m}}+
(-1)^{\#_m\#_{n}}\dq^{n}\frac{\partial f_{n}}{\partial q^{m}}\,,}\label{dtpmH}
\end{eqnarray}
where we made use of the definition (\ref{Hamiltonian}) of the canonical Hamiltonian and of the constraint (\ref{prctrs}).
The equation (\ref{dtpmH}) leads to ${\cal M}$ linear constraints on the ${\cal M}$ variables $\dq^{m}$. Once we know the complete solution of $\dq^{m}$, as we will do in Sec.\ref{Solving}, the other equations (\ref{velocitydqm}) and (\ref{dtpaH}) determine the time evolution of the remaining variables $p_{a}, q^{a}$.\\

For an equivalent but more compact description of the Hamiltonian dynamics for the variables $(q^A,p_a,\dq^m,t)$, we introduce the ``total Hamiltonian'' defined by
\be
H_{T}(q^A,p_B,u^m,t):=H(q^{A},p_a,t)+\phi_{m}(q^A,p_B,t)\,u^{m}\,,
\label{totH}
\ee
where $H_T$ is indeed a function on the `total' phase space with coordinates $(q^B,p_A)$ but
demand that
\be
\ba{ll}
\D{\left.\frac{\partial H_{T}}{\partial p_{A}}\right|_{V}=(-1)^{\#_{A}}\dq^{A}}\,,~~~~&~~~~
\D{\left.\frac{\partial H_{T}}{\partial q^{A}}\right|_{V}=-\frac{\partial L(q,\dq)}{\partial q^{A}}\,.}
\ea
\label{preHdynamics}
\ee
Combining (\ref{prctrs}) and (\ref{preHdynamics}) we identify
\be
u^{m}=(-1)^{\#_{m}}\dq^{m}\,,
\label{EXP}
\ee
hence $\phi_{m}u^{m}=\dq^{m}\phi_{m}\,$.
We also have
\be
\ba{ll}
\D{\left.H_{T}\right|_{V}=\left.H\right|_{V}}\,,~~~~~~~~&~~~~~~~~
\D{\left.\frac{\partial H_{T}}{\partial u^{m}}\right|_{V}=0\,.}
\ea
\ee
~\\

In summary we note that there exist two one-to-one maps:
\be
\left(\ba{l}q^{A}\\
{}\\
p_b\\
{}\\
\dq^{\hat m}\ea\right)~\longleftrightarrow~
\left(\ba{l}
q^{A}\\
{}\\
\dq^{B}\ea\right)~\longleftrightarrow~
\left(\ba{c}q^{A}~\\
{}\\
p_b~\\
{}\\
u^{m}
\ea\right)\,.
\label{11maps}
\ee
The Lagrangian dynamics is equivalent to the Hamiltonian one with the primary constraints.
In the former system, the dynamical variables are by definition $\{q^{A},\dq^{B}\}$, while in the latter the set of independent variables can be chosen to be  $\left\{q^{A},p_b\,,\,u^{m}\right\}$ for convenience.\\

As we will shortly show in Subsection \ref{Solving}, in the Hamiltonian dynamics there is a systematic way of identifying  the variables $\left\{\,u^{m}\right\}$. Once the most general solution $u^m(q,p,t)$ preserving the primary constraints is obtained, as in (\ref{mostu}), the Hamiltonian dynamics determines the time evolution of the other variables $\{q^{A},p_b\}\,$, with the restriction on $V$. After the generalization to the `total Hamiltonian system', it can govern the dynamics of the $2N$-dimensional whole phase space, with variables $\{q^{A},p_{B}\}\,$, free from any restriction.

\subsection{From Hamiltonian to Lagrangian}\label{HtoL}

We start\footnote{This subsection is provided for completeness, and may be skipped at the first reading.} with a given Hamiltonian $H(p,q)$  on the $2\N$-dimensional phase space of coordinates $(p_{A}, q^{B})\,$, and $\M$ independent  arbitrary but fixed primary constraints, $\phi_m(p,q,t)=0\,$,  which can be solved as $p_m=f_m(q^{A},p_b,t)$ to express $\M$ of the momenta in terms of the positions and the $\N-\M$ other momenta. There can be  some freedom in choosing different  sets of the independent momenta, $\{p_a\}$.  The relevant phase space reduces to a $(2\N-\M)$-dimensional hyperspace $V$, as in (\ref{hyperV}), which will be further restricted to its sub-manifold $\V\subset V$ if there occurs `secondary constraints' (see Subsection \ref{Solving}).  \\

Introducing $\M$ new variables $u^{m}\,$, we define the total Hamiltonian,
\be
\D{H_{T}(p,q,u,t)=H(p,q,t)+\sum_{m=1}^{\M}\,\phi_{m}(p,q,t)u^{m}\,.}
\label{totalH}
\ee
The action principle is derived from the action
\begin{equation}
S[\,q^A,p_B,u^m]=\int dt\,\Big(p_A\,\dq^A-H_T\Big)\,.
\end{equation}
For instance, this leads to
\be
\tq^{A}(q,p_{i},u^{l},t)\equiv\D{\left.(-1)^{\#_{A}}\frac{\partial H_{T}}{\partial p_{A}}\right|_{V}=(-1)^{\#_{A}}\left.\left(
\frac{\partial H}{\partial p_{A}}+
\frac{\partial \phi_{m}}{\partial p_{A}}\,u^{m}\right)\right|_{V}
\,,}
\label{defdq}
\ee
where $|_{V}$ means that we substitute $p_m$ by $p_m=f_m(q^A,p_b,t)$ after taking the partial derivatives.

We  assume  that there exists an inverse map, $(q^{A},\tq^{B})\rightarrow(q^{A},p_b,u^m)$, to write
\be
\ba{lll}
p_a(q,\tq,t)\,,~~~~&~~~~p_m=f_m\Big(q,p_b(q,\tq,t),t\Big)\,, ~~~~&~~~~ u^m(q,\tq,t)\,.
\ea
\label{invHL}
\ee
Provided with these functions, $p_{A}(q,\tq,t)$, we define
\be
L(q,\tq,t):=\left.\Big(\tq^{A}p_{A}-H(p,q,t)\Big)\right|_{V}=
\left.\Big(\tq^{A}p_{A}-H_{T}(p,q,u,t)\Big)\right|_{V}\,,
\ee
to recover the Lagrangian dynamics,
\begin{eqnarray}
{}&&\D{\frac{\partial L(q,\tq,t)}{\partial\tq^{A}}}=\D{\left.\left(p_{A}+(-1)^{\#_{A}\#_{B}}\tq^{B}
\frac{\partial p_{B}}{\partial\tq^{A}}-\frac{\partial p_{B}}{\partial\tq^{A}}\frac{\partial H_{T}}{\partial p_{B}}\right)\right|_{V}}=p_{A}(q,\tq,t)\,,\label{recover1}\\
{}&&\nonumber\\
{}&&\D{\frac{\partial L(q,\tq,t)}{\partial q^{A}}}=\D{\left.\left((-1)^{\#_{A}\#_{B}}\tq^{B}
\frac{\partial p_{B}}{\partial q^{A}}-\frac{\partial p_{B}}{\partial q^{A}}\frac{\partial H_{T}}{\partial p_{B}}-\frac{\partial H_{T}}{\partial q^{A}}\right)\right|_{V}}=\D{-\left.\frac{\partial H_{T}(p,q,u,t)}{\partial q^{A}}\right|_{V}\,,}\nonumber\\
{}&&\label{recover2}
\end{eqnarray}
along with the $\M$ constraints, $p_m=f_m(q^A,p_b,t)\,$. \\

This analysis shows the equivalence between the Hamiltonian and the Lagrangian formalism,
up to  the technical assumption on the existence of the inverse map (\ref{invHL}).\footnote{Notice that the equivalence between the Lagrangian and total Hamiltonian system can be shown in a large number of ways, see \textit{e.g.} the exercises 1.2 and 1.4 of the book \cite{Henneaux:1992ig} for some alternatives. The latter exercise is based on the general result about the elimination of ``auxiliary fields'' (in the present case, the `auxiliary' fields are $p_b$ and $u^m$) \textit{via} their own equations of motion~\label{auxiliaryf}.}\\
{}\\
{}\\
{}\\

\newpage
\section{Total Hamiltonian dynamics}\label{TotaL}

\subsection{Poisson bracket}

On the $2\N$-dimensional phase space, with coordinates $(q^{A}, p_{B})$, equipped with the ${\mathbb Z}_{2}$-grading,
we define the  Poisson Bracket as
\be
\D{\left[F,G\right\}_{P.B.}=(-1)^{\#_{A}\#_{F}}\frac{\partial F}{\partial q^{A}}\frac{\partial G}{\partial p_{A}}-(-1)^{\#_{A}(\#_{F}+1)}\frac{\partial F}{\partial p_{A}}\frac{\partial G}{\partial q^{A}}\,.}
\label{PB}
\ee
The Poisson bracket can be rewritten in a more compact form,
\be
\D{\left[~~~,~~~\right\}_{P.B.}=(-1)^{\#_{A}}\frac{\overleftarrow{\partial} ~~~}{\partial q^{A}}\,\frac{\overrightarrow{\partial}~~~ }{\partial p_{A}}~-~\frac{\overleftarrow{\partial} ~~~}{\partial p_{A}}\,\frac{\overrightarrow{\partial}~~~ }{\partial q^{A}}\,,}
\label{Poisson}
\ee
where the arrows indicate the direction the derivatives act.
It satisfies the graded skew-symmetry property,
\be
\D{\left[F,G\right\}_{P.B.}=-(-1)^{\#_{F}\#_{G}}\left[G,F\right\}_{P.B.}\,,}
\label{skewPB}
\ee
the Leibniz rule,
\be
\ba{l}
\D{\left[F,GK\right\}_{P.B.}=\left[F,G\right\}_{P.B.}K+(-1)^{\#_{F}\#_{G}}
G\left[F,K\right\}_{P.B.}\,,}\\
{}\\
\D{\left[FG,K\right\}_{P.B.}=F\left[G,K\right\}_{P.B.}+(-1)^{\#_{G}\#_{K}}
\left[F,K\right\}_{P.B.}G\,,}
\label{Leibniz}
\ea
\ee
and the Jacobi identity,
\be
\D{\left[\left[F,G\right\}_{P.B.},K\right\}_{P.B.}=
\left[F,\left[G,K\right\}_{P.B.}\right\}_{P.B.}-(-1)^{\#_{F}\#_{G}}
\left[G,\left[F,K\right\}_{P.B.}\right\}_{P.B.}\,,}
\ee
or equivalently,
\be
\D{(-1)^{\#_{F}\#_{H}}\left[F,\left[G,H\right\}\right\}_{P.B.}
+(-1)^{\#_{G}\#_{F}}\left[G,\left[H,F\right\}\right\}_{P.B.}
+(-1)^{\#_{H}\#_{G}}\left[H,\left[F,G\right\}\right\}_{P.B.}=0\,.}
\label{Jacobi}
\ee
~\\

Let ${}^\dagger$ denote the Hermitian conjugation such that $(a\,b)^\dagger\,=\,b^\dagger \,a^\dagger\,$, \textit{i.e.} ${}^\dagger$ is an involution on the algebra of functions on the phase space.
Reality condition on the phase space reads,
\be
\ba{ll}
q^{A}{}^{\dagger}=q^{A}\,,~~~~~&~~~~~ p_{A}{}^{\dagger}=(-1)^{\#_{A}}p_{A}\,,
\ea
\ee
because the symplectic form $\dq^Ap_A$ must be real and
\be (\dq^A p_A)^\dagger=p_A^\dagger(\dq^A)^\dagger=(-1)^{\#_{A}}(\dq^A)^\dagger p_A^\dagger\,.\ee
Hence we have
\be
\ba{ll}
\D{\left(\frac{\partial F}{\partial q^{A}}\right)^{\dagger}=(-1)^{\#_{A}(\#_{F}+1)}\,
\frac{\partial (F^{\dagger})}{\partial q^{A}}\,,}~~~~&~~~~
\D{\left(\frac{\partial F}{\partial p_{A}}\right)^{\dagger}=(-1)^{\#_{A}\#_{F}}
\frac{\partial (F^{\dagger})}{\partial p_{A}}\,,}
\ea
\ee
and hence,
\be
\D\Big(\left[F,G\right\}_{P.B.}\Big)^{\dagger}=(-1)^{\#_{F}\#_{G}}
[F^{\dagger},G^{\dagger}\}_{P.B.}=-[G^{\dagger},F^{\dagger}\}_{P.B.}\,.
\label{ccPB}
\ee

\subsection{Time derivatives - preliminary}

With a given  total Hamiltonian,
\be
\D{H_{T}(p,q,u(p,q,t),t)=H(p,q,t)+\sum_{m=1}^{\M}\,\phi_{m}(p,q,t)u^{m}(p,q,t)\,,}
\ee
and the generalization motivated at the end of Subsection \ref{LtoHK}, we let the following formulae govern
the dynamics of the  \textit{full} phase space,
\be
\ba{ll}
\D{\dq^{A}=(-1)^{\#_{A}}\frac{\partial H_{T}}{\partial p_{A}}\,,}~~~~~&~~~~~
\D{\dot{p}_{A}=-\frac{\partial H_{T}}{\partial q^{A}}\,,}
\ea
\label{Hd}
\ee
where  the variables $u^{m}(p,q,t)$ are to be understood as  functions of $p,q,t$, of which the explicit forms are  not yet specified. Note that with the restriction on the hypersurface $V$, \textit{i.e.} putting   $\phi_{m}=0$ after taking the derivatives, the above equations reduce to (\ref{defdq},\ref{recover1},\ref{recover2}) which already indicates some equivalence between the total Hamiltonian dynamics and the Lagrangian dynamics.
\footnote{It is worthwhile to note that the above equations (\ref{Hd}) indeed govern the full dynamics of all the coefficients $[p_{A}]_{J}$ and $[q^{A}]_{J}$ of the Grassmann algebra {(see Appendix \ref{subsectionGrassmann}).}}\\

The time derivative of an arbitrary quantity $F(p,q,t)$ takes a simple form\footnote{One may generalize Eq.(\ref{HdF}) by adding  an arbitrary quantity proportional to the constraints to the right hand side,
\be
\frac{{\rm{d}}F}{{\rm{d}}t}=\left[F,H_{T}\right\}_{P.B.}+\frac{\partial F}{\partial t}+\phi_{m}C^{m}(F)\,.
\label{gTH}
\ee
See (\ref{generaliationTH}) and also the Dirac bracket (\ref{DHF}) for further discussion. The Dirac bracket gives an alternative dynamics which coincides with the dynamics of the Poisson bracket \textit{only} on $\V$, but it is more suitable for quantization.} in terms of the Poisson bracket (\ref{PB}),
\be
\D{\frac{{\rm{d}}F(p,q,t)}{{\rm{d}}t}=\left[F,H_{T}\right\}_{P.B.}+\frac{\partial F}{\partial t}\,.}
\label{HdF}
\ee

The crucial viewpoint we adopt here is the following: \textit{Though the above equations supplemented by the primary constraints are equivalent to the Lagrangian formalism} (as shown in Subsection \ref{LtoHK}) \textit{we consider  them to be more fundamental. Namely, without restriction on the hypersurface, we let them govern the entire  phase space.} Then, we try  to make sure that the dynamics can be consistently truncated to the hypersurface. Namely we will look for the necessary and sufficient conditions to maintain the constraints throughout the time evolution. By imposing so, it may well be the case that we determine  some of the unknown variables $u^{m}$  completely, at least the values on the hypersurface, and obtain further consistency conditions or `secondary constraints'. In the latter case, both the primary constraints and the secondary constraints should be imposed to define the hypersurface, say $\V$. In this way, we indeed make our Hamiltonian dynamics be consistent with the Lagrangian dynamics \textit{if one considers the space of physical states to live on the constraint surface $\V\,$}.

\subsection{Preserving the constraints - primary  and secondary constraints}\label{Solving}

On the hypersurface $V$, we have
\be
\D{\left(\frac{{\rm{d}}~}{{\rm{d}}t}F(p,q,t)\right)
\,\simeq\,\left[F,H\right\}_{P.B.}+\left[F,\phi_{m}\right\}_{P.B.}u^{m}+\left(\frac{\partial F}{\partial t}\right)\,,}
\label{tde2}
\ee
{where we introduced the notation $\simeq$ for the ``weak equality'' defined via the equivalence relation
\be
f\simeq g\quad\Longleftrightarrow\quad f|_V=g|_V\quad\Longleftrightarrow\quad f=g+\phi_m z^m\,.
\label{weakequ}
\ee
We remind the reader that $|_{V}$ means the restriction on the primary constraint surface $V$, or the expression of $(p,q)$ in terms of the variables $x^i$ as in (\ref{onV}), after taking the partial derivatives. In other words, the symbol $\simeq$ stands for ``equal on the primary constraint surface''. This symbol makes the distinction with the ``strong equality'', which is the usual equality throughout all phase space,  and proves to be  convenient in order to avoid repetitions of the restriction on $V$ for every term in a lengthy expression.} On the left hand side, we get from (\ref{Fm}) that
\be
\D{\left.\left(\frac{{\rm{d}}~}{{\rm{d}}t}F(p,q,t)\right)\right|_{V}=
\left.\left(\frac{{\rm{d}}~}{{\rm{d}}t}\Big(\left.F(p,q,t)\right|_{V}+\phi_{m}F^{m}\Big)
\right)\right|_{V}=
\frac{{\rm{d}}~}{{\rm{d}}t}\left(\left.F(p,q,t)\right|_{V}\right)+\dot{\phi}_{m}F^{m}}\,.
\ee
~\\

For the consistency with the Lagrangian formalism, the time derivative of the primary constraints must vanish on $V$,
\be
\ba{ll}
\D{\frac{{\rm{d}}\phi_{m}}{{\rm{d}}t}
=\left(\left[\phi_{m},H\right\}_{P.B.} +\frac{\partial\phi_{m}}{\partial t}+
\left[\phi_{m},\phi_n
\right\}_{P.B.}u^n\right)\simeq 0\,,}~~~&~~~~(m,n=1,2,\cdots,\M)\,,
\ea
\label{cons}
\ee
which precisely corresponds to (\ref{dtpmH}).
We focus on the following supermatrix defined on $(q^{A},p_b, t)$,
\be
\D{\Omega_{mn}:=\left.\left[\phi_{m},\phi_n\right\}_{P.B.}\right|_{V}=-(-1)^{\#_{m}\#_n}
\left.\left[\phi_n,\phi_{m}\right\}_{P.B.}\right|_{V}=\left(\ba{cc}
A_{-}\,&\,\Psi\\
-\Psi^{T}\,&\,iA_{+}
\ea\right)_{ m\,n}\,,}
\label{defOmega}
\ee
where $A_{\pm}$ are symmetric or anti-symmetric ($A_{\pm}^{T}=\pm A_{\pm}$) bosonic matrices while $\Psi$ is a fermionic matrix. Now Eq.(\ref{cons}) can be taken as a set of linear equations with the $\M$ unknown  variables $u^{m}$ ($1\leq m\leq \M$),
\be
\D{\left[\phi_{n},H\right\}_{P.B.}+\frac{\partial\phi_{n}}{\partial t}\simeq \,-\Omega_{nm}u^{m}\,,}
\label{LinearOmega}
\ee
where, surely, the lefthand-side and $\Omega$ are given by fixed functions on $V$ of either $(x,t)$ or, equivalently, $(p_b,q^{A},t)$.\footnote{For a given set of the \textit{generating elements of the Grassmann algebra} $\Lambda_{\hat{N}}$ (as in Appendix \ref{subsectionGrassmann}) the above formula (\ref{LinearOmega}) can be, in principle, completely analyzed.
Furthermore, for the consistency of the Lagrangian mechanics, there must be a solution thereof.
Indeed, we always implicitly assumed that we disregard any inconsistent Lagrangian like $L=q$ which would lead to $\delta L/\delta q=1=0$.
In the same spirit, we can also expect that, if necessary,  there may occur some more extra constraints on $V$, that is `secondary constraints'. }\\

Let us analyze the process more concretely. Without loss of generality, assuming  that all the primary constraints are real,
\be
\ba{ll}
(\phi_{m})^{\dagger}=\phi_{m}\,,~~~~~~&~~~~~~(u^{m})^{\dagger}=(-1)^{\#_{m}}u^{m}\,.
\ea
\label{realitypu}
\ee
the supermatrix is anti-Hermitian, $\Omega_{ml}=-(\Omega_{lm})^{\dagger}$, so that
all the matrices are real, $A_{\pm}=A_{\pm}^{\ast}$, $\Psi^{\ast}=\Psi$. The forthcoming analysis is rather technical and complicate since one includes fermions. The main results are summarized in Subsection \ref{summary} to which the reader may jump directly in a first reading.\\

Under the real linear transformation,\footnote{Note that, in general, $(L^{T})^{m}{}_{k}=(-1)^{\#_{m}(\#_{m}+\#_{k})}L_{k}{}^{m}$, $(L^{\ast})_{k}{}^{m}=(-1)^{\#_{m}(\#_{k}+\#_{m})}(L_{k}{}^{m})^{\ast}$, see Eq.(\ref{Transpose}).}
\be
\ba{l}
(L^{\ast})_{k}{}^{m}=(-1)^{\#_{m}(\#_{k}+\#_{m})}(L_{k}{}^{m})^{\ast}=L_{k}{}^{m}\,,\\
{}\\
\phi_m~~\longrightarrow~~ L_m{}^k\phi_k=\phi_k(L^{T})^k{}_{m}\,,\\
{}\\
u^m~~\longrightarrow~~ (\,(L^{T})^{-1}\,)^m{}_{k}u^{k}=(-1)^{\#_{m}+\#_{k}}u^{k}(L^{-1})_{k}{}^{m}\,,
\ea
\ee
the contraction $\phi_mu^m$ is invariant, the reality condition (\ref{realitypu}) is preserved,  and  the supermatrix transforms as
\be
\left.\left[\phi,\phi\right\}_{P.B.}\right|_{V}~~\longrightarrow~~ L\left.\left[\phi,\phi\right\}_{P.B.}\right|_{V}L^{T}\,,
\ee
where we have set $\left.L\right|_{V}=L$ for simplicity.
As shown in (\ref{Lemma4}), one can always transform any anti-Hermitian  supermatrix into the following `canonical' form by a real linear transformation,\footnote{Contrary to the usual complex number valued Hermitian matrix, a Hermitian supermatrix may not be completely diagonalizable. However, if the Hermitian supermatrix is nondegenerate,  it is diagonalizable. See our Lemma 4 in (\ref{Lemma4}).}
\be
L\Omega L^{T}=\left(
\ba{cccc}
\,b_{-}\,&\,0\,&\,0\,&\,0\,\\
\,0\,&\,s_{-}\,&\,0\,&\,\psi\,\\
\,0\,&\,0\,&\,ib_{+}\,&\,0\,\\
\,0\,&\,-\psi^{T}\,&\,0\,&is_{+}\,
\ea
\right)\,,
\ee
where all the matrices are real, $b_{\pm}=b_{\pm}^{\ast}$,\, $s_{\pm}=s_{\pm}^{\ast}$,\,  $\psi=\psi^{\ast}$~; ~$b_{\pm}$ are nondegenerate  bosonic matrices so that $ b_{\pm}{}^{-1}$  exist~;
~$s_{\pm}$ are bosonic products of fermions (\textit{i.e.} even number of products of fermions); ~and $b_{\pm}=\pm b_{\pm}^{T}$, $\,s_{\pm}=\pm s_{\pm}^{T}$\,. It may be the case  that $s_{\pm}$ and/or $\psi$ vanish.\\

With the decomposition of the index $m$ into $c,\,\dot{c}$ for bosonic variables  and  $\alpha,\dot{\alpha}$ for fermionic ones, which should be obvious from the inspection of Eqs.(\ref{c1})-(\ref{c4}), the consistency condition (\ref{cons}) now splits  into
\begin{eqnarray}
&&\D{\left[\phi_{c},H\right\}_{P.B.}+\frac{\partial\phi_{c}}{\partial t}\simeq -(b_{-})_{cd}u^{d}\,,}\label{c1}\\
&&{}\nonumber\\
&&\D{\left[\phi_{{\alpha}},H\right\}_{P.B.}+
\frac{\partial\phi_{{\alpha}}}{\partial t}\simeq -i(b_{+})_{\alpha\beta}u^{\beta}
\,,~~~~~~~~~~~~~}\label{c2}
\end{eqnarray}
and
\begin{eqnarray}
&&\D{\left[\phi_{\dot{c}},H\right\}_{P.B.}+\frac{\partial\phi_{\dot{c}}}{\partial t}\simeq -(s_{-})_{\dot{c}\dot{d}}u^{\dot{d}}
-\psi_{\dot{c}\dot{\beta}}u^{\dot{\beta}}\,,}\label{c3}\\
&&{}\nonumber\\
&&\D{\left[\phi_{\dot{\alpha}},H\right\}_{P.B.}+
\frac{\partial\phi_{\dot{\alpha}}}{\partial t}\simeq \psi_{\dot{c}\dot{\alpha}}u^{\dot{c}}-
i(s_{+})_{\dot{\alpha}\dot{\beta}}u^{\dot{\beta}}\,.}\label{c4}
\end{eqnarray}

The meaning of the first two equations, (\ref{c1}) and (\ref{c2}),  is clear. Since $b_{\pm}$ are nondegenerate, they  fix the unknown variables, $u^{c}$, $u^{\alpha}$, completely as  functions of $(q^{A},p_b,t)$ or $(x,t)$ on the hypersurface $V$\,. \\

The analysis of  the  last two equations, (\ref{c3}) and (\ref{c4}), is somewhat tricky. Before the full analysis, we first focus on  the purely bosonic systems, which was the case studied by Dirac~\cite{Dirac}.

\begin{itemize}
\item  \textit{Bosonic systems.}\\
In the bosonic systems, the equations (\ref{c2}) and (\ref{c4}) simply do not appear, and the  essential relations are
\be
\D{\left(\ba{l}
{\left[\phi_{c},H\right\}_{P.B.}+\frac{\partial\phi_{c}}{\partial t}}\\
{}\\
{\left[\phi_{\dot{c}},H\right\}_{P.B.}+
\frac{\partial\phi_{\dot{c}}}{\partial t}}\ea\right)
\simeq -\left(\ba{cc}(b_{-})_{{c}{d}}~&~0\\
{}&{}\\
0~&~0\ea\right)
\left(\ba{c}
u^{d}\\
{}\\
u^{\dot{d}}\ea\right)\,.}
\ee
Since $b_{-}$ is nondegenerate,  the  variables $u^{c}$ are completely determined on $V$ in terms of the variables $(p_a,q^B,t)$, while the other variables $u^{\dot{c}}$
remain as locally free variables at this stage. The vanishing of the second row in the left-hand-side can give some new, namely `secondary', constraints. In this case, the primary and secondary constraints define together a smaller hypersurface, say $V^{\prime}\subset V$, and some of the $p_a$'s can be expressed in terms of others. Then the already determined variables $u^{c}$ should be further restricted on $V^{\prime}$, making the first row hold on $V^{\prime}$ too. We note that the number of secondary  constraints, say $n$,  are not greater that the number of the \textit{yet} free variables, $m\equiv\dim\{u^{\dot{c}}\}$, $~~n\leq m$.

The next step is to consider the time derivatives of the  secondary constraints, analogously to (\ref{cons}). Regarding them as  linear equations in the variables $u^{\dot{c}}$ and taking  some linear transformations to the canonical form, (\ref{Lemma3}),
\be
\ba{ll}
\left(\ba{l}\left.\times\right|_{V^{\prime}}\\{}\\
\left.\times^{\prime}\right|_{V^{\prime}}\ea\right)=
\left(\ba{ll}1_{k\times k}~&~0_{k\times (m-k)}\\
{}&{}\\
0_{(n-k)\times k}~&~0_{(n-k)\times (m-k)}\ea\right)
\left(\ba{l}
u^{c^{\prime}}\\{}\\u^{\dot{c}^{\prime}}\ea\right)\,,~~~~&~~~0\leq k\leq n\leq m\,,
\ea
\ee
one can determine  the variables $u^{c^{\prime}}$  completely on $V^{\prime}$, and there may appear new, namely ``tertiary'', constraints $\left.\times^{\prime}\right|_{V^{\prime}}\neq 0$. Again the number of the tertiary constraints,  are not greater that the number of the surviving locally free variables $u^{\dot{c}^{\prime}}\,$, since $(n-k)\leq (m-k)\,$.
The procedure may go on, but it should terminate at certain point, since the total number of constraints should not exceed the dimension of the whole phase space for any consistent dynamics. By a slight abuse of terminology, one refers to all these new constraints as `secondary'.

Eventually, we end up with  a set of constraints,
\be
\Big\{\,~\phi_{h}= 0~\,,~1\leq h\leq \M+\M^\prime~\Big\}\,,
\ee
of which the ranges $1\leq h\leq\M$ and $\M+1\leq h\leq\M+\M^\prime$ respectively correspond to the primary and secondary constraints.  They define the hypersurface, $\V$,
\be
\V:=\Big\{\,(p,q)~\big|~\phi_{h}= 0\,,~1\leq h\leq \M+\M^\prime~\Big\}\,.
\ee

All the constraints $\phi_{h}$ ($1\leq h\leq\M+\M^\prime$) are static on $\V$, in the sense that {the restriction on $\V$ is preserved by the time evolution,} $\,\left.\dot{\phi}_{h}\right|_{\V}=0\,$. In other words, they satisfy
\be
\ba{l}
-\D{\left.\left(\left[\phi_{h},H\right\}_{P.B.} +\frac{\partial\phi_{h}}{\partial t}\right)\right|_{\V}=\sum_{m=1}^{\M}\,\left.\Big(
\left[\phi_{h},\phi_{m}
\right\}_{P.B.}\,u^{m}\Big)\right|_{\V}}\\
{}\\
\Longleftrightarrow~~~
\left(\ba{l}\times\\{}\\0\ea\right)=
\left(\ba{ll}~1~&~0~\\{}&{}\\~0~&~0~\ea\right)
\left(\ba{l}
u^{m^{\prime}}\\{}\\u^{{m}^{\prime\prime}}\ea\right)~~~~\mbox{:~~in ~the ~canonical~ form~on~}\,\V\,.
\ea
\ee

Some of the $u^{m}$ variables, \textit{i.e.} the $u^{m^{\prime}}\,$'s, are completely determined\footnote{Strictly speaking, what we have determined are the variables on the hypersurfaces, $\{\left.u^{m^{\prime}}\right|_{\V}\}$.  For the generic dynamics  in the full phase space, we may either employ them literally as they are, or use the continuously extended functions which have nontrivial dependence on the orthogonal directions to the hypersurfaces. In any case the dynamics on $\V$ is the same.} on $\V$ as functions of $(p,q,t)$, while the others, the $u^{m^{\prime\prime}}\,$'s, if any, remain  as locally free (that is, arbitrarily time dependent variables). The latter correspond to the zero eigenvectors of the $(\M+\M^\prime)\times \M$ matrix, $\left.
\left[\phi_{h},\phi_{m}\right\}_{P.B.}\right|_{\V}\,.$\\

\item  \textit{Generic systems.}\\
Now we return to the equations, (\ref{c3}) and  (\ref{c4}), which are relevant to the generic systems of  bosons and fermions,
\be
\D{\left(\ba{l}
\D{\left[\phi_{\dot{a}},H\right\}_{P.B.}+\frac{\partial\phi_{\dot{a}}}{\partial t}}\\
{}\\
\D{\left[\phi_{\dot{\alpha}},H\right\}_{P.B.}+
\frac{\partial\phi_{\dot{\alpha}}}{\partial t}}\ea\right)
\simeq-\left(\ba{cc}(s_{-})_{\dot{a}\dot{b}}~&~\psi_{\dot{a}\dot{\beta}}\\
{}&{}\\
-\psi_{\dot{b}\dot{\alpha}}~&~i(s_{+})_{\dot{\alpha}\dot{\beta}}\ea\right)
\left(\ba{l}u^{\dot{b}}\\
{}\\
u^{\dot{\beta}}\ea\right)\,.}
\label{bfc}
\ee
In general, the complete analysis of the above formulae  is always possible, if we introduce  explicitly the Grassmann algebra basis of (\ref{basisL}). By expanding all the quantities  in terms of the basis accompanied with the real or complex  number coefficients, \textit{i.e.} $[q^{A}]_{J}$,  $[p_{A}]_{J}$,  one can  convert them into the  linear equations in $[u^{\dot{a}}]_{J}$, $[u^{\dot{\alpha}}]_{J}$, over $\mathbb{R}$ or $\mathbb{C}$. The linear equations can be completely analyzed, essentially in the same way as  in the  previous  bosonic case. After all the finitely  repeated procedures, the results will be parallel~:
some of the coefficients,  $[u^{\dot{a}}]_{J}$, $[u^{\dot{\alpha}}]_{J}$,   are completely determined in terms  of $([q^{A}]_{J}$,  $[p_{A}]_{J}$, $t$), while others remain as locally free  parameters, implying that the time evolution in the phase space is not deterministic.
There may appear secondary constraints in terms of the coefficients,  $[q^{A}]_{J}$,  $[p_{A}]_{J}$.\\

However, in practice we  favor  the  Lagrangian  systems which do not require any explicit use of the basis for the Grassmann algebra. In such `good' systems of both bosons and fermions, all the expressions can be written collectively in terms of $(p,q,u,t)$, rather than $([p]_{J},[q]_{J},[u]_{J},t)$, as if in the bosonic system. We summarize the results in the following separate subsection.

\end{itemize}

\subsection{Hamiltonian dynamics after analyzing  the constraints - \textit{summary}}\label{summary}

In all the bosonic systems as well as all the `good' systems for bosons and fermions, in the sense that the explicit use of the basis of the Grassmann algebra is not required, we have the following generic situation:

\begin{itemize}
\item The Hamiltonian, $H(p,q,t)$, as well as the primary constraints, $\phi_{m}(p,q,t)$, are  given as functions  on the $2\N$-dimensional full phase space with coordinates $(p_{A}, q^{B})$ with $1\leq A,B\leq\N\,$.  \\
In particular, the primary constraints define a $(2\N-\M)$-dimensional hypersurface
\be
V=\{(p,q)\,|\,\phi_{m}(p,q,t)=0,~~1\leq m\leq \M\}\,.
\ee

\item The total Hamiltonian is a sum of the canonical Hamiltonian and linear combinations of the primary constraints,
\be
\D{H_{T}(p,q,u(p,q,t),t)=H(p,q,t)+\sum_{m=1}^{\M}\,\phi_{m}(p,q,t)\,u^{m}(p,q,t)\,.}
\ee

\item The dynamics of the whole $2\N$-dimensional phase space is subject to
\be
\ba{ll}
\D{\dq^{A}=(-1)^{\#_{A}}\frac{\partial H_{T}}{\partial p_{A}}\,,}~~~~~&~~~~~
\D{\dot{p}_{A}=-\frac{\partial H_{T}}{\partial q^{A}}\,,}
\ea
\ee
so that the time derivative of an arbitrary function, say $F(p,q,t)$,  on the $2\N$-dimensional phase space  reads
\be
\D{\frac{{\rm{d}}F(p,q,t)}{{\rm{d}}t}=\left[F,H_{T}\right\}_{P.B.}+\frac{\partial F}{\partial t}\,.}
\label{HD1}
\ee
where the variables $u^{m}(p,q,t)$  are not yet specified, but by  looking for the necessary and sufficient conditions to maintain the
{hypersurface $V\,$, in order to} be consistent with the dynamics,  we may determine some of them completely. In general we encounter the following situation:

\item  The whole primary constraint surface $V$  may not be consistent with the dynamics, in the sense that it may not be preserved by the time evolution. It may well be the case that only a subset of $V$, say $\V\subseteq V$, is preserved. The constraint surface $\V$ is specified by the primary as well as the secondary constraints,
\be
\V=\{(p,q)\,|\,\phi_{h}(p,q,t)\approx 0,~~1\leq h\leq \M+\M^\prime\}\,,
\ee
where $\{\phi_{h}\}$ denote the complete set of constraints indexed by $h\,$, such that $\M+1\leq h\leq \M+\M^\prime$ correspond to the secondary constraints. We introduced the notation $\approx$ for the other ``weak equality'', that is defined as `equal on the constraint surface $\V$',
\be f\approx g\quad\Longleftrightarrow\quad f|_\V=g|_\V\quad\Longleftrightarrow\quad f=g+\phi_h z^h\,.\label{weakequ2}\ee
{We emphasize the important distinction between the two weak equalities, because (\ref{weakequ}) implies (\ref{weakequ2}) but the converse is not always true since $\V\subseteq V\,$.}
All the constraints, in principle, can be solved to express any point on $\V$ by  $2\N-\M-\M^\prime$ independent variables, say $y^{\hat{\imath}}$ (with $1\leq\hat{\imath}\leq 2\N-\M-\M^\prime$),
\be
\V=\left\{\big(p\,,\,q\big)=\big(f(y,t)\,,\,g(y,t)\big)\right\}\,,
\label{oncalV}
\ee
which provide a local coordinate chart on $\V$. On the other hand, $\M+\M^\prime$ of all $\phi_{h}$'s can be taken as complementary coordinates for the orthogonal directions to $\V$ in the full phase space.  The entire $2\N$-dimensional phase space has then two sets of coordinate charts,
\be
\ba{lll}
{}&
\left(p_{A}\,,\,q^{B}\right)\,,~~&~~1\leq A,B\leq \N\\
{}&{}\\
\Longleftrightarrow &\,\left(y^{\hat{\imath}},\,\phi_{h}\right)\,,~~& ~~1\leq\hat{\imath}\leq 2\N-\M-\M^\prime\,,~~ 1\leq h\leq \M+\M^\prime\,.
\ea
\label{newco}
\ee

\item There exists at least one set of solutions for $\{\,u^{m},~1\leq m\leq\M\,\}$ and a $(\M+\M^\prime)\times (\M+\M^\prime)$ supermatrix $T^{h^{\prime}}{}_{h}$  satisfying  for all the  constraints,
\be
\ba{ll}
\D{\dot{\phi}_{h}=\left[\phi_{h},H+\phi_{m}u^{m}\right\}_{P.B.} +\frac{\partial\phi_{h}}{\partial t} =\phi_{h^{\prime}}T^{h^{\prime}}{}_{h}\,,}~~&~~1\leq h\leq\M+\M^\prime\,,
\ea
\label{dotphi}
\ee
or equivalently
\be
\D{\left[\phi_{h},H\right\}_{P.B.} +\frac{\partial\phi_{h}}{\partial t}\approx
-\left[\phi_{h},\phi_{m}\right\}_{P.B.}u^{m}\,.}
\label{inheq}
\ee
{where $\approx$ indicates that the equality strictly holds after} the restriction on $\V$ or, equivalently, putting $\big(p\,,\,q\big)=\big(f(y,t)\,,\,g(y,t)\big)$, after taking the derivatives.\\

The most general solution of Eq.(\ref{inheq}) reads
\be
u^{m}(p,q,t)=U^{m}(p,q,t)+V^{m}{}_{i}(p,q,t)\,\free^{i}(t)\,,
\label{mostu}
\ee
where $\free^{i}(t)$ (with $1\leq i\leq\cal I\leq \M$) are arbitrary time dependent functions,  $U^{m}(p,q,t)$ is  one particular solution, and    $V^{m}{}_{i}(p,q,t)$ span a basis of the kernel of the $(\M+\M^\prime)\times\M$ supermatrix $\left.\left[\phi_{h},\phi_{m}\right\}_{P.B.}\right|_{\V}\,$,
\be
\ba{ll}
\D{\sum_{m=1}^{\M}\,\left[\,\phi_{h},\phi_{m}
\right\}_{P.B.}\,V^{m}{}_{i}\approx 0\,,}~~&~~\mbox{for~all~~}1\leq h\leq\M+\M^\prime\,.
\ea
\label{1stc}
\ee

\item When $\phi_{m}=p_{m}-f_{m}(q,p_{a},t)$, substituting the most general {solution (\ref{mostu}) for $u^m$ into the expression of the velocity $$\dq^{A}=\dq^{A}\Big(q^B,p_{a},(-1)^{\#_{m}}u^{m}(p,q,t),t\Big)$$ given by (\ref{velocitydqm}),  the total Hamiltonian reads simply:}
\be
H_{T}(p_A,q^B,t)\,=\,\dq^{A}\,p_{A}\,-\,L(q^B,\dq^A,t)\,.
\label{simpletotalH}
\ee

\item As in (\ref{gTH}), one can generalize the total Hamiltonian dynamics  by adding terms proportional to the constraints. In particular, still preserving  the Poisson bracket structure of the time evolution, (\ref{HD1}), - which is essential for  the  quantization - one can modify  the total Hamiltonian alone by adding freely  terms  quadratic (or higher) in the constraints,  both  primary and  secondary,
\be
\ba{lll}
H_{T}=H+\phi_{m}u^{m}~~&~~~\Longrightarrow~~~&~~ \D{H_{T}=H+\phi_{m}u^{m}+\half\phi_{h}\phi_{h^{\prime}}w^{hh^{\prime}}}\,.
\ea
\label{generaliationTH}
\ee
where $1\leq h,h^{\prime}\leq \M+\M^\prime$, while $1\leq m\leq\M$, and $w^{hh^{\prime}}=(-1)^{\#_{h}\#_{h^{\prime}}}w^{h^{\prime}h}$ are newly introduced  local parameters,  being   arbitrary time dependent  functions. The modification does not affect our previous analysis at all, and the resulting dynamics remains the same on the hypersurface $\V\,$.

\item Other characteristic features of the total Hamiltonian dynamics are discussed in the following subsections.
\end{itemize}

\subsection{First-class and second-class}

We define a dynamical quantity, say $F(p,q,t)$ a function on phase space, to be ``first-class", if it has zero Poisson brackets with all the  constraints, both primary and secondary, on $\V$,
\be
\ba{ll}
\D{\Big[F,\phi_{h}\Big\}_{P.B.}\approx 0\,,}~~~~&~~~1\leq h\leq \M+\M^\prime\,.
\ea
\ee
Otherwise it is said to be ``second-class". Physically, this distinction is extremely important, particularly for the constraints, as will be explained later on in this subsection.\\

\subsubsection{Main properties of first-class quantities}

We enunciate some useful properties of first-class quantities:
\begin{itemize}
\item \textit{If the function $F$ on phase space is first-class, then the Poisson bracket $\left[F,\phi_{h}\right\}_{P.B.}$ must be a linear combination of the constraints,}
\be
\D{\Big[F,\phi_{h}\Big\}_{P.B.}=f_{h}{}^{h^{\prime}}\phi_{h^{\prime}}\,.}
\label{Ffirstcl}
\ee

\item \textit{If  $F$ is first-class, then for any arbitrary dynamical variable, say $G(p,q,t)$,}
\be
\D{\left.\Big[F,\left. G\,\right|_{\V}\Big\}_{P.B.}\right|_{\V}=
\left.\Big[F,G \Big\}_{P.B.}\right|_{\V}\,.}
\label{VV}
\ee
\proof{For an arbitrary analytic function, $G(p,q,t)$, with the new coordinates for the $2\N$-dimensional phase space, $(y^{\hat{\imath}},\,\phi_{h})$ from (\ref{newco}),  we define $\widetilde{G}(y,\phi,t):=G(p,q,t)$.
Then from the property of being first-class and the Leibniz rule (\ref{Leibniz}) of the Poisson bracket, one can derive the result,
\be
\left.\Big[F,G \Big\}_{P.B.}\right|_{\V}\!
=\!\left.\Big[F, \widetilde{G}(y,\phi,t)\Big\}_{P.B.}\right|_{\V}\!
=\!\left.\Big[F, \widetilde{G}(y,0,t)\Big\}_{P.B.}\right|_{\V}\!
=\!\left.\Big[F,\left. G\,\right|_{\V}\Big\}_{P.B.}\right|_{\V}\,.
\ee
}

\item  \textit{The Poisson bracket of two first-class quantities is also first-class.}
\proof{This can be shown from the Jacobi identity (\ref{Jacobi}). For two first-class quantities, say $F_{1}$ and $F_{2}$,
\be
\ba{l}
\D{\left.\Big[\left[F_{1},F_{2}\right\}_{P.B.},
\phi_{h}\Big\}_{P.B.}\right|_{\V}}\\
{}\\
=\D{\left.\left.
\Big[F_{1},\big[F_{2},\phi_{h}\big\}_{P.B.}\Big\}_{P.B.}\right|_{\V}-(-1)^{\#_{F_{1}}\#_{F_{2}}}
\Big[F_{2},\left[F_{1},\phi_{h}\right\}_{P.B.}\Big\}_{P.B.}\right|_{\V}}\\
{}\\
=\D{\left.\left.
\left[F_{1},f_{2h}{}^{h^{\prime}}\phi_{h^{\prime}}\right\}_{P.B.}\right|_{\V}-
(-1)^{\#_{F_{1}}\#_{F_{2}}}
\left[F_{2},f_{1h}{}^{h^{\prime}}\phi_{h^{\prime}}\right\}_{P.B.}\right|_{\V}}\\
{}\\
=0\,.
\ea
\label{fff}
\ee
}
\end{itemize}

\subsubsection{First-class constraints as gauge symmetry generators}\label{firstclassgauge}

From Eq.(\ref{1stc}),  the complete set of primary first-class constraints is given by
\be
\ba{ll}
\phi^{1{\rm st}}_{i}:=\phi_{m}V^{m}{}_{i}\,,~~~~&~~~~
\D{\left[\phi_{h},\phi_{i}^{1{\rm st}}\right\}_{P.B.}\approx 0\,.}
\ea
\label{prfirstclctr}
\ee
In virtue of (\ref{generaliationTH}) and (\ref{prfirstclctr}) the total Hamiltonian reads now
\be
H_{T}=H^{\prime}+\phi^{1{\rm st}}_{i}\free^{i}(t)+\half\phi_{h}\phi_{h^{\prime}}w^{hh^{\prime}}\,,
\label{totham}
\ee
where $H^\prime$ is defined as
\be
H^{\prime}:=H+\phi_{m}U^{m}\,,
\label{TOTH}
\ee
and satisfies, for any $1\leq h\leq\M+\M^\prime$,
\be
\D{\left[\phi_{h},H^{\prime}\right\}_{P.B.}\approx
\left[\phi_{h},H_{T}\right\}_{P.B.}\approx-
\left.\frac{\partial\phi_{h}}{\partial t}\right|_{\V}\,.}
\ee
Thus, when the constraints do not have any explicit time dependence,  both $H^{\prime}$ and $H_{T}$ are  first-class, and up to quadratic terms in the constraints the total Hamiltonian is a sum of the first-class Hamiltonian $H^\prime$ plus a linear combination of the primary, first-class constraints with arbitrary functions of time as coefficients. Notice that such a decomposition is not unique since $U^m$ can be any solution of the inhomogeneous equation (\ref{inheq}).\\

The  time derivative of an arbitrary function $F(p,q,t)$, \textit{c.f.} (\ref{HD1}), on the $2\N$-dimensional phase space  can be rewritten as
\be
\D{\frac{{\rm{d}}F(p,q,t)}{{\rm{d}}t}=\left[F,H^{\prime}\right\}_{P.B.}+
\left[F,\phi^{1{\rm st}}_{i}\right\}_{P.B.}\free^{i}(t)+\frac{\partial F}{\partial t}\,.}
\label{HD2}
\ee
As a result of the appearance of the arbitrary time dependent functions $\free^{i}(t)\,$, the dynamical variables at future times are not completely determined by the initial dynamical variables. \\

We should recall the following viewpoint by Dirac~\cite{Dirac}: ``all those values for the $p$'s and $q$'s at a certain time which can evolve from one initial state must correspond to the same physical state at that time''. A natural definition of the space of physical states is thus as the set of initial variables that should be given at some given moment of time, say $t_0$, in order to determine completely the time evolution via the equations of motion.
The presence of arbitrary functions of time, $\free^i(t)\,$, in the total Hamiltonian $H_T$ signals that the phase space contains some unphysical degrees of freedom. Indeed two different choices of arbitrary functions, say $\free_1^i$ and $\free_2^i\,$, would lead to distinct total Hamiltonians and thus different time change of a dynamical variable, say $F$. After some time interval $\delta t\,$, the evolutions of $F$ would differ by
\be
\delta F=\left[F,\phi^{1{\rm st}}_{i}\right\}(\free_2^i-\free_1^i)\,\delta t\,.
\label{gtransfgen}
\ee
The key philosophy we stick to is the standard one that \textit{different choices of the local gauge parameters correspond to different total Hamiltonian systems, which nevertheless should be taken equivalent, describing the same physics.} Following the viewpoint advocated in subsection \ref{gsym} for Lagrangian systems, this means that Eq.(\ref{gtransfgen}) defines an ambiguity in the time evolution that should be physically irrelevant. In other words, the transformation (\ref{gtransfgen}) is a gauge symmetry. In modern terminology, this implies that:
\begin{itemize}
	\item[(i)] \textit{a physical state is represented by an equivalence class, where one mods out by the gauge symmetries}, therefore
	\item[(ii)] \textit{the space of physical states must be understood as the quotient of the constraint surface by the gauge orbits} and
	\item[(iii)] \textit{an observable is a gauge invariant function on the constraint surface.}
\end{itemize}
The space of physical states is also a symplectic manifold\footnote{The section 1.4.2 of \cite{Henneaux:1992ig} is devoted to the subtle counting of physical degrees of freedom in the Hamiltonian context, which  is equal to half the dimension of the symplectic manifold.} and is sometimes called ``reduced phase space.'' As a short dictionary for physicists on mathematical jargon, we may say that the former quotient space is usually called ``symplectic reduction'' by mathematicians while they would introduce gauge transformations via a ``Lie group action'' and might refer to Noether's theorem as the ``moment map'' (see \textit{e.g.} \cite{Souriau}\,).

\vspace{2mm}A remarkable property is that the Poisson bracket $\{H^\prime,\phi^{1{\rm st}}_{i}\}$ between the first class Hamiltonian $H^\prime$ and any primary first class constraint $\phi^{1{\rm st}}_{i}$ is also a gauge symmetry generator. This can be shown by comparing the time evolution successively determined by (i) the total Hamiltonian $H_T$ during an interval $\delta t_1$ and after by the first-class Hamiltonian $H^\prime$ during an interval $\delta t_2$, or (ii) the same operations, but in the reverse order. The net difference must be a gauge transformation since $H_T$ and $H^\prime$ define the same evolution of physical states. By using the Jacobi identity, one may check explicitly that this gauge transformation is indeed generated by $\{H^\prime,\phi^{1{\rm st}}_{i}\}\,$.\\

As one can see in (\ref{gtransfgen}), \textit{the primary first-class constraints generate gauge symmetries}.
A natural question is whether the converse is true: are all gauge symmetries generated by primary first-class constraints? In full generality, the answer is no. This can be understood from the fact that the Poisson bracket $\{H^\prime,\phi^{1{\rm st}}_{i}\}$ is also a gauge symmetry generator. From Eq.(\ref{Ffirstcl}) we know that this Poisson bracket is a linear combination of first-class constraints, but it is not guaranteed that only primary constraints appear. Therefore, some secondary constraints may also generate gauge transformations. Then another question arises: do all the secondary first-class constraints generate gauge symmetries? Dirac conjectured that the answer would be yes.
But, again, in full generality the answer is no, although in most physical applications the answer is yes.\footnote{A counterexample of the Dirac conjecture is given in subsection 1.2.2 of the book \cite{Henneaux:1992ig}. A proof of the Dirac conjecture under some hypotheses is provided in its subsection 3.3.2.} This is the reason why first-class quantities have such a distinct status.

\subsection{Extended Hamiltonian dynamics}\label{extended}

One can generalize the total Hamiltonian system further to the so-called ``extended Hamiltonian'' system.  We define the extended Hamiltonian in a similar way to the total one except that the former includes \textit{all} first-class constraints (the primary as well as the secondary, tertiary, \textit{etc}),
\be
H_{E}(p,q,t,\free^{i},\free^{i^\prime},w):=H_{T}+\phi^{1{\rm st}}_{i^\prime}\free^{i^\prime}(t)=H^{\prime}(p,q,t)+\phi^{1{\rm st}}_{i}\free^{i}(t)+\phi^{1{\rm st}}_{i^\prime}\free^{i^\prime}(t)+\half\phi_{h}\phi_{h^{\prime}}w^{hh^{\prime}}\,,
\ee
where $\phi^{1{\rm st}}_{i^\prime}$ (with $1\leq i^\prime\leq{\cal I}^\prime\leq \M^\prime$) denotes the secondary first-class constraints.

The extended Hamiltonian is usually preferred because, from the Hamiltonian point of view, the distinction between primary and secondary constraints is actually irrelevant. The distinction becomes important only if one wants to make contact with the Lagrangian formulation, as in Subsection \ref{HtoL}. So, one may let $H_{E}$ governs the whole dynamics rather than $H_{T}$,
\be
\D{\frac{{\rm d}F}{{\rm d}t}=[F,H_{E}\}_{P.B.}+\frac{\partial F}{\partial t}\,.}
\ee
Compared to the total Hamiltonian dynamics, the constraint surface $\V$, is still preserved but a generic other object, say $F$ (rather than the constraint $\phi_{h}$), undergoes a different time evolution from the total Hamiltonian dynamics, even on $\V\,$. Indeed,
$\left.[F,\phi^{1{\rm st}}_{i^\prime}\}_{P.B.}\right|_{\V}\neq 0$ in general. Thus, unlike the total Hamiltonian dynamics, the extended Hamiltonian dynamics is in general different from the Lagrangian dynamics. Still, if the Poisson bracket of a quantity $F$ with any secondary first-class constraint
is zero, then its evolution on $\V$ are the same.\\

With an arbitrary local parameter $\varepsilon(t)\,$, if there exists a gauge symmetry generator $\Q^{1{\rm st}}=Q^{1{\rm st}}\,\varepsilon(t)$
which takes any solution $(q^{A},p_B)(t)$ of the extended Hamiltonian dynamics to another by
\be
\delta q^{A}=[\,q^{A},  Q^{1{\rm st}}\}_{P.B.}\,\varepsilon(t)\,,\quad \delta p_{B}=[\,p_{B},  Q^{1{\rm st}}\}_{P.B.}\,\varepsilon(t)\,,
\ee
then at any time, say $t_{0}$, one can start to transform the solution to another without changing the initial data $(q^{A},p_B)(t_{0})$ by simply setting $\varepsilon(t_{0})=0$.  Thus, again the future dynamical variables are not uniquely determined by the initial data. From the point of view of the gauge symmetries, the main difference between the total and extended Hamiltonian dynamics is that the latter assumes the Dirac conjecture applies.
In this case, a dynamical quantity $F(q^A,p_B,t)$ on the phase space defines an observable if and only if its Poisson bracket with any first-class constraint vanishes weakly
\be
\left[\,F,\phi^{1{\rm st}}_{i^{\prime\prime}}\right\}\approx 0\,,\quad(1\leq i^{\prime\prime}\leq {\cal I}+{\cal I}^\prime)\,.
\ee
Notice that an observable has been defined as a function on the constraint surface, so one should identify two functions  that coincide on $\V\,,$ \textit{i.e.} the observable corresponding to the first-class quantity $F$ is the equivalence class for the weak equality.
The conclusion is that the physical quantities (that is, the observables) undergo the same evolution under the total and extended Hamiltonian dynamics.\footnote{For more comments on the relationship between the total and extended Hamiltonian formalisms the reader may look, \textit{e.g.} at \cite{Dresse:1990dj}.}

\subsection{Time independence of the Poisson bracket}

For a given set $\{\free^{i}(t)\}$ of the local functions, the  dynamical variables $(p,q)$  follow a unique and invertible trajectory in the phase space such that there exists a   one to one map between $(p,q)$ and the initial data,
\be
\ba{llll}
p_{A}(t,p_{0},q_{0})\,,~~~&~p_{A}(t_{0},p_{0},q_{0})=p_{0A}\,;~~~~&~~~~q^{B}(t,p_{0},q_{0})\,,
~~~&~q^{B}(t_{0},p_{0},q_{0})=q_{0}^{B}\,.
\ea
\ee

A crucial fact follows, proven in Eq.(\ref{tiPproof}):
\textit{The Poisson bracket is independent of time},
\be
\D{\left[~~~,~~~\right\}_{P.B.}=\left.\left[~~~,~~~\right\}_{P.B.}\right|_{t_{0}}}\,,
\ee
or
\be
\D{(-1)^{\#_{A}}\frac{\overleftarrow{\partial} ~~~}{\partial q^{A}}\,\frac{\overrightarrow{\partial}~~~ }{\partial p_{A}}~-~\frac{\overleftarrow{\partial} ~~~}{\partial p_{A}}\,\frac{\overrightarrow{\partial}~~~ }{\partial q^{A}}=
(-1)^{\#_{A}}\frac{\overleftarrow{\partial} ~~~}{\partial q_{0}^{A}}\,\frac{\overrightarrow{\partial}~~~ }{\partial p_{0A}}~-~\frac{\overleftarrow{\partial} ~~~}{\partial p_{0A}}\,\frac{\overrightarrow{\partial}~~~ }{\partial q_{0}^{A}}\,.}
\label{tiP}
\ee
Namely  the time evolution generated by the total Hamiltonian is a symplectic transformation. \\

\subsection{Other remarks on the total Hamiltonian formalism}

\begin{itemize}
\item A useful identity.\\
For an arbitrary function $F(p,q,t)\,$, we have the following identity,\footnote{The above relation (\ref{VddV})  should be compared with
\[
\D{\frac{\partial~}{\partial t}\Big(\left.F\right|_{\V}\Big)=
\left.\frac{\partial F}{\partial t}\right|_{\V}+\frac{\partial p_{A}}{\partial t}\left.\left(\frac{\partial F}{\partial p_{A}}\right)\right|_{\V}+\frac{\partial q^{A}}{\partial t}\left.\left(\frac{\partial F}{\partial q^{A}}\right)\right|_{\V}}\,.
\]}
\be
\D{\frac{{\rm d}~}{{\rm d}t}\Big(F(p,q,t)\Big|_{\V}\Big)=\left.\left(\frac{{\rm d}~}{{\rm d}t}F(p,q,t)\right)\right|_{\V}}\,,
\label{VddV}
\ee
proven in Eq.(\ref{VddVproof}).
In words, the following two actions, taking the time derivative and restricting on $\V$, commute with each other. Intuitively, this is obvious since we imposed that $\V$ be preserved under the time evolution.

\item The time derivatives of the primary first-class constraints $\phi^{1{\rm st}}_{i}$ are of the general form, using Eqs. (\ref{dotphi}) and (\ref{totham}),
\be
\ba{ll}
\multicolumn{2}{l}{\D{\frac{{\rm d}{\phi}^{1{\rm st}}_{i}}{{\rm d}t}
:=\phi_{h}T^{h}{}_{i}=\left(\left[\phi^{1{\rm st}}_{i},H^{\prime}\right\}_{P.B.}+\frac{\partial \phi^{1{\rm st}}_{i} }{\partial t}\right)\,+\,\free^{1{\rm st}}_{\bot i}\,,}}\\
{}&{}\\
~\free^{1{\rm st}}_{\bot i}:=\left[\phi^{1{\rm st}}_{i},\phi^{1{\rm st}}_{j}\right\}_{P.B.}\free^{j}\,,
~~~~&~~~~\free^{i}\free^{1{\rm st}}_{\bot i}=0\,,\\
{}&{}\\
\multicolumn{2}{l}{\D{\Big[\free^{1{\rm st}}_{\bot i},\phi_{h}\Big\}_{P.B.}\approx 0~~~~\mbox{for\,~all~~~}h=1,2,\cdots, \M+\M^\prime\,.}}
\ea
\label{dotphi1st}
\ee
Namely the time derivative of any primary, first-class constraint decomposes into two parts, one  being independent of the local gauge parameters $\free^{i}(t)\,$, and the other one being first-class and orthogonal to the local gauge parameter.
\item In the static case, where there is no explicit time dependence in $\phi_{h}(p,q)$ and $H^{\prime}(p,q)$ defined in (\ref{TOTH}),  the time derivative of any primary as well as secondary first-class constraint, $\phi^{1{\rm st}}$,  is first-class\footnote{Actually, this is an example of the fact that the Poisson bracket of two first-class quantities is also first-class (\ref{fff}).} too, since $\D{\dot{\phi}^{1{\rm st}}=\left[\phi^{1{\rm st}},H_{T}\right\}_{P.B.}}\,$ so that
\be
\D\left[\dot{\phi}^{1{\rm st}},\phi_{h}\right\}_{P.B.}=
\left[\phi^{1{\rm st}},\left[H_{T},\phi_{h}\right\}\right\}_{P.B.}-
\left[H_{T},\left[\phi^{1{\rm st}},\phi_{h}\right\}\right\}_{P.B.}\approx 0\,.
\label{staticdot1st}
\ee
Hence
, in the static case, for all the first-class constraints, both the primary ones, $\phi^{1{\rm st}}_{i}$,
and secondary ones, $\phi^{1{\rm st}}_{i^\prime}$, we can write
\be
\ba{l}
\D{\dot{\phi}^{1{\rm st}}_{i}=\left[\phi^{1{\rm st}}_{i},H_{T}\right\}_{P.B.}=\phi^{1{\rm st}}_{j}\T^{j}{}_{i}+\phi^{1{\rm st}}_{i^\prime}\T^{i^\prime}{}_{i}\,,}\\
{}\\
\D{\dot{\phi}^{1{\rm st}}_{i^\prime}=\left[\phi^{1{\rm st}}_{i^\prime},H_{T}\right\}_{P.B.}=\phi^{1{\rm st}}_{j^\prime}\T^{j^\prime}{}_{i^\prime}+\phi^{1{\rm st}}_{j}\T^{j}{}_{i^\prime}\,.}
\ea
\label{TTTT}
\ee

\item Combining the above two results in the static case, $[{\phi}^{1{\rm st}}_{i},H^{\prime}\}_{P.B.}$ and $\free^{1{\rm st}}_{\bot i}$ are separately first-class constraints.
\end{itemize}

\newpage
\section{Symmetry in the total Hamiltonian system}\label{SymmetrY}

\subsection{Symmetry from the Lagrangian system - revisited}\label{preSTH}

In the Lagrangian formalism,\footnote{This subsection is parallel to Sec.\ref{LtoHK} where the Hamiltonian corresponds to the time translational symmetry generator.} the notion of symmetry corresponds to a change of variables\footnote{{In order to avoid confusion with some subsequent notations, we slightly changed the convention by using `prime' instead of `tilde' to denote the new variables.}} $q^{A}\rightarrow q^{\prime A}$ which leaves the Lagrangian invariant up to the total derivative term  as in (\ref{symmetry}),
\be
L(q^{\prime},\dq^{\prime},t)=L(q,\dq,t)+\D{\frac{{\rm d}K}{{\rm d}t}\,.}
\ee
As a consequence, the symmetry takes one solution of the Euler-Lagrange equations to a new one as in (\ref{qQEOM}),
\be
\ba{lll}
\D{\frac{\partial L(q,\dq,t)}{\partial q^{A}}\equiv\frac{{\rm d}\,}{{\rm d}t}\left(\frac{\partial L(q,\dq,t)}{\partial \dq^{A}}\right)}~~~&\Longrightarrow&~~~
\D{\frac{\partial L(q^{\prime},\dq^{\prime},t)}{\partial q^{\prime A}}\equiv\frac{{\rm d}\,}{{\rm d}t}\left(\frac{\partial L(q^{\prime},\dq^{\prime},t)}{\partial \dq^{\prime A}}\right)\,.}
\ea
\ee
As discussed in Section \ref{subsecPRE}, if the infinitesimal symmetry transformation $q^{A}\rightarrow q^{A}+\delta q^{A}(q,\dq,t)$  depends  on $q^{A}$, $\dq^{A}$, $t$ only, then the quantity  $\delta K$ in $\delta L=\frac{{\rm d}~}{{\rm d}t}\delta K$ must also depend only on $q^{A}$, $\dq^{A}$, $t\,$, and hence so is the corresponding Noether charge:
\be
Q(q,\dq,t)=\delta q^{A}(q,\dq,t)p_{A}(q,\dq,t)-\delta K(q,\dq,t)\,.
\ee

\subsubsection{Change of variables}

{Henceforth in the present subsection, \textit{i.e.} until Eq.(\ref{pQ}), we take $(q,p_{a},\dq^{\hat{m}},t)$ as the independent variables for any quantity which carries a hat symbol. For instance, we set
\be
\ba{l}
\widehat{p}_{_A}:=\left(\frac{\partial L(q,\dq,t)}{\partial\dq^{A}}\right)_{\dq^{\hat a}\,=\, h^{\hat{a}}(q^B,\,p_{a},\,\dq^{\hat m},\,t)}\,,~~~~~~\widehat{p}_{a}
:=p_{a}\,,~~~~~~\widehat{p}_{m}:=f_{m}(q^B,p_{a},t)\,,\\
{}\\
\Deltaq^{A}(q^B,p_{a},\dq^{\hat{m}},t):=\delta q^{A}\Big(q^B,h^{\hat{a}}(q,p_{a},\dq^{\hat{n}},t),\dq^{\hat{m}},t\Big)\,,\\
{}\\
\DeltaK(q^B,p_{a},\dq^{\hat{m}},t):=\delta  K\Big(q^B,h^{\hat{a}}(q,p_{a},\dq^{\hat{n}},t),\dq^{\hat{m}},t\Big)\,,
\label{bfq}
\ea
\ee
where we substituted $\dq^{\hat{a}}$ by $h^{\hat{a}}(q^B,p_{a},\dq^{\hat{m}},t)$ according to the relation in Eq.(\ref{primary1}).
Notice that
\be
\frac{\partial \widehat{p}_B}{\partial \dq^{\hat{m}}}=0\,,\label{widehatp}
\ee
as follows from (\ref{primary1}).
Furthermore, in agreement with Eq.(\ref{deltap0}) we define  another  function  depending on $(q,p_{a},\dq^{\hat{m}},t)$,
\be
\ba{ll}
\D{\Deltap_{A}(q,p_{a},t,\dq^{\hat{m}})}\!\!&\D{:
=\left[\delta\!\left(\frac{\partial L}{\partial \dq^A}\right)
-\frac{\partial (\delta q^{B})}{\partial\dq^{A}}\left(
\frac{{\rm{d}\,}}{{\rm{d}t}}\left(\frac{\partial L}{\partial\dq^{B}}\right)-\frac{\partial L}{\partial q^{B}}\right)\right]_{\dq^{\hat a}\,=\, h^{\hat{a}}(q,\,p_{a},\,\dq^{\hat m},\,t)}}\\
{}&{}\\
{}&\D{\,\,=\left[
\frac{\partial\,(\delta K)}{\partial q^{A}}-\frac{\partial\,(\delta q^{B})}{\partial q^{A}}\frac{\partial L}{\partial\dq^{B}}\right]_{\dq^{\hat a}\,=\, h^{\hat{a}}(q,\,p_{a},\,\dq^{\hat m},\,t)}\,,}
\ea
\label{Deltap}
\ee
which has been defined in such a way that the r.h.s. is independent of the accelerations, as it should in the Hamiltonian formalism to come.
Similarly, notice that the evaluation of (\ref{KL1}) at $\dq^{\hat a}\,=\, h^{\hat{a}}(q,\,p_{a},\,\dq^{\hat m},\,t)\,$, gives
\be
\left(\frac{\partial (\delta K)}{\partial \dq^A}\right)_{\dq^{\hat{a}}= h^{\hat{a}}}\,=\,\left(\frac{\partial (\delta q^B)}{\partial \dq^A}\right)_{\dq^{\hat{a}}= h^{\hat{a}}}\,\widehat{p}_B\,,
\label{KKL1}
\ee
From the next equation (\ref{dpqm}) until the equation (\ref{pQ}), the partial derivatives acting on any quantity with a hat symbol, say
\be
\widehat{A}(q^B,p_{a},\dq^{\hat{m}},t):=\Big(A(q,\dq,t)\Big)_{\dq^{\hat{a}}= h^{\hat{a}}(q,p_{a},\dq^{\hat{m}},t)}\,\,,
\ee
is taken regarding
$(q^A,p_{B},\dq^{\hat{m}},t)$ as independent variables, while for unhatted quantities the independent variables are $(q^A,\dq^{B},t)$.
Concretely, this means that one should make use of the chain rule, so that
\begin{eqnarray}
\frac{\partial\widehat{A}}{\partial q^A}&=&\Big(\frac{\partial A}{\partial q^A}\Big)_{\dq^{\hat{a}}= h^{\hat{a}}}\,+\,\frac{\partial h^{\hat{b}}}{\partial q^A}\,\Big(\frac{\partial A}{\partial \dq^{\hat{b}}}\Big)_{\dq^{\hat{a}}= h^{\hat{a}}}\,,\nonumber\\
\frac{\partial\widehat{A}}{\partial \dq^A}&=&\frac{\partial h^{\hat{b}}}{\partial \dq^A}\,\Big(\frac{\partial A}{\partial \dq^{\hat{b}}}\Big)_{\dq^{\hat{a}}= h^{\hat{a}}}\,+\,\delta_A^{\hat{m}}\,\Big(\frac{\partial A}{\partial \dq^{\hat{m}}}\Big)_{\dq^{\hat{a}}= h^{\hat{a}}}\,.
\label{chrule}
\end{eqnarray}
On-shell, the variation (\ref{Deltap}) is equal to the variation of the momenta. Moreover, it satisfies
\be
\ba{ll}
\D{\frac{\partial\,\Deltap_{A}(q,p_{a},\dq^{\hat{m}},t)}{\partial\dq^{\hat{m}}}}&
\D{=\frac{\partial\,\Deltaq^{B}(q,p_{a},\dq^{\hat{m}},t)}{\partial\dq^{\hat{m}}}\left(
\frac{\partial^{2}L(q,\dq,t)}{\partial\dq^{B}\partial q^{A}}\right)_{\dq^{\hat a}\,=\, h^{\hat{a}}(q,\,p_{a},\,\dq^{\hat m},\,t)}}\\
{}&{}\\
{}&\D{=(-1)^{\#_{A}\#_{n}}\frac{\partial\,\Deltaq^{n}(q,p_{a},t,\dq^{\hat{m}})}{
\partial\dq^{\hat m}}\frac{\partial f_{n}(q,p_{a},t)}{\partial q^{A}}\,.}
\ea
\label{dpqm}
\ee
The proof of these two equalities is provided in Appendix \ref{someProofs}.}
Similarly to the corresponding steps, we also have the following identities:
\be
\ba{l}
\D{(-1)^{\#_{A}\#_{B}}\frac{\partial\Deltaq^{B}}{\partial\dq^{\hat m}}\frac{\partial h^{\hat a}}{\partial p_{A}}
\frac{\partial^{2}L}{\partial h^{\hat a}\partial\dq^{B}}=0\,,}\\
{}\\
\D{(-1)^{\#_{A}\#_{B}}\frac{\partial\Deltaq^{B}}{p_{C}}\frac{\partial h^{\hat a}}{\partial q^{A}}
\frac{\partial^{2}L}{\partial h^{\hat a}\partial\dq^{B}}=
(-1)^{(\#_{A}+\#_{B})\#_{C}}\frac{\partial h^{\hat a}}{\partial q^{A}}
\frac{\partial\,\delta q^{B}}{\partial h^{\hat a}}\frac{\partial\hat{p}_{B}}{\partial p_{C}}\,.}
\ea
\label{int2}
\ee
~\newline

The above substitution  induces  a Noether charge depending  on  $(q,p_{a},t)$:
\be
\widehat{Q}(q^A,p_{a},t)=\Deltaq^{A}\widehat{p}_{A}-\DeltaK\,.
\label{Qhat}
\ee
From (\ref{KKL1}) one can easily verify that the Noether charge is indeed a function of $q^{A}$, $p_{a}$ and $t$ only (\textit{i.e.} it is independent of $\dq^{\hat{m}}$),
\be
\D{\frac{\partial\widehat{Q}~}{\partial \dq^{\hat{m}}}=\left(\frac{\partial h^{\hat{a}}}{\partial \dq^{\hat{m}}}\frac{\partial (\delta q^{A})}{\partial \dq^{\hat{a}}}+
\frac{\partial (\delta q^{A})}{\partial \dq^{\hat{m}}}\right)\widehat{p}_{A}
-\left(\frac{\partial h^{\hat{a}}}{\partial \dq^{\hat{m}}}\frac{\partial (\delta K)}{\partial \dq^{\hat{a}}}+\frac{\partial (\delta K)}{\partial \dq^{\hat{m}}}\right)=0\,.}
\label{Qpdqm}
\ee
Furthermore, with  (\ref{KKL1})  we get
\be
\ba{ll}
\D{\frac{\partial\widehat{Q}(q,p_{a},t)}{\partial p_{a}}}&\D{=
\frac{\partial h^{\hat{a}}}{\partial p_{a}}\frac{\partial (\delta q^{A})}{\partial \dq^{\hat{a}}}\,\widehat{p}_{A}+(-1)^{\#_{a}}\Deltaq^{a}+
(-1)^{\#_{a}\#_{m}}\Deltaq^{m}\frac{\partial  f_{m}}{\partial p_{a}}-\frac{\partial h^{\hat{a}}}{\partial p_{a}}
\frac{\partial (\delta K)}{\partial \dq^{\hat{a}}}}\\
{}&{}\\
{}&\D{=(-1)^{\#_{a}}\Deltaq^{a}+(-1)^{\#_{a}\#_{m}}\Deltaq^{m}\frac{\partial  f_{m}}{\partial p_{a}}\,,}
\ea
\label{Qpp}
\ee
and  with (\ref{Deltap}),
\be
\ba{ll}
\D{\frac{\partial\widehat{Q}(q,p_{a},t)}{\partial q^{A}}}&\D{=\left(\frac{\partial(\delta q^{B})}{\partial q^{A}}+\frac{\partial h^{\hat{a}}}{\partial q^{A}}\frac{\partial (\delta q^{B})}{\partial h^{\hat{a}}}\right)\widehat{p}_{B}+(-1)^{\#_{A}\#_{m}}\Deltaq^{m}\frac{\partial f_{m}}{\partial q^{A}}-\frac{\partial(\delta K)}{\partial q^{A}}-\frac{\partial h^{\hat{a}}}{\partial q^{A}}\frac{\partial (\delta K)}{\partial h^{\hat{a}}}}\\
{}&{}\\
{}&\D{=-\Deltap_{A}+(-1)^{\#_{A}\#_{m}}\Deltaq^{m}\frac{\partial f_{m}}{\partial q^{A}}\,.}
\ea
\label{pQ}
\ee
{In order for the Noether charge $\widehat{Q}(q,p_{a},t)$ to generate the symmetry transformations $(\Deltaq,\Deltap)$ \textit{via} the Poisson bracket in the corresponding Hamiltonian system, the derivatives of $f_m$ in the r.h.s. of (\ref{Qpp}) and (\ref{pQ}) should be absent.} Instead, in a spirit similar to the total Hamiltonian, we  will define a ``total Noether charge'' which will be  a function of $(p_{A},q^{B},\dq^{{m}},t)$ rather than\footnote{Note that the equation (\ref{velocitydqm}) implies that $\frac{\partial~}{\partial\dq^{m}}=\frac{\partial\dq^{\hat m}}{\partial\dq^{m}}\frac{\partial~}{\partial\dq^{\hat m}}=(-1)^{\#_{\hat m}(1+\#_{m})}
\frac{\partial\phi_{m}}{\partial p_{\hat m}}\frac{\partial~}{\partial\dq^{\hat m}}$.} $(p_{A},q^{B},\dq^{\hat{m}},t)$.

\subsubsection{Total Noether charge}

{Let us denote by $\v^A$ the explicit expression of $\dq^A$ in terms of the variables $q^B\,$, $p_{a}\,$, $\dq^{{m}}$ and $t\,$, as in (\ref{velocitydqm}).}
Substituting the velocities $\dq$ by their explicit form $\v(q,p_{a},\dq^{{m}},t)$ henceforth, we
take $(q^B,p_{a},\dq^{{m}},t)$ as the independent variables for any quantity which carries  a tilde symbol. In other words, we set
\be
\ba{l}
\widetilde{p}_{A}:=\left(\frac{\partial L(q,\dq,t)}{\partial\dq^{A}}\right)_{\dq^A\,=\,\v^A(q,p_{a},\dq^{{m}},t)}=\widehat{p}_{A}\,,
~~~~~~~\widetilde{p}_{a}:=p_{a}\,,~~~~~\widetilde{p}_{m}:=f_{m}(q,p_{a},t)\,,\\
{}\\
\DeltatK(q,p_{a},\dq^{{m}},t):=\delta  K(q,\v,t)=\DeltaK(q,p_{a},\v^{\hat m},t)\,,\\
{}\\
\Deltatq^{A}(q,p_{a},\dq^{{m}},t):=\delta q^{A}(q,\v,t)=\Deltaq^{A}(q,p_{a},\v^{\hat{m}},t)\,,\\
{}\\
\Deltatp_{A}(q,p_{a},\dq^{{m}},t):=\Deltap_{A}(q,p_{a},\v^{\hat{m}},t)\,.
\ea
\label{vartilde}
\ee
The total Noether charge is then defined as
\be
\ba{ll}
\widetilde{Q}_{T}(p,q,\dq^{{m}},t)&:
=\,\Deltatq^{A}(q,p_{a},\dq^{{m}},t)\,p_{A}-\DeltatK(q,p_{a},\dq^{{m}},t)\\
{}&{}\\
{}&\,\,=\,\widehat{Q}(q,p_{a},t)+\Deltatq^{n}(q,p_{a},\dq^{m},t)\Big(p_{n}-f_{n}(q,p_{a},t)\Big)\,.
\label{QTdef}
\ea
\ee
Contrary  to the quantity $\widehat{Q}(q,p_{a},t)$ in (\ref{Qhat}), in the above expression of  $\widetilde{Q}_{T}(p,q,t,\dq^{{m}})$ the constrained momenta $p_{m}$  have not been substituted by the primary {constraints $p_{m}=f_{m}(q,p_a,t)\,$, since it is the untilded momenta which multiplies $\Deltatq\,$.} It follows from {Eqs.(\ref{Qpdqm})-(\ref{pQ})} that\footnote{Similar equations to (\ref{pQT}) can be straightforwardly obtained either for the case where we take $(q,p_{a},t,\dq^{\hat{m}})$ as independent variables or for the case where all the $\dq^{m}={(-1)^{\#_{m}}u^{m}}$ are completely determined in terms of $(q,p_{a},t)$ and  free parameters $\free^{i}(t)$ after solving all the constraints~(\ref{mostu}).}
\be
\ba{ll}
\multicolumn{2}{l}{\D{\frac{\partial \widetilde{Q}_{T}(p,q,t,\dq^{{m}})}{\partial \dq^{{m}}}=\frac{\partial(\Deltatq^{n})}{\partial \dq^{{m}}}\Big(p_{n}-f_{n}(q,p_{a},t)\Big)}\,,}\\
{}&{}\\
\multicolumn{2}{l}{\D{\frac{\partial \widetilde{Q}_{T}(p,q,t,\dq^{{m}})}{\partial p_{A}}=(-1)^{\#_{A}}\Deltatq^{A}+
\frac{\partial(\Deltatq^{n})}{\partial p_{A}}\Big(p_{n}-f_{n}(q,p_{a},t)\Big)}\,,}\\
{}&{}\\
\D{\frac{\partial \widetilde{Q}_{T}(p,q,t,\dq^{{m}})}{\partial q^{A}}}=&\D{-\Deltatp_{A} +\frac{\partial(\Deltatq^{n})}{\partial q^{A}}\Big(p_{n}-f_{n}(q,p_{a},t)\Big)\,.}
\ea
\label{pQT}
\ee
In particular,  in terms of the Poisson bracket we have
\be
\ba{l}
\D{[\,\widetilde{Q}_{T},q^{A}\}_{P.B.}=-\Deltatq^{A}-[\,q^{A},\Deltatq^{m}\}_{P.B.}
\Big(p_{m}-f_{m}(q,p_{a},t)\Big)\,,}\\
{}\\
\D{[\,\widetilde{Q}_{T},p_{A}\}_{P.B.}=-\Deltatp_{A}-[\,p_{A},\Deltatq^{m}\}_{P.B.}
\Big(p_{m}-f_{m}(q,p_{a},t)\Big)\,.}
\ea
\ee

Note that the expressions are consistent with the  integrability relations \textit{e.g.~}$\partial_{P_A}\partial_{\dq^{m}}\widetilde{Q}_{T}=(-1)^{\#_A\#_{m}}\partial_{\dq^{m}}\partial_{P_A}\widetilde{Q}_{T}$, thanks to (\ref{dpqm}) and (\ref{int2}).
Obviously on the primary constraint surface $V$,  where $p_{m}=f_{m}(q,p_{a},t)\,$, the above relations get simplified: the first relation in (\ref{pQT}) means {that $\widetilde{Q}_T$ becomes independent of $\dq^{{m}}$ on $V\,$,} and  the other  equations  lead to
\be
\ba{ll}
\D{[\,\widetilde{Q}_{T},q^{A}\}_{P.B.}\simeq-\Deltatq^{A}\,,}~~~&~~~\D{[\,\widetilde{Q}_{T},p_{A}\}_{P.B.}\simeq-\Deltatp_{A}\,.}
\ea
\label{pQT2}
\ee
In the Hamiltonian formalism {to come, the relations (\ref{pQT2}) will be interpreted as the property that
the total Noether charge $\widetilde{Q}_T$ generates} the infinitesimal symmetry transformations on $V$  and on-shell.\footnote{The interplay between symmetries and conserved charges in the Lagrangian \textit{vs} total Hamiltonian formalisms is briefly discussed in various exercises of the book \cite{Henneaux:1992ig} as particular cases of very general results on the elimination of auxiliary fields (mentioned in Footnote \ref{auxiliaryf}). More precisely, see \textit{e.g.} Exercises 3.17, 3.25, 3.28 and 18.15 of \cite{Henneaux:1992ig}. An analogous derivation of such results for the extended Hamiltonian formalism should follow the general procedures introduced in \cite{Dresse:1990dj}. As mentioned in the introduction, in the present text we prefer a more direct and pedestrian approach.}\\

{By making use, first of Eq.(\ref{pQT2}) and then of Eq.(\ref{Deltap}), one can show that the primary constraint surface $V$ is preserved under the infinitesimal symmetry transformations, at least on-shell. Indeed,\footnote{In each line of (\ref{QTVV}) and (\ref{QTV}), the velocity should be replaced by $\v(q,p_{a},\dq^{{m}},t)$ similarly to (\ref{velocitydqm}) and the expression is  independent of the acceleration \textit{via}  cancelation due to (\ref{ddqL}), as it must be.}
\be
\ba{l}
\D{[\,\widetilde{Q}_{T},p_{m}{-f_{m}}(q,p_{a},t)\}_{P.B.}}\,\,\,\textstyle{\!\simeq -\Deltatp_m
+\Deltatq^{A}\frac{\partial f_{m}}{\partial q^{A}}+\Deltatp_{a}\frac{\partial f_{m}}{\partial p_{a}}}\\
{}\\
\textstyle{\!\simeq-\left\{\delta\!\left(\!\frac{\partial L\,}{\partial\dq^{m}}\!\right)
-\frac{\partial (\delta q^{B})}{\partial\dq^{m}}\!
\left[\frac{{\rm{d}\,}}{{\rm{d}t}}\!
\left(\!\frac{\partial L}{\partial\dq^{B}}\!\right)\!-\!\frac{\partial L\,}{\partial q^{B}}\right]\right\}
+\Deltatq^{A}\frac{\partial f_{m}}{\partial q^{A}}+
\left\{\delta\!\left(\!\frac{\partial L}{\partial\dq^{a}}\!\right)
-\frac{\partial (\delta q^{B})}{\partial\dq^{a}}\!
\left[\frac{{\rm{d}\,}}{{\rm{d}t}}\!
\left(\!\frac{\partial L\,}{\partial\dq^{B}}\!\right)\!-\!\frac{\partial L}{\partial q^{B}}\right]\right\}
\frac{\partial f_{m}}{\partial p_{a}}}\,.
\ea
\label{QTVV}
\ee
Furthermore, we notice that the identity
\be
\delta f_m(q^A,p_a,t)=\Deltatq^{A}\frac{\partial f_{m}}{\partial q^{A}}+
\delta\!\left(\!\frac{\partial L}{\partial\dq^{a}}\!\right)\frac{\partial f_{m}}{\partial p_a}\,.
\ee
combined with the expression of momenta (\ref{ep}) when \small{$A=m\,$,}
\be
\frac{\partial L(q,\dq,t)}{\partial \dq^{m}}=f_{m}(q,\dq,t)\,,
\ee
leads to
\be
\left[\,\widetilde{Q}_T,p_m{-f_{m}}(q,p_{a},t)\right\}_{P.B.}
\simeq
\left(\frac{\partial (\delta q^{B})}{\partial\dq^{m}}-(-1)^{(\#_{a}+\#_{m})\#_{a}}\frac{\partial f_{m}}{\partial p_{a}}\frac{\partial (\delta q^{B})}{\partial\dq^{a}}\right)\left[\frac{{\rm{d}\,}}{{\rm{d}t}}\!
\left(\!\frac{\partial L}{\partial\dq^{B}}\!\right)\!-\!\frac{\partial L\,}{\partial q^{B}}\right]\equiv 0\,.
\label{QTV}
\ee
}

\subsubsection{Phase space variables}

As discussed in Sec.\ref{summary},  after solving the constraints, all the {functions $\dq^{m}={(-1)^{\#_{m}}u^{m}}(p,q,t)$ are completely determined in terms of time and the phase space variables, together with the local free parameters $\free^{i}(t)\,$, as in Eq.(\ref{mostu}). Therefore, all the velocitites have been removed so that the only independent variables are phase space variables. In other words, in this sense we are again working in the genuine Hamiltonian formalism. Substituting the general solution given in (\ref{mostu}) into (\ref{vartilde})} reduces the infinitesimal transformations and the total Noether charge to be functions of $(p,q,t)$:
\be
\ba{l}
\Delta q^{A}(p,q,t):=\Deltatq^{A}\left(q,p_{a},\dq^{m}(p,q,t),t\right)\,,\\
{}\\
\Delta p_{A}(p,q,t):=\Deltatp_{A}\left(q,p_{a},\dq^{m}(p,q,t),t\right)\,,\\
{}\\
Q_{T}(p,q,t):=\widetilde{Q}_{T}\left(q,p_{a},\dq^{m}(p,q,t),t\right)
=\Delta q^{A}p_{A}-\DeltatK\left(q,p_{a},\dq^{m}(p,q,t),t\right)\,,
\ea
\label{DeltaqpQ}
\ee
which satisfy
\be
\ba{l}
\D{[{Q}_{T},q^{A}\}_{P.B.}=-\Delta q^{A}-[q^{A},\Delta q^{m}\}_{P.B.}
\Big(p_{m}-f_{m}(q,p_{a},t)\Big)\,,}\\
{}\\
\D{[{Q}_{T},p_{A}\}_{P.B.}=-\Delta p_{A}-[p_{A},\Delta q^{m}\}_{P.B.}
\Big(p_{m}-f_{m}(q,p_{a},t)\Big)\,.}
\ea
\label{QTD}
\ee

The total Hamiltonian is equal to $H_{T}:=\v^{A}p_{A}-L(q,\v,t)$ according to (\ref{simpletotalH}), where the expression of $u^{m}$ given by (\ref{mostu}) is substituted. {Now, we get on the primary constraint surface $V$ that
\be
[\,Q_{T},H_{T}\}_{P.B.}
\simeq
[\,Q_{T},\v^{A}\}_{_{P.B.}}\,p_{A}-\v^{A}\Delta p_{A}+\Delta q^{A}\frac{\partial L(q,\v,t)}{\partial q^{A}}-
[\,Q_{T},\v^{A}\}_{P.B.}\frac{\partial L(q,\v,t)}{\partial\v^{A}}\,,
\ee
due to (\ref{QTD}).
But the expressions of the momenta (\ref{ep}) show that the sum
\be
[\,Q_{T},\v^{A}\}_{_{P.B.}}\left(p_{A}-\frac{\partial L(q,\v,t)}{\partial\v^{A}}\right)\,=[\,Q_{T},\v^{\,m}\}_{_{P.B.}}\left(p_{m}-\frac{\partial L(q,\v,t)}{\partial\v^{m}}\right)\simeq0\,,
\ee
vanishes on the primary constraint surface $V\,$. Therefore,
\be
\ba{l}
[Q_{T},H_{T}\}_{P.B.}\\
{}\\
\D{\simeq
-\,\v^{A}\Delta p_{A}+\Delta q^{A}\frac{\partial L(q,\v,t)}{\partial q^{A}}}\\
{}\\
\D{\simeq-\,\v^{A}\left(\frac{\partial\,\delta K(q,\v,t)}{\partial q^{A}}
-\frac{\partial\,\delta q^B(q,\v,t)}{\partial q^{A}}\frac{\partial L(q,\v,t)}{\partial\v^{B}}\right)
+\Delta q^{A}\frac{\partial L(q,\v,t)}{\partial q^{A}}}\\
{}\\
\D{\simeq \frac{\partial\,\delta K(q,\v,t)}{\partial t}
-\frac{\partial\,\delta q^A(q,\v,t)}{\partial t}\frac{\partial L(q,\v,t)}{\partial\v^{A}}}\\
{}\\
\D{\simeq \frac{\partial(\delta K)}{\partial t}
-\frac{\partial\,(\delta q^A)}{\partial t}{\,}p_{A}\,.}
\ea
\ee
where we made use of Eq.(\ref{Deltap}) to get the third line and of Eq.(\ref{KLqdot}) to obtain the fourth line.
In terms of the very definition of the total Noether charge (\ref{DeltaqpQ}), we have thus shown that the Noether charge is conserved on the primary constraint surface,
\be
\D{[\,Q_{T},H_{T}\}_{P.B.}+\frac{\partial Q_{T}}{\partial t}\simeq 0\,,}
\label{QTHT}
\ee
It is worth noting that this result is off-shell and parallel to the off-shell invariance of the action under the symmetry transformation. Of course, on-shell $dQ_{T}/dt\equiv[\,Q_{T},H_{T}\}_{P.B.}+\frac{\partial Q_{T}}{\partial t}\simeq 0$ to be compared with the on-shell conservation (\ref{Noetherconserved}) of the Noether charge. Another way of expressing (\ref{QTHT}) is to say that $[\,Q_{T},H_{T}\}_{P.B.}+\frac{\partial Q_{T}}{\partial t}$ is a linear combination of the constraints. In other words, the total Noether charge generates a transformation which preserves the Hamiltonian on the primary constraint surface.}\\

Furthermore, from the last formula in (\ref{pQT}), not only $\widetilde{Q}_{T}$ but also $Q_{T}$ preserves
the primary constraints on-shell as in (\ref{QTV}). {More precisely,
\be
\left[\,Q_T,\phi_m(q,p_{a},t)\right\}_{P.B.}
\simeq 0\,,\quad\mbox{on-shell.}
\ee
Therefore, the time evolution of this condition also vanishes on the primary constraint surface and on-shell, \textit{i.e.}
\be
\frac{{\rm d}}{{\rm d}t}\left[\,Q_T,\phi_m(q,p_{a},t)\right\}_{P.B.}=\Big[\,\left[\,Q_{T}\,,\,\phi_{m}\right\}_{P.B.},H_{T}\Big\}_{P.B.}
+\frac{\partial~}{\partial t}\left[\,Q_{T},\phi_{m}\right\}_{P.B.}\simeq 0\,,\quad\mbox{on-shell.}
\label{label}
\ee
For} the secondary constraints which essentially stem from
$\left[\phi_{m}\,,H_{T}\right\}_{P.B.}+\frac{\partial \phi_{m}}{\partial t}$, we notice
\be
\ba{l}
\textstyle{\Big[Q_{T},\left[\phi_{m}\,,H_{T}\right\}_{P.B.}+\frac{\partial \phi_{m}}{\partial t}\Big\}_{P.B.}}\\
{}\\
\textstyle{= \Big[\,\left[\,Q_{T}\,,\,\phi_{m}\right\}_{P.B.},H_{T}\Big\}_{P.B.}
+\frac{\partial~}{\partial t}\left[\,Q_{T},\phi_{m}\right\}_{P.B.}+
\Big[\phi_{m}\,,\left[\,Q_{T}\,,H_{T}\right\}_{P.B.}+\frac{\partial Q_{T}}{\partial t}\Big\}_{P.B.}}\,,
\ea
\ee
and hence, from (\ref{QTHT}) and (\ref{label}) we deduce that \textit{not only the primary constraints but also the secondary constraints are  preserved on-shell by $Q_{T}$ if~  $\left[\,Q_{T},H_{T}\right\}_{P.B.}+\frac{\partial Q_{T}}{\partial t}$ corresponds to a first class constraint. }  In this case, $Q_{T}$ is  first class on-shell. As we see in the next subsection, the condition further implies that $Q_{T}$ preserves the solution space too.\\

\subsection{Symmetry in the total Hamiltonian system}

In this subsection, motivated by the results in the previous subsection where we studied the general properties of the total Noether charge which originates from a symmetry in a Lagrangian system, we discuss the symmetry in the total Hamiltonian system directly without referring to any Lagrangian system. In order to make the analysis concise, we introduce a single letter, $x^{M}$,  $1\leq M\leq 2\N$, to denote both $p$ and $q$,
\be
\ba{ll}
x^{A}=q^{A}\,,~~~~~&~~~~~x^{\N+A}=p_{A}\,.
\ea
\ee
We define a $2\N\times 2\N$ constant non-degenerate graded skew-symmetric matrix by
\be
\D{J^{MN}=\left[\,x^{M},x^{N}\right\}_{P.B.}=-(-1)^{\#_{M}\#_{N}}J^{NM}
=\left(\ba{cc}0~&~(-1)^{\#_{A}}\delta^{A}{}_{B}\\
-\delta_{A}{}^{B}~&~0\ea\right)\,,}
\label{skewJ}
\ee
which gives
\be
\D{\left[\,F,G\,\right\}_{P.B.}=\frac{\overleftarrow{\partial}F}{\partial x^{M}}
J^{MN}\frac{\overrightarrow{\partial}G}{\partial x^{N}}\,.}
\ee
In particular, $\left[x^{M},G\right\}_{P.B.}=J^{MN}\partial_{N}G$,
where $\displaystyle{
\partial_{N}=\frac{\overrightarrow{\partial}~~}{\partial x^{N}}}$.\\

\subsubsection{Definition of symmetry transformations}

Now we \textsl{define} a symmetry of the total Hamiltonian system
as a coordinate transformation on the jet space that
\begin{enumerate}
  \item \textit{depends on the phase space only}, (\textit{i.e.} it should not depend on $\dot{x}^M$, $\ddot{x}^M$, \textit{etc}),
\be
x^{M}~~~\longrightarrow~~~ x^{\prime M}(x,t)\,;
\ee
	\item \textit{preserves the symplectic structure}
\be
\displaystyle{J^{MN}=\frac{\overleftarrow{\partial}x^{\prime M}}{\partial x^{K}}
\,J^{KL}\,\frac{\overrightarrow{\partial}x^{\prime N}}{\partial x^{L}}=
\left[x^{\prime M},x^{\prime N}\right\}_{P.B.}\,;}
\label{syminv}
\ee
  \item \textit{takes any physical solution to another which means the preservation of both the on-shell relations and the constraints}.  Namely, if $x(t)$ is a solution of the time evolution governed by a total {Hamiltonian $H_{T}(x,t;v,w)\,$, given by (\ref{totham}),}
\be
\D{\dot{x}^{M}=\left[x^{M},H_{T}(x,t;v,w)\right\}_{P.B.}=J^{MN}\partial_{N}H_{T}(x,t;v,w)\,,}
\label{solx}
\ee
then so must be $x^{\prime M\!}(x(t),t)$ for the same total Hamiltonian, up to the change of the local parameters $\free^i(t)$ and $w^{hh^\prime}(t)$,
\be
\D{\dot{x}^{\prime M}=\dot{x}^{N}\partial_{N}x^{\prime M}+\partial_{t} x^{\prime M} =J^{MN}\partial^{\prime}_{N}H_{T}(x^{\prime},t;\free^{\prime},w^{\prime})\,.}
\label{solxx}
\ee
Furthermore, such a symmetry must  preserve the constraint surface $\cal V$ on-shell,
\be
\phi_h(x^\prime,t)\approx c^{g}_{\,\,\,h}(x,t)\phi_{g}(x,t)~~~~~~~\mbox{on-shell.}
\label{defpresV}
\ee
\end{enumerate}
~\\


Infinitesimally, the second requirement (\ref{syminv}) reads
\be
\displaystyle{J^{-1}_{ML}\partial^{}_{N}(\delta x^{L})=(-1)^{\#_{M}\#_{N}}
J^{-1}_{NL}\partial^{}_{M}(\delta x^{L})\,.}
\ee
In other words, the `super' one-form $\delta x_M:=J^{-1}_{ML}\delta x^{L}$ is closed, hence exact. Therefore, there exists a ``generating function'' $\Q_H(x)$ on the phase space such that, \textit{c.f.} (\ref{QTD}),
\be
\delta x^{M}=\left[\,x^{M},\Q_H\right\}_{P.B.}\,.
\label{generating}
\ee
Conversely, any such transformation leaves the symplectic structure invariant.  \\

In comparison to the above definition of symmetry transformations, which refers to a specific given total Hamiltonian, a ``canonical transformation,'' $x\rightarrow x^{\prime}(x,t)$  is defined as a coordinate transformation on phase space such that for \textit{every} total Hamiltonian $H_{T}(x,t)$ there must exist another  (\textit{not} necessarily equal) total Hamiltonian $H^{\prime}_{T}(x^{\prime},t)$ obeying
\be
\D{\dot{x}^{\prime M}=\dot{x}^{N}\partial_{N}x^{\prime M}+\partial_{t} x^{\prime M}
=J^{MN}\partial^{\prime}_{N}H^{\prime}_{T}(x^{\prime},t)\,.}
\label{condCanTr}
\ee
As the time evolution is generated by the total Hamiltonian, $\dot{x}^{N}=J^{NK}\partial_{K}H_{T}(x,t)\,$, the condition (\ref{condCanTr}) is equivalent {to
\be
\D{\left[x^{\prime M},x^{\prime N}\right\}_{P.B.}
\partial^{\prime}_{N}
H_{T}(x,t)+\partial_{t} x^{\prime M}
=J^{MN}\partial^{\prime}_{N}H^{\prime}_{T}(x^{\prime},t)\,,}
\label{analysisof}
\ee
since
\be
\displaystyle{\left[x^{\prime M},x^{\prime N}\right\}_{P.B.}=\frac{\overleftarrow{\partial}x^{\prime M}}{\partial x^{K}}
\,J^{KL}\,\frac{\overrightarrow{\partial}x^{\prime N}}{\partial x^{L}}
\,\,.}
\ee
The analysis of Eq.(\ref{analysisof})} leads essentially to an integrability condition on the left hand side for arbitrary $H_{T}(x,t)$, in order to be consistent with $\partial^{\prime}_{M}\partial^{\prime}_{N}H^{\prime}_{T}=(-1)^{\#_{M}\#_{N}}
\partial^{\prime}_{N}\partial^{\prime}_{M}H^{\prime}_{T}$. Namely with the notation $x^{\prime}_{M}:=J^{-1}_{MN}x^{\prime N}$, the integrability condition reads
\be
\ba{l}
\displaystyle{\left(\partial^{\prime}_{M}\!\left[x^{\prime}_{N},x^{\prime K}\right\}_{P.B.}\!\right)\!
\partial^{\prime}_{K}H_{T}+(-1)^{\#_{M}\#_{N}}\left[x^{\prime}_{N},x^{\prime K}\right\}_{P.B.}\!\partial^{\prime}_{K}
\partial^{\prime}_{M}H_{T}+\partial^{\prime}_{M}x^{K}\partial_{K}\partial_{t}x^{\prime}_{N}}\\
{}\\
\displaystyle{\!=(-1)^{\#_{M}\#_{N}}\!\left(\partial^{\prime}_{N}\!\left[x^{\prime}_{M},x^{\prime K}\right\}_{P.B.}\!\right)\!
\partial^{\prime}_{K}H_{T}+\left[x^{\prime}_{M},x^{\prime K}\right\}_{P.B.}\!\partial^{\prime}_{K}
\partial^{\prime}_{N}H_{T}+(-1)^{\#_{M}\#_{N}}
\partial^{\prime}_{N}x^{K}\partial_{K}\partial_{t}x^{\prime}_{M}\,.}
\ea
\ee
This must hold for arbitrary  $H_{T}(x,t)$ and hence we have three independent relations:
\begin{eqnarray}
&&\left(\partial^{\prime}_{M}\!\left[x^{\prime}_{N},x^{\prime K}\right\}_{P.B.}\!\right)
\partial^{\prime}_{K}H_{T}=(-1)^{\#_{M}\#_{N}}\partial^{\prime}_{N}\!\left[x^{\prime}_{M},x^{\prime K}\right\}_{P.B.}\!\partial^{\prime}_{K}H_{T}\,,\label{cat1}\\
{}\nonumber\\
&&\left(\left[x^{\prime}_{N},x^{\prime K}\right\}_{P.B.}\!\right)\partial^{\prime}_{K}
\partial^{\prime}_{M}H_{T}=(-1)^{\#_{M}\#_{N}}\left[x^{\prime}_{M},x^{\prime K}\right\}_{P.B.}\!\partial^{\prime}_{K}\partial^{\prime}_{N}H_{T}\,,\label{cat2}\\
{}\nonumber\\
&&\partial^{\prime}_{M}x^{K}\partial_{K}\partial_{t}x^{\prime}_{N}=
(-1)^{\#_{M}\#_{N}}\partial^{\prime}_{N}x^{K}\partial_{K}\partial_{t}x^{\prime}_{M}\,.\label{cat3}
\end{eqnarray}
Firstly, the second relation (\ref{cat2}) with the quadratic choice $H_{T}=x^{\prime P}x^{\prime Q}$ shows that $\left[x^{\prime}_{N},x^{\prime K}\right\}_{P.B.}$ is proportional to $\delta_{N}^{~K}$ or
\be
\left[x^{\prime M},x^{\prime N}\right\}_{P.B.}=f(x,t)\,J^{MN}\,.
\label{fxt}
\ee
Secondly, Eq.(\ref{cat1}) with the linear choice $H_{T}=x^{\prime P}$ further reveals that $f(x,t)$ is independent of $x\,$, \textit{i.e.}
$$\partial_{M}f(x,t)=0\,.$$ Finally, the last relation (\ref{cat3}) shows that there exists a bosonic function $\Phi(x^{\prime},t)$ satisfying $\partial_{t}x^{\prime}_{M}=\partial_{M}^{\prime}\Phi(x^{\prime},t)$. Using this  and from  (\ref{skewPB}),  (\ref{fxt}) we note that  the explicit time derivative of $f(t)$ vanishes as
\be
\ba{ll}
\partial_{t}f(t)J^{-1}_{NM}&=\left[\partial_{M}^{\prime}\Phi,x^{\prime}_{N}\right\}_{P.B.}
+\left[x^{\prime}_{M},\partial_{N}^{\prime}\Phi\right\}_{P.B.}\\
{}&{}\\
{}&=-(-1)^{\#_{M}\#_{N}}\left[x^{\prime}_{N},x^{\prime L}\right\}_{P.B.}\partial^{\prime}_{L}\partial_{M}^{\prime}\Phi
+\left[x^{\prime}_{M},x^{\prime L}\right\}_{P.B.}\partial^{\prime}_{L}\partial_{N}^{\prime}\Phi\\
{}&{}\\
{}&=-(-1)^{\#_{M}\#_{N}} f\partial^{\prime}_{N}\partial_{M}^{\prime}\Phi+ f\partial^{\prime}_{M}\partial_{N}^{\prime}\Phi\\
{}&{}\\
{}&=0\,.
\ea
\ee
Thus, from Eq.(\ref{fxt}) one notices that canonical transformations leave the symplectic structure invariant up to a constant
$\left[x^{\prime M},x^{\prime N}\right\}_{P.B.}=\mbox{constant}\times J^{MN}$. {Shortly, \textit{up to rescalings, canonical transformations are symplectic transformations}.}\\

Finally we note that, if we
require the preservation of {the ${\mathbb Z}_2$-graduation and of the reality properties $(\delta x^{M})^{\dagger}=\delta(x^{M}{}^{\dagger})$,} we may set $\Q_H$ to be bosonic and Hermitian  $\Q_{H}=\Q_{H}^{\dagger}$, so that
\be
(\delta x^{M})^{\dagger}=\left[x^{M}{}^{\dagger},\Q_{H}\right\}_{P.B.}=\delta(x^{M}{}^{\dagger})\,.
\ee

\subsubsection{Criteria for symmetry generators}

In order to clarify the criteria for the generating function $\Q_{H}$ in (\ref{generating}) to  meet the remaining  conditions (\ref{solxx}) and  (\ref{defpresV}) as to be a  symmetry generator,   we investigate the infinitesimal version of them which are given by\footnote{From the Leibniz rule of the Poisson bracket (\ref{Leibniz}), it is worth to note an  identity  for an arbitrary  function $F(x,t)$,
\[\left[\Q_{H}\,,\,F(x,t)\right\}_{P.B.}=\left[\Q_{H}\,,\,x^{M}\right\}_{P.B.}\partial_{M}F(x,t)\,.\]
}
\be
\ba{ll}
\D{\left[\,\delta x^{M},H_{T}\right\}_{P.B.}+\partial_{t} (\delta x^{M}})&
\D{=J^{MN}\left(\delta x^{L}\partial_{L}\partial_{N}H_{T}+(\partial_{N}
\phi_{i}^{1{\rm st}})\,\delta \free^{i}+(\partial_{N}\phi_{h})\phi_{h^{\prime}}\delta w^{hh^{\prime}}\right)}\\
{}&{}\\
{}&\D{=\delta x^{L}\partial_{L}\left[x^{M},H_{T}\right\}_{P.B.}+\left[x^{M},
\phi_{i}^{1{\rm st}}\delta \free^{i}+\half\phi_{h}\phi_{h^{\prime}}\delta w^{hh^{\prime}}\right\}_{P.B.}\,,}
\ea
\label{symcon}
\ee
and for $1\leq h\leq\M+\M^{\prime}$,
\be
\delta x^M\partial_M\phi_h=\left[x^{M},\Q_H\right\}_{P.B.}\partial_M\phi_h=\left[\Q_H,\phi_{h}\right\}_{P.B.}
\approx 0~~~~~~~\mbox{on-shell.}
\label{symcon2}
\ee
The latter simply implies that $\Q_H$ must be first-class on-shell. {The former condition (\ref{symcon}) must hold for arbitrary solutions of the total Hamiltonian. In particular,} it should hold at the  initial time, say $t_{0}$, at which the initial data $x_{0}$ can be taken arbitrarily. Thus the  condition (\ref{symcon}) should be interpreted  \textit{off-shell}, and the general solution $\delta x^{M}(x,t)$ of the partial differential equation  (\ref{symcon}) may lead to all the symmetries in a given total Hamiltonian system. Rather,  we translate the condition (\ref{symcon})  in terms of the symmetry generator $\Q_H$, as done in (\ref{symcon2}),
\be
\D{\Big[\left[x^{M},\Q_{H}\right\},H_{T}\Big\}_{P.B.}\! =-\Big[\Q_{H},\left[x^{M},H_{T}\right\}\Big\}_{P.B.}\!+\!\left[x^{M},\,\phi_{i}^{1{\rm st}}\delta \free^{i}
+\half\phi_{h}\phi_{h^{\prime}}\delta w^{hh^{\prime}}-\partial_{t}\Q_{H}\right\}_{P.B.}}\,.
\ee
Due to  Jacobi identity this is equivalent to
\be
\D{\Big[x^{M}\,,~\left[\Q_{H},H_{T}\right\}_{P.B.}+\partial_{t}\Q_{H}-\phi_{i}^{1{\rm st}}\delta \free^{i}
-\half\phi_{h}\phi_{h^{\prime}}\delta w^{hh^{\prime}}\Big\}_{P.B.}=0\,.}
\label{tHsp}
\ee
This condition should hold for  every  $x^{M}$, $1\leq M\leq 2\N$. Therefore,
\be
\D{\left[\Q_{H},H_{T}\right\}_{P.B.}+\partial_{t}\Q_{H}-\phi_{i}^{1{\rm st}}\delta \free^{i}
-\half\phi_{h}\phi_{h^{\prime}}\delta w^{hh^{\prime}}=f(t)\,,}
\ee
where $f(t)$ is an arbitrary time dependent, but phase-space independent function. This function can be removed by a redefinition of  the generator,
\be
\Q_{H}~~~~~\longrightarrow~~~~~\D{\Q_{H}+\int_{t_{0}}^{\,t}{{\rm d}t^{\prime}}\,f(t^{\prime})\,,}
\ee
as the shift has no effect on the symmetry transformation, $\delta x^{M}=[x^{M},\Q_{H}\}_{P.B.}$.\\
~\\

Thus, \textit{the necessary and sufficient condition for an on-shell first class quantity $\Q_{H}(p,q,t)$ to be a symmetry generator of a given total Hamiltonian $H_T(p,q,t;v,w)$ reads}
\be
\D{
\left[\Q_{H},H_{T}\right\}_{P.B.}+\frac{\partial\Q_{H}}{\partial t}=\phi_{i}^{1{\rm st}}\delta \free^{i}+\half\phi_{h}\phi_{h^{\prime}}\delta w^{hh^{\prime}}\,,}
\label{tHs}
\ee
which is consistent with (\ref{QTHT}).  {The usual Hamiltonian version of the Noether theorem in unconstrained systems states that any conserved charge ($d\Q_H/dt=0$) is a symmetry generator. The formula (\ref{tHs}) is the corresponding generalization to constrained systems.}

\subsection{Solutions}

\begin{itemize}

\item \textit{Every quantity, which is first-class and conserved on-shell, is a symmetry generator.}\\
Indeed, the fact that $\Q_{H}$ remains constant  under the time evolution reads
\be
\D{\frac{{\rm d}\Q_{H}}{{\rm d} t}\equiv\left[\Q_{H},H_{T}\right\}_{P.B.}+\frac{\partial\Q_{H}}{\partial t}=0\,,}
\ee
which is stronger than (\ref{tHs}).

\item \textit{The total Noether charge $Q_{T}$}~(\ref{DeltaqpQ}) \textit{originating from a Lagrangian system is a symmetry generator if  $~\left[Q_{T},H_{T}\right\}_{P.B.}+\frac{\partial Q_{T}}{\partial t}$ corresponds to a first class constraint.} In this case, the full expressions  of the right hand sides of  (\ref{DeltaqpQ}) should correspond to $\delta x^{M}$ and it follows that $Q_{T}$ is  first class on-shell, as we saw in Sec.\ref{preSTH}.\\

\item \textit{\bf{Static examples}}:

In the static case, there is no explicit time dependence in the constraints $\phi_{h}(p,q)$ and the quantity $H^{\prime}(p,q)$ defined in (\ref{TOTH}). The time derivative of any first-class constraint ${\phi}^{1{\rm st}}_{i}$ is then first-class too, as shown in (\ref{staticdot1st}).

\begin{itemize}

\item \textit{In the static case, the total Hamiltonian itself corresponds to a symmetry generator}
\be
\D{\frac{{\rm d}H_{T}}{{\rm d} t}\equiv\frac{\partial H_{T}}{\partial t}=0\,.}
\ee
~\\

\item \textit{If there is no secondary first-class constraint in the given system (so that all the first class constraints are linear in} $\phi^{1{\rm st}}_{i}$), \textit{then}
\be
\D{\dot{\phi}^{1{\rm st}}_{i}=\left[\phi^{1{\rm st}}_{i},H_{T}\right\}_{P.B.}=\phi^{1{\rm st}}_{j}{\cal T}^{j}{}_{i}\,.}
\ee
\textit{Hence} $\phi^{1{\rm st}}_{i}\varepsilon^{i}(t)$ \textit{corresponds to a gauge symmetry generator}
\textit{with  arbitrary time dependent functions} $\varepsilon^{i}(t)$, \textit{i.e.}
\be
\Q_{H}=\phi^{1{\rm st}}_{i}\varepsilon^{i}(t)\,,
\label{gs1}
\ee
\textit{satisfying} (\ref{tHs}).\\

  \item \textit{Alternatively if  the time derivative of ${\phi}^{1{\rm st}}_{i}$ is quadratic  in the constraints} $\phi_{h}$ (and hence first-class)
\be
\D{\dot{\phi}^{1{\rm st}}_{i}=\left[\phi^{1{\rm st}}_{i},H_{T}\right\}_{P.B.}=\half
\phi_{h}\phi_{h^{\prime}}{\cal T}^{hh^{\prime}}{}_{i}\,,}
\label{qphi}
\ee
\textit{then again $\phi^{1{\rm st}}_{i}\varepsilon^{i}(t)$ corresponds to a gauge symmetry generator,}
\be
\Q_{H}=\phi^{1{\rm st}}_{i}\varepsilon^{i}(t)\,.
\label{gs2}
\ee

\item Combining the above two cases we have more general solutions. Namely, \textit{if  the time derivative of ${\phi}^{1{\rm st}}_{a}$ is a sum of terms linear in  $\phi^{1{\rm st}}_{b}$ and quadratic  in  $\phi_{h}$},
\be
\D{\dot{\phi}^{1{\rm st}}_{a}=\left[\phi^{1{\rm st}}_{a},H_{T}\right\}_{P.B.}=\phi^{1{\rm st}}_{b}{\cal T}^{b}{}_{a}+\half
\phi_{h}\phi_{h^{\prime}}{\cal T}^{hh^{\prime}}{}_{a}\,,}
\label{gs3con}
\ee
\textit{then $\phi^{1{\rm st}}_{a}\varepsilon^{a}(t)$ corresponds to a gauge symmetry generator,}
\be
\Q_{H}=\phi^{1{\rm st}}_{a}\varepsilon^{a}(t)\,.
\label{gs3}
\ee
For example, we consider the Lagrangian $L(x,y,\dot x,\dot y)=\half \,e^{y}\dot{x}^{2}$, whose equations of motion leave $y$ arbitrary (so $y$ is pure gauge) but fix $x$ in time $x=x_{0}$. This Lagrangian produces the Hamiltonian $H=\half\, e^{-y}p_{x}^{2}$, one primary first-class constraint $\phi^{1{\rm st}}=p_{y}$,  one secondary first-class constraint $p_{x}$, and the total Hamiltonian $H_{T}=\half\, e^{-y}p_{x}^{2}+p_{y}\free(t)\,$. This is a counterexample to Dirac's conjecture (see \textit{e.g.} the subsection 1.2.2 of \cite{Henneaux:1992ig}) because the \textit{secondary} first-class constraint $p_x$ does not generate any gauge symmetry as $x$ is fixed by the equations of motion.
{However, the time derivative of the \textit{primary} first-class constraint satisfies Eq.(\ref{qphi}),
\be
\dot{p}_y=\left[p_{y},H_{T}\right\}_{P.B.}=\half\, e^{-y}
(p_x)^{2}\,,
\ee
and generates arbitrary shifts of the pure gauge variable $y\,$.}
~\\

\item  \textit{A linear combination of the primary and secondary first-class constraints, $\phi^{1{\rm st}}_{a}$ and $\phi^{1{\rm st}}_{s}$, can be a gauge symmetry generator,}
\be
\Q_{H}=\phi^{1{\rm st}}_{a}\varepsilon^{a}(t)+\phi^{1{\rm st}}_{s}\varepsilon^{s}(t)\,,
\label{gs4}
\ee
\textit{if the local functions, $\varepsilon^{a}(t)$ and $\varepsilon^{s}(t)$ satisfy,  with } (\ref{TTTT}),
\be
\D{\frac{{\rm d}\varepsilon^{s}(t)}{{\rm d}t}+\T^{s}{}_{r}\varepsilon^{r}(t)+\T^{s}{}_{a}\varepsilon^{a}(t)=0\,.}
\label{gs4con}
\ee
For example we consider the Maxwell theory of which the  Lagrangian and the Hamiltonian read\footnote{This example is also handled in the section 19.1.1 of the reference \cite{Henneaux:1992ig}.}
\be
\ba{l}
\D{{\cal L}=-\textstyle{\frac{1}{4}}F_{\mu\nu}F^{\mu\nu}\,,}\\
{}\\
\D{H=H^{\prime}=\int {\rm d}^{{\scriptscriptstyle{D-1}}}x~\left[\textstyle{\frac{1}{4}}F_{ij}F_{ij}+
\half\Pi^{i}\Pi^{i}-A_{0}\partial_{i}\Pi{i}\right]\,,}\\
{}\\
\D{H_{T}=\int {\rm d}^{{\scriptscriptstyle{D-1}}}x~\left[\textstyle{\frac{1}{4}}F_{ij}F_{ij}+
\half\Pi^{i}\Pi^{i}-A_{0}\partial_{i}\Pi^{i}+v(x)\Pi^{0}\right]\,.}
\ea
\ee The gauge symmetry is one example of (\ref{gs4}) and (\ref{gs4con})  as
\be
\D{\Q_{H}=\int {\rm d}x^{{\scriptscriptstyle{D-1}}}~\left[\varepsilon(x)\,\partial_{i}\Pi^{i}- \partial_{0}\varepsilon(x)\,\Pi^{0}\right]\,,}
\ee
where $\Pi^{\mu}=F^{\mu 0}$ are the gauge invariant canonical momenta for $A_{\mu}$, and  $\Pi^{0}$ is the primary first-class constraint, while $\partial_{i}\Pi^{i}$ is the secondary first-class constraint. There appears no other constraint. \\

\item In the extended Hamiltonian formalism of Sec.\ref{extended}, every first-class constraint corresponds to a gauge symmetry, if the system is static. Namely
$\phi^{1{\rm st}}_{i}\varepsilon^{i}(t)+\phi^{1{\rm st}}_{i^\prime}\varepsilon^{i^\prime}(t)$ is  a gauge symmetry  generator with arbitrary time dependent functions, $\varepsilon^{i}(t)$ and $\varepsilon^{i^\prime}(t)$,
\be
\Q_{E}=\phi^{1{\rm st}}_{i}\varepsilon^{i}(t)+\phi^{1{\rm st}}_{i^\prime}\varepsilon^{i^\prime}(t)\,.
\label{gs5}
\ee
\end{itemize}
As one can see, the generators of local (\textit{i.e.} gauge) symmetries are linear combinations of the constraints, therefore they vanish weakly in contradistinction with the generators of global symmetries. In this sense, only global symmetries lead to non-trivial conserved charges.
\end{itemize}

\subsection{Dynamics with the arbitrariness  - gauge symmetry}

We remind the reader that the key philosophy we stick to is that different choices of the local gauge parameters correspond to  different  total Hamiltonian systems, which nevertheless  should be  taken  equivalent,  describing the same physics (see Subsection \ref{firstclassgauge}).

A somewhat less drastic - though equivalent - perspective is to consider only one total Hamiltonian throughout the time evolution,  with a single set of local gauge parameters. The local functions should be continuous all the time but infinitely differentiable, \textit{i.e.} $C^{\infty}$, only piecewise in time.  This discontinuity in the derivatives corresponds to changes of local gauge parameters at different times.    As long as the local functions  $\free^{i}(t)$ are continuous, one can change them arbitrarily  at any moment. The continuity guarantees the continuity of the first order time derivative of the dynamical variables $(\dot{p},\dot{q})$. However, in the Hamiltonian dynamics, there is no reason to require the continuities  for the higher order time derivatives.\\

Explicitly, expressing  the dynamical variable, $F$, at a future time, $t+dt$, as a power expansion  of $dt$ around  the present time, $t$,  we have
\be
\ba{ll}
F(t+dt)&=\D{\left.F\left(p(t^{\prime}),q(t^{\prime}),t^{\prime}\right)\right|_{t^{\prime}=t+dt}}\\
{}&{}\\
{}&\D{=F+dt\dot{F}+\half\, dt^{2}\ddot{F}+{\cal O}(dt^{3})}\\
{}&{}\\
{}&\D{=F+dt\left(\frac{\partial F}{\partial t}+\Big[F,H_{T}\Big\}_{P.B.}\right)}\\
{}&{}\\
{}&\D{~~~+\half\, dt^{2}
\left(\frac{\partial^{2} F}{\partial t^{2}}+2\left[
\frac{\partial F}{\partial t},H_{T}\right\}_{P.B.}+\left[
F,\frac{\partial H_{T}}{\partial t}\right\}_{P.B.}+
\Big[\left[{F},{H}_{T}\right\},{H}_{T}\Big\}_{P.B.}\right)}\\
{}&{}\\
{}&=\D{\hat{F}+dt\left[\hat{F},\hat{H}_{T}\right\}_{P.B.}
+\half dt^{2}\left[[\hat{F},\hat{H}_{T}\},\hat{H}_{T}\right\}_{P.B.}+{\cal O}(dt^{3})}\,,
\ea
\label{evo}
\ee
where we have assumed that $\ddot{F}$ exists or $\free^{i}(t)$ is differentiable,  and we have set
\be
\ba{l}
\hat{F}=\D{F+dt\frac{\,\partial F}{\partial t}+\half\, dt^{2}\,\frac{\,\partial^{2} F}{\partial t^{2}}}\,,\\
{}\\
\hat{H}_{T}=\D{H_{T}+\half\,dt\frac{\,\partial H_{T}}{\partial t}=H^{\prime}+{\phi}^{1{\rm st}}_{i}\free^{i}+\half\,dt\left(\frac{\,\partial H^{\prime}}{\partial t}+\frac{\,\partial{\phi}^{1{\rm st}}_{i}}{\partial t}\,\free^{i}+ \phi^{1{\rm st}}_{i}\frac{{\rm d}\free^{i}}{{\rm d}t}\right)\,.}
\ea
\label{evo2}
\ee
~\\

The coefficients $\free^i(t)$ are completely arbitrary and at our disposal. We recall that different choices
of the local parameters mean  different total Hamiltonian systems, which nevertheless  should be regarded equivalent,
\textit{i.e.} describing the same physics.
For two different choices of the coefficients, say $\free$ and $\free+\Delta \free$,  the dynamical variable at the future time differs by
\be
\ba{ll}
\Delta F(t+dt)=&
\D{dt\left[\hat{F},\phi^{1{\rm st}}_{i}\right\}_{P.B.}\Delta \free^{i}}\\
{}&{}\\
{}&+\D{
\half dt^{2}\left[\hat{F},\frac{\partial {\phi}^{1{\rm st}}_{i}}{\partial t}\right\}_{P.B.}\!\!\Delta \free^{i}+\half dt^{2}\left[\hat{F},{\phi}^{1{\rm st}}_{i}\right\}_{P.B.}\!\!\frac{{\rm{d}}\Delta \free^{i}}{{\rm{d}}t}}\\
{}&{}\\
{}&\D{+\half dt^{2}\left[\left[\hat{F},{\phi}^{1{\rm st}}_{i}\right\},{H}^{\prime}\right\}_{P.B.}\Delta \free^{i}+\half dt^{2}
\left[\left[\hat{F},{H}^{\prime}\right\},{\phi}^{1{\rm st}}_{i}\right\}_{P.B.}\Delta \free^{i}}\\
{}&{}\\
{}&+\D{\half dt^{2}\left[\left[\hat{F},{\phi}^{1{\rm st}}_{i}\right\},{\phi}^{1{\rm st}}_{j}\right\}_{P.B.}(\free^{i}\Delta \free^{j}+\free^{j}\Delta \free^{i}+\Delta \free^{j}\Delta \free^{i})}\\
{}&{}\\
{}&\D{~+~{\cal O}\left(dt^{3},\Delta \free\right)}\,.
\ea
\label{DeltaF1}
\ee
Thus, the leading order in the difference appears   at the first order in $dt$ or the velocity, when $\Delta \free(t)\neq 0$. Namely, different velocities for the same initial configuration can still correspond to the same physical state.  This may\footnote{For $\Q_{H}=\phi^{1{\rm st}}_{i}\varepsilon^{i}(t)$ to be  actually a symmetry generator, meaning it preserves the solution space, $\{p(t), q(t)\}$, some extra conditions should be satisfied as (\ref{gs3con}) or (\ref{gs4}). }  correspond to the choice of the Noether charge, $Q_{T}=\phi^{1{\rm st}}_{a}\varepsilon^{a}(t)$ such that $\varepsilon^{i}(t)=0$ and $\dot{\varepsilon}^{i}(t)=\Delta \free^{i}$ at time `$\,t\,$'.\\

On the other hand, if at time $t$ one has $\Delta \free(t)=0$, then  the time derivative $(\dot{p},\dot{q})$ of all the dynamical variables are the same in the two different total Hamiltonian systems, and the first nontrivial difference appears  at the order of $dt^{2}$ or the  `acceleration',
\be
\D{\Delta F(t+dt)~\sim~\half \,dt^{2}\left[\hat{F},{\phi}^{1{\rm st}}_{i}\right\}_{P.B.}\!\!\frac{{\rm{d}}\Delta \free^{i}}{{\rm{d}}t}.}
\ee
Again,  this may correspond to the choice of the Noether charge, $Q_{T}=\phi^{1{\rm st}}_{i}\varepsilon^{i}(t)$ such that $\varepsilon^{i}(t)=0$,  $\dot{\varepsilon}^{i}(t)=0$ and $\ddot{\varepsilon}^{i}(t)=\frac{{\rm d}{\Delta \free}^{i}}{{\rm d}t}$ at time $t$.

\newpage
\section{Dirac quantization for second class constraints}\label{Diracqu}

\subsection{Dirac bracket}

On the $2\N$-dimensional phase space with variables $q^{A}$ and $p_{B}$ ($1\leq A,B\leq\N$), we consider a set of functions $\rho_{s}(p,q)$ (where $1\leq s\leq \dim\{\rho_{s}\}\leq 2\N\,$) such that the following supermatrix is non-degenerate,
\be
\Omega_{st}:=[\rho_{s},\rho_{t}\}_{P.B.}=-(-1)^{\#_{s}\#_{t}}\Omega_{ts}\,,
\ee
or its inverse exists,
\be
\ba{lll}
\Omega_{st}\,(\Omega^{-1})^{tu}=\delta_{s}{}^{u}~&~~~\Longleftrightarrow~~~&~(\Omega^{-1})^{ st}\,\Omega_{tu}=\delta^{s}{}_{u}\,.
\ea
\ee
We note
\be
(\Omega^{-1})^{ st}=-(-1)^{\#_{s}\#_{t}+\#_{s}+\#_{t}\,}(\Omega^{-1})^{ ts}\,,
\label{D1}
\ee
and for an arbitrary quantity, $F$,
\be
[F,(\Omega^{-1})^{ st}\}_{P.B.}=-(-1)^{\#_{F}(\#_{s}+\#_{u})}(\Omega^{-1})^{ su}[F,\Omega_{uv}\}_{P.B.}(\Omega^{-1 })^{vt}\,.
\label{D2}
\ee

We \textit{define} the ``Dirac bracket'' associated with the functions $\rho_s$ as,
\be
{}
[F,G\}_{\Dirac}:=
[F,G\}_{P.B.}\,-\,[F,\rho_{s}\}_{P.B.}\Omega^{-1 st}[\rho_{t},G\}_{P.B.}\,.
\label{Diracbra}
\ee

Some crucial identities  follow. We first note that for an arbitrary object, $F$,
\be
\ba{ll}
[\rho_{s},F\}_{\Dirac}=0\,,~~~~&~~~~[F,\rho_{s}\}_{\Dirac}=0\,.
\ea
\label{rhoDzero}
\ee
This property is the \textit{raison d'\^etre} of the Dirac bracket. It means that one may impose $\rho_s=0$ either before or after computing the Dirac bracket, whichever one prefers.
Just like the Poisson bracket, the Dirac bracket  satisfies the symmetric property,
\be
\D{\left[F,G\right\}_{\Dirac}=-(-1)^{\#_{F}\#_{G}}\left[G,F\right\}_{\Dirac}\,,}
\ee
and the Leibniz rule,
\be
\ba{l}
\D{\left[F,GK\right\}_{\Dirac}=\left[F,G\right\}_{\Dirac}K+(-1)^{\#_{F}\#_{G}}
G\left[F,K\right\}_{\Dirac}\,,}\\
{}\\
\D{\left[FG,K\right\}_{\Dirac}=F\left[G,K\right\}_{\Dirac}+(-1)^{\#_{G}\#_{K}}
\left[F,K\right\}_{\Dirac}G\,.}
\label{LeibnizD}
\ea
\ee
Moreover,  from the Jacobi identity for the Poisson bracket, (\ref{Jacobi}) and (\ref{D1}, \ref{D2}), one can verify  the Jacobi identity for the Dirac bracket  after some tedious calculations,
\be
\D{\left[\left[F,G\right\}_{\Dirac},H\right\}_{\Dirac}=
\left[F,\left[G,H\right\}_{\Dirac}\right\}_{\Dirac}-(-1)^{\#_{F}\#_{G}}
\left[G,\left[F,H\right\}_{\Dirac}\right\}_{\Dirac}\,.}
\label{JacobiD}
\ee
In mathematical terms, one says that the Dirac bracket obeys to the axioms of a graded Poisson bracket.

\subsection{Total Hamiltonian dynamics with Dirac bracket}
One can prove by contradiction that the Poisson bracket between all the second class constraints is non-degenerate. For constrained systems, the Dirac bracket is defined for all the second-class constraints by identifying $\rho_{s}$ with
$\phi^{2{\rm nd}}_{s}\,$. It is convenient to let the Dirac bracket governs the dynamics rather than the Poisson bracket,
\be
\D{\frac{{\rm{d}}F(p,q,t)}{{\rm{d}}t}=\left[F,H_{T}\right\}_{\Dirac}+\frac{\partial F}{\partial t}\,,}
\label{DHF}
\ee
because in such case one may impose the second-class constraints before computing the evolution of the system. It follows that one may omit  the second-class primary constraints while adding to the Hamiltonian or to the Noether charge in the definition of the total Hamiltonian (\ref{totH}) or the  total Noether charge (\ref{QTdef}).  We note that
\be
\left[F,H_{T}\right\}_{P.B.}-\left[F,H_{T}\right\}_{\Dirac}=
[F,\phi^{2\rm{nd}}_{s}\}_{P.B.}\Omega^{-1 st}[\phi^{2\rm{nd}}_{t},H_{T}\}_{P.B.}\,.
\ee
Hence, as long as the second-class constraints have no explicit time dependence, the Poisson brackets $[\phi^{2\rm{nd}},H_{T}\}_{P.B.}$ in the right-hand-side vanish on the constraint hypersurface $\V\,$. In such case, the dynamics with the Dirac bracket and the other with the Poisson bracket, are identical on $\V$. Namely both reduce to the same Lagrangian dynamics.

\begin{itemize}

\item {\bf First order kinetic terms}\\
In most of the cases, the momenta $p_{\alpha}$ for the fermions are linear in the spinor field $\psi^{\alpha}$, resulting in the primary second-class constraints,
\be
\ba{lll}
\phi^{2\rm{nd}}_{\alpha}:=p_{\alpha}-L_{\alpha\beta}\psi^{\beta}\,,~~~~&~~~~
L_{\alpha\beta}=L_{\beta\alpha}\,,~~~~&~~~~
{}[\phi^{2\rm{nd}}_{\alpha},\phi^{2\rm{nd}}_{\beta}\}_{P.B.}=2L_{\alpha\beta}\,.
\ea
\label{PBf}
\ee
Using Eq.(\ref{Diracbra}), the Dirac bracket of an unconstrained bosonic system coupled with fermions read
\be
\D{\left[F,G\right\}_{\Dirac}=\sum_{{\rm bosons}}\!\!\left(\frac{\partial F}{\partial q^{b}}\frac{\partial G}{\partial p_{b}}-\frac{\partial F}{\partial p_{b}}\frac{\partial G}{\partial q^{b}}\right)\,+\,\sum_{{\rm fermions}}\half(-1)^{\#_{F}}L^{-1\alpha\beta}
\frac{\partial F}{\partial \psi^{\alpha}}
\frac{\partial G}{\partial \psi^{\beta}}\,,}
\label{DBFG}
\ee
where the factor $1/2$ comes from the factor two in the last equation of (\ref{PBf}).
Notice that it is important in Eq.(\ref{DBFG}) to treat the partial derivatives $(q^{b},p_{b},\psi^{\alpha},\phi^{2{\rm nd}}_{\beta})$ as the independent variables rather than $(q^{b},p_{b},\psi^{\alpha},p_{\alpha})$, and hence
\be
\ba{lll}
{}\left[\phi^{2{\rm nd}}_{\alpha},G\right\}_{\Dirac}=0\,,~~&~~~
{}\left[\psi^{\alpha},\psi^{\beta}\right\}_{\Dirac}=-\half L^{-1 \alpha\beta}\,,~~&~~~
{}\left[p_{\alpha},\psi^{\beta}\right\}_{\Dirac}=-\half \delta_{\alpha}{}^{\beta}\,,
~\mathit{etc}.
\ea
\label{DPR}
\ee
The last expression should be compared with $\left[p_{\alpha},\psi^{\beta}\right\}_{P.B.}=- \delta_{\alpha}{}^{\beta}$.
This `halfness' of the Dirac bracket is not related to the fermionic character of $\psi$, instead it is typical for the system with first order Lagrangians (such as the variational principle of the Hamiltonian formulation itself). If $\psi$ is complex, then the `halfness' of the quantization is `doubled' and one recovers the standard naive rule of the canonical quantization for a Dirac spinor.

\item {\bf Quantization}\\
The quantization can be straightforwardly performed by \textit{replacing the Dirac bracket by the super-commutator with a factor $-i$},
\be
\ba{lll}
{}\left[~~,~~\right\}_{\Dirac}~~&~~\Longrightarrow~~&~~-i\left[~~,~~\right\}\,,
\ea
\label{Dquantization}
\ee
which gives the standard convention,  $[\,q^{A},p_{B}\}=+\,i\,\delta^{A}{}_{B}$.
The point is that, from $$[\phi^{2{\rm nd}}_{s},F\}_{\Dirac}=0\quad \forall F\,,$$ the second-class constraints are central, even after the quantization.
Therefore, one can simultaneously \textit{impose the second-class constraints on the physical states.} This would not be possible if one had naively performed the correspondence rule in terms of the Poisson bracket itself. The second-class constraints should be represented by identically vanishing operators on the physical Hilbert space. In practice, this may be realized by solving explicitly the constraints in terms of some set $\{y^w\}$ of independent variables
\be
\phi^{2{\rm nd}}_{s}(x^M,t)=0\,\quad\Longleftrightarrow\quad x^M=f^M(y^v)
\ee
and try to represent the algebra $[y^v,y^w\}_{\Dirac}=g^{vw}(y)$ on the Hilbert space of functions of the $y$'s only.

Although this way of quantizing second-class constrained systems looks pretty straightforward and conceptually clear (one imposes all the constraints), in most practical cases, second-class constraints are most tedious because in general either we are not able to invert the matrix $\Omega_{st}$ and the Dirac bracket is not known explicitly, so that nothing can be done at all, or we are not able to find a faithful representation of the Dirac bracket algebra.\footnote{More comments on the quantization of second-class constraints can be found in the section 13.1 of \cite{Henneaux:1992ig}. Notice that some systems admit only the Dirac method of quantization and not the so-called ``reduced phase space'' method \cite{Plyushchay:1993hs}
.} Somehow surprisingly, first-class constraints are preferable because there is an algortihmic - though involved and subtle - way to quantize the theory in terms of the Poisson bracket (which is easy to represent).
\end{itemize}

\newpage
\section{BRST quantization for first class constraints}\label{BRSTquant}

The BRST procedure is motivated through the Faddeev-Popov construction. Here we review the essential features of them in a self-contained manner.

\subsection{Integration over a Lie group - Haar measure}

For a Lie group $\G$ of dimension $\NG\,$, we parameterize its elements $g$ by the coordinates $\theta^{a}$ ($1\leq a\leq\NG$) of the corresponding Lie algebra of a basis $\{T_{a}\}$,
\be
\ba{ll}
\D{g(\theta)=e^{i\theta^{a}T_{a}}\,,}~~~~&~~~~[T_{a},T_{b}]=iC_{ab}^{\,c}T_{c}\,.
\ea
\ee
We also define  a set of $\NG$ functions $\zeta^{a}(\theta_{1}^{b},\theta^{c}_{2})$ from the multiplication,
\be
g(\,\theta_{1})\,g(\,\theta_{2})=g\left(\zeta(\theta_{1},\theta_{2})\right)\,.
\ee
From the Baker-Campbell-Hausdorff formula,
\be
\D{\ln\left(e^{x}e^{y}\right)=x+y+\half[x,y]+\mbox{higher~order~commutators}\,,}
\label{BCH}
\ee
we obtain explicitly,
\be
\D{\zeta^{a}(\theta_{1},\theta_{2})=\theta^{a}_{1}+\theta_{2}^{a}-
\half\,\theta^{b}_{1}\theta_{2}^{c}C_{bc}^{\,a}+\mbox{higher~order~terms}\,.}
\label{BCH2}
\ee
~\\
The left invariant measure for the integration over the Lie group $\G$ is denoted by
\be
\D{\calD_{L} g:=\calD\theta W_{L}(\theta)=\prod_{a=1}^{\NG}\,{\rm{ d}}\theta^{a}~W_{L}(\theta)\,.}
\ee
By definition, it must satisfy the property of left invariance, \textit{i.e.} for an arbitrary fixed element $g_{0}\in \G\,$ and any function on the group  $F(g)$,
\be
\D{\int\calD_{L}g\,F(g)=\int\calD_{L}g\,F(g_{0}g)\,.}
\ee
Hence, the following identity must hold for any $\theta_{0}$ and $\theta$,
\be
\D{W_{L}(\theta)=\det\left(\frac{\partial\zeta^{a}(\theta_{0},\theta)}{\partial \theta^{b}}\right)W_{L}\left(\zeta(\theta_{0},\theta)\right)\,.}
\ee
Some simple choices like $\theta=0$ or $\theta_{0}=-\theta$ give explicitly
\be
\D{W_{L}(\theta)=W_{L}(0)\left.\det\left(\frac{\partial \zeta^{a}(\theta_{0},\theta)}{\partial \theta^{b}}\right)\right|_{\theta_{0}=-\theta}
=W_{L}(0)\left.{\det}^{-1}\left(\frac{\partial \zeta^{a}(\theta,\vartheta)}{\partial \vartheta^{b}}\right)\right|_{\vartheta=0}\,.}
\ee
Similarly one can define the right invariant measure $\calD_{R}g=\calD\theta W_{R}(\theta)$ to obtain
\be
\D{W_{R}(\theta)=W_{R}(0)\left.\det\left(\frac{\partial \zeta^{a}(\theta,\theta_{0})}{\partial \theta^{b}}\right)\right|_{\theta_{0}=-\theta}
=W_{R}(0)\left.{\det}^{-1}\left(\frac{\partial \zeta^{a}(\vartheta,\theta)}{\partial \vartheta^{b}}\right)\right|_{\vartheta=0}\,.}
\ee
~\\

Now we are going to show that both measures can be set   equal.
The chain rule for $\theta^\prime:=\zeta(\theta,v)\,$ gives
\be
\D{\left.\left(\frac{\partial \zeta^{a}(\zeta(\theta,\vartheta),\theta_0)}{\partial \vartheta^{c}}\right)\right|_{\vartheta=0}
=
\left.\left(\frac{\partial \zeta^{a}(\theta^\prime,\theta_{0})}{\partial \theta^{\prime b}}\right)\right|_{\theta^\prime=\theta}
\left.\left(\frac{\partial \zeta^{b}(\theta,\vartheta)}{\partial \vartheta^{c}}\right)\right|_{\vartheta=0}\,,}
\label{chainr}
\ee
because $\theta^\prime=\theta$ when $v=0\,$.
The associativity property explicitly reads
\be
\zeta(\theta_{1},\theta_{2},\theta_{3}):=
\zeta(\zeta(\theta_{1},\theta_{2}),\theta_{3})=
\zeta(\theta_{1},\zeta(\theta_{2},\theta_{3}))\,.
\ee
Therefore, evaluating Eq.(\ref{chainr}) at $\theta_0=-\theta\,$, we obtain
\be
\D{\left.\det\left(\frac{\partial \zeta^{a}(\theta,\vartheta,-\theta)}{\partial \vartheta^{c}}\right)\right|_{\vartheta=0}=\left.\det\left(\frac{\partial \zeta^{a}(\theta,\theta_{0})}{\partial \theta^{b}}\right)\right|_{\theta_{0}=-\theta}\left.\det\left(\frac{\partial \zeta^{b}(\theta,\vartheta)}{\partial \vartheta^{c}}\right)\right|_{\vartheta=0}\,.}
\label{chainrule}
\ee
From
\be
\ba{ll}
\D{g(\zeta(\theta,\vartheta,-\theta))=g(\theta)g(\vartheta)g(\theta)^{-1}=
e^{i\vartheta^{a}T_{\theta a}}\,,}~~~&~~~\D{T_{\theta a}:=g(\theta)T_{a}g(\theta)^{-1}=:(M_{\theta})_{ a}{}^{b}T_{b}\,,}
\ea
\ee
it follows that
\be
\zeta^{a}(\theta,\vartheta,-\theta)=\vartheta^{b}(M_{\theta})_{ b}{}^{a}\,,
\label{refref}
\ee
Consequently,
\be
\D{\tr\!\left(T_{a}T_{b}\right)=(M_{\theta})_{ a}{}^{c}(M_{\theta})_{ b}{}^{d}\,\tr\!\left(T_{c}T_{d}\right)\,.}
\ee
Thus, as long as $\D{\tr\!\left(T_{a}T_{b}\right)}$ is invertible as a $\NG\times\NG$ matrix, (\textit{e.g.} when $T_a$ are in the adjoint representation of a semisimple\footnote{If it is compact, then one can further take $\D{\tr\!\left(T_{a}T_{b}\right)\propto\delta_{ab}}$.} Lie group) we have $\left(\det M_{\theta}\right)^{2}=1$. Furthermore, from the continuity at $\theta= 0$ we disregard the possibility of $\det M_{\theta}$ being minus one. Thus, we obtain from (\ref{refref})
\be
\D{\det\left(\frac{\partial \zeta^{a}(\theta,\vartheta,-\theta)}{\partial \vartheta^{c}}\right)=\det M_{\theta}=+1\,.}
\ee
Inserting this relation on the left-hand-side of (\ref{chainrule})
and setting $W_{L}(0)= W_{R}(0)$, one has shown that the left and right invariant measures may be chosen identical for the Lie groups where $\det\Big(\tr(T_{a}T_{b})\Big)\neq 0\,$:
\be
W_{\Haar}(\theta):=W_{L}(\theta)=W_{R}(\theta)\,.
\ee
The corresponding measure is known as ``Haar measure''\cite{Haar} and satisfies
\be
\D{\int\calD g\,F(g)=\int\calD g\,F(g_{0}g)=\int\calD g\,F(gg_{0})=
\int\calD\theta\,W_{\Haar}(\theta)F(g(\theta))\,.}
\ee
\\

In the case of the local gauge symmetry in field theories with the gauge group $\G$, the parameters $\theta^{a}(x)$ are in fact arbitrary local functions. We  assume that there exists a countable complete  set in the commutative algebra of local functions,
\be
\ba{lll}
\{\,f_{n}(x)\,\}\,,~~~~&~~~~f_{n}(x)f_{m}(x):=d_{nm}^{\,l}f_{l}(x)\,,~~~~&~~~~\langle f_{n}(x)|f_{m}(x)\rangle=\delta_{nm}\,,
\ea
\ee
where $d_{nm}^{\,l}$ are the structure constants of the algebra with the pointwise product.
Then we can write $\theta^{a}(x)=\theta^{an}f_{n}(x)$ so that
\be
\ba{lll}
\theta^{a}(x)T_{a}=\theta^{an}T_{an}(x)\,,~~~&~~~T_{an}(x):=T_{a}f_{n}(x)\,,~~~&~~~
[T_{an}(x),T_{bm}(x)]=iC^{\,c}_{ab}d_{nm}^{\,l}T_{cl}(x)\,.
\ea
\ee

As the  structure constants for the set $\{T_{an}(x)\}$ are independent of the coordinates $x$, the above relations  induce a novel   group $\hat{\G}=\D{\left\{\,\hat{g}(\hat{\theta}^{an})\,\right\}}$, which is defined by the representation  $g(\theta(x))$, with the parameters  $\hat{\theta}^{an}$. Namely, though the representation is given for some fixed coordinates system $x$, there exists an abstract group independent of the coordinate choice.  We have
\be
\ba{ll}
\hat{g}(\hat{\theta}_{1})\hat{g}(\hat{\theta}_{2})=\hat{g}\left(\hat{\zeta}(\hat{\theta}_{1},
\hat{\theta}_{2})\right)\,,~~~~&~~~~\hat{\zeta}^{an}(\hat{\theta}_{1},
\hat{\theta}_{2}):=\langle f_{n}(x)|\zeta^{a}(\theta_{1}(x),\theta_{2}(x))\rangle\,.
\ea
\ee
The group $\hat{\G}$ is infinite-dimensional since it is a local (\textit{i.e.} position dependent) group.

Now we are ready to straightforwardly apply the left/right invariant measure to this local group:
\be
\ba{l}
\D{\int\calD \hat{g}\,F(\hat{g})=\int\calD \hat{\theta}\,\hat{W}_{\Haar}(\hat{\theta})\,F(\hat{g}(\hat{\theta}))=
\int\calD \hat{g}\,F(\hat{g}_{0}\hat{g})=\int\calD \hat{g}\,F(\hat{g}\hat{g}_{0})\,,}\\
{}\\
\D{\hat{W}_{\Haar}(\hat{\theta})=\left.\det{}^{-1}\left(\frac{\partial\hat{\zeta}^{am}
(\hat{\vartheta},\hat{\theta})}{\partial \hat{\vartheta}^{bn}}\right)\right|_{\hat{\vartheta}=0}
=\left.\det{}^{-1}\left(\frac{\partial\hat{\zeta}^{am}
(\hat{\theta},\hat{\vartheta})}{\partial \hat{\vartheta}^{bn}}\right)\right|_{\hat{\vartheta}=0},~~~~~~~
\calD\hat{\theta}=\prod_{a,n}{{\rm d}\hat{\theta}^{an}}\,.}
\ea
\label{Haarg}
\ee
The situation in  gauge field theories is that $F(\hat{g})$ is given by a spacetime integral of a functional $F(g,\partial_{\mu} g)$ which depends on the local group element $g$ and its spacetime derivatives $\partial_{\mu} g$,
\be
\D{F(\hat{g})=\int {\rm d}^{D}x\,F(g,\partial_{\mu} g)\,.}
\ee
In the remaining of the paper, for short notation we drop the hat symbol and simply denote \textit{e.g.}
\be
\D{\int\calD g\,F[{g}]=\int\calD \hat{g}\,F(\hat{g})\,.}
\ee

\subsection{Faddeev-Popov method}

We consider a dynamical system where a finite-dimensional Lie group $\G$,  acts on the dynamical variables  which we denote collectively  by $\Phi$.  For each element, ${g}\in{\G}$, we define a map
(\textit{i.e.} here a gauge transformation),
\be
\D{{g}~:~\Phi~{\longrightarrow}~\Phi_{{g}}\,.}
\ee
For later purpose, we write the successive gauge transformation in the following order,
$\Phi\rightarrow\Phi_{{g}_{1}}\rightarrow\Phi_{{g}_{2}{g}_{1}}$. In other words, one has a left action of ${\G}$ on the space of dynamical variables.\\

We introduce a Lie algebra valued functional $h[\Phi]=h^{a}[\Phi]T_{a}$ of $\Phi$ (which may depend on its derivatives as well).  We assume that $h[\Phi]$ is non-degenerate under the gauge transformations as\footnote{It is not necessary to require the non-degeneracy for $\theta\neq 0$.}
\be
\D{\left.\det\left(\frac{\partial h^{a}[\Phi_{g(\theta)}]}{\partial\theta^{b}}\right)\right|_{\theta=0}\neq 0\,,}
\ee
in which case it is called the ``gauge-fixing functional''.
~\\
~\\
For an arbitrary function $B(h)$ of $h^{a}$, the Faddeev-Popov functional $\calB[\Phi]$ of $\Phi$ reads
\be
\D{\calB[\Phi]:=B(h[\Phi])\left.\det\left(\frac{\partial h^{a}[\Phi_{g(\theta)}]}{\partial\theta^{b}}\right)\right|_{\theta=0}\,.}
\ee
By making use of the definition (\ref{Haarg}) of the Haar measure, one finds the following crucial identity,
\be
\ba{ll}
\D{\int\calD g\,\calB[\Phi_{g}]}&=\D{\int\calD g\,B(h[\Phi_{g}])\left.\det\left(\frac{\partial h^{a}[\Phi_{g(\vartheta)\, g}]}{\partial\vartheta^{b}}\right)\right|_{\vartheta=0}}\\
{}&{}\\
{}&=\D{\int\calD\theta\,\left.\det\left(\frac{\partial \zeta^{b}(\theta,\theta_{0})}{\partial \theta^{c}}\right)\right|_{\theta_{0}=-\theta}\,B(h[\Phi_{g(\theta)}])
\left.\det\left(\frac{\partial h^{a}[\Phi_{g(\zeta(\vartheta,\theta))}]}{\partial\vartheta^{b}}\right)\right|_{\vartheta=0}}\\
{}&{}\\
{}&=\D{\int\calD\theta\,\left.\det\left(\frac{\partial
h^{a}[\Phi_{g\left(\zeta(\theta,\theta_{0},-\theta_{0})\right)}]}{\partial \theta^{c}}\right)\right|_{\theta_{0}=-\theta}\,B(h[\Phi_{g(\theta)}])}\\
{}&{}\\
{}&=\D{\int\calD\theta\,\det\left(\frac{\partial
h^{a}[\Phi_{g(\theta)}]}{\partial \theta^{b}}\right)\,B(h[\Phi_{g(\theta)}])}\\
{}&{}\\
{}&=\D{\int\calD h\,B(h)\,,}
\ea
\label{crucialid}
\ee
where \be\calD h:=\prod\limits_{a=1}^{\NG}{\rm d}h^{a}\ee
Thus, \textsl{ the integral of $\calB[\Phi_{g}]$ over $\G$ is just a $c$-number,  being independent of $\Phi$ or the  choice of the gauge-fixing functionals $h^{a}[\Phi]$,   provided that  the integral domain for $h$ is $\Phi$ independent.} \\

Now we  consider a gauge invariant functional,
\be
\F(\Phi)=\F(\Phi_{g})\,,\label{FP1}
\ee
and further assume that the path integral measure is gauge invariant,
\be
\ba{ll}
\D{\calD\Phi=\calD\Phi_{g}\,,}~~~~&~~~~
\D{\det\!\left(\frac{\,\calD\Phi_{g}}{\,\calD\Phi\,}\right)=1\,,}
\label{FP2}
\ea
\ee
which is always satisfied for any dynamical system where  the fields are in the adjoint representation or there are equal number of fundamental and anti-fundamental fields, transformed by $g$ and $g^{-1}$ respectively. \\

\textsl{The Faddeev-Popov path integral method prescribes to multiply the gauge invariant functional integrand by the Faddeev-Popov functional,}
\be
\ba{lll}
\D{\int\!\calD\Phi~\F(\Phi)}&\longrightarrow&
\D{\int\!\calD\Phi~\F(\Phi)\,\calB[\Phi]=\int\!\calD\Phi~\F(\Phi)B(h[\Phi])
\left.\det\left(\frac{\partial h^{a}[\Phi_{g(\theta)}]}{\partial\theta^{b}}\right)\right|_{\theta=0}\,,}
\ea
\ee
\textsl{to satisfy}
\be
\D{\int\calD g\int\calD\Phi\,\F(\Phi)\calB[\Phi]=
\int\calD g\int\calD\Phi\,\F(\Phi)\calB[\Phi_{g}]=\left(\int\calD h\,B(h)\right)\int\calD\Phi\,\F(\Phi)\,,}
\ee
where we made use of (\ref{crucialid}), (\ref{FP1}) and (\ref{FP2}). Equivalently we have
\be
\D{\int\calD\Phi\,\F(\Phi)\calB[\Phi]=\left(\int\calD h\,B(h)\right)\,\frac{~~
{\D{\int}}\calD\Phi\,\F(\Phi)~~}{\D{\int}\calD g~1}~\propto~\int\calD\Phi\,\F(\Phi)\,.}
\ee
~~\\

In  gauge field theories, the volume integral of the gauge group is often  divergent, and the Faddeev-Popov method \cite{Faddeev:1967fc}
provides a regularization scheme for that by adding the Faddeev-Popov factor to the original gauge invariant action. In the path integral, the gauge invariant functional is given by the exponential of the gauge symmetric action,
\be
\ba{ll}
\F[\Phi]=\D{e^{\coeff\,\calS[\Phi]}}\,,~~~~&~~~~\calS[\Phi]=\calS[\Phi_{g}]\,.
\ea
\ee
Furthermore, in this context, the arbitrary function $B(h)$ is reformulated as a Fourier transformation,
\be
\D{B\left(h[\Phi]\right)=\int\calD k\,e^{\coeff\,\left(-V(k)\,+\, k_{a}h^{a}[\Phi]\,\right)}}\,,
\ee
where the fields $k_a$ are said to be ``auxiliary'' and $V(k)$ is proportional to the logarithm of the Fourier transform ${\tilde B}(k)$ of the function $B(h)$ (anyway, the coice of $V$ is as arbitrary as the one of $B$).
Moreover, the Faddeev-Popov determinant can be written by introducing a pair of fermionic scalar fields, called ``ghosts'', $\bar{\w}_{a}$ and $\w^{b}$ ($1\leq a,b\leq \NG$),
\be
\D{\left.\det\left(\frac{\partial h^{a}[\Phi_{g(\theta)}]}{\partial\theta^{b}}\right)\right|_{\theta=0}=\int\calD\bar{\w}\calD \w\,e^{\coeff\,\bar{\w}_{a}\Delta^{a}{}_{b}[\Phi]\w^{b}}}~,
\ee
where we set\footnote{All these relations are valid for theories in the Minkowskian spacetime. For the Euclidean theories we only need to replace the factor, $\coeff$  by $-1$, and consider the Laplace transformation rather than the Fourier transformations.}
\be
\D{\Delta^{a}{}_{b}[\Phi]:=
\left.\frac{\partial h^{a}[\Phi_{g(\theta)}]}{\partial\theta^{b}}\right|_{\theta=0}}\,.
\ee
Note that  $\bar{\w}_{a}$ and  $\w^{b}$ are not necessarily complex conjugate to each other. We further assign the \textit{ghost number}, $+1$ for $\w^{a}$, $-1$ for $\bar{\w}_{b}$, and $0$ for the other  fields $\Phi$ and $k_{a}$.\\

Combining them all, the  Faddeev-Popov action reads
\be
\D{\calS_{\FP}[\Phi,\w,\bar{\w},k]=\calS[\Phi]-V(k)+ k_{a}h^{a}[\Phi]+\bar{\w}_{a}\Delta^{a}{}_{b}[\Phi]\w^{b}\,.}
\label{FPaction}
\ee
~\\

For example, in the Yang-Mills theory (on a curved spacetime), if we take the Lorentz gauge,
\be
h[A]:=\nabla_{\mu}A^{\mu}\,,
\label{LorentzG}
\ee
then we can set
\be
\displaystyle{\calS_{\FP}[\Phi,\w,\bar{\w},k]=\calS_{\scriptscriptstyle{YM}}[\Phi]+\int{\rm d}x^{D}\sqrt{g}\,\tr\Big(-V(k)+ k\,h[A]-\nabla^{\mu}\bar{\w}D_{\mu}\w\Big)\,.}
\label{FPYM}
\ee
~\\
Another example is the gauged Hermitian one-matrix model. Diagonal gauge choice leads to a  Faddeev-Popov  determinant which is nothing but the Vandermonde determinant \cite{Klebanov:1991qa}.

\subsection{BRST symmetry}

Remarkably, even after choosing a gauge\footnote{For example, the choice $V(k)=0$ leads to a delta function to fix the gauge as $h^{a}[\Phi]=0$.}, the path integral still does have a symmetry related to the gauge invariance. Indeed, the Faddeev-Popov action (\ref{FPaction}) possesses a fermionic nilpotent rigid symmetry known as  ``BRST symmetry'', after its discoverers, Becchi-Rouet-Stora \cite{Becchi:1975nq} and, independently, Tyutin \cite{Tyutin:1975qk}.\\

To discuss the BRST symmetry it is useful to  note that
the infinitesimal transformation associated with the Lie algebra element $\delta g=i\,\vartheta^{a}T_{a}$ is given by
\be
\D{\delta\Phi=\left.\frac{{\rm d}~}{{\rm d}s}\Phi_{g(s\vartheta)}\right|_{s=0}\,,}
\ee
and, from (\ref{BCH}), the Faddeev-Popov matrix $\Delta^{a}{}_{b}[\Phi]$ transforms as,
\be
\ba{ll}
\D{\delta\Delta^{a}{}_{b}[\Phi]}&\D{=
\left.\frac{{\rm d}~}{{\rm d}s}\Delta^{a}{}_{b}[\Phi_{g(s\vartheta)}]\right|_{s=0}}\\
{}&{}\\
{}&\D{=\left.\frac{{\rm d}~}{{\rm d}s}\left(\frac{\partial h^{a}[\Phi_{g\left(\zeta(\theta,s\vartheta)\right)}]}{\partial\theta^{b}}
\right)\right|_{\theta=0,\,s=0}}\\
{}&{}\\
{}&\D{=\left.\left(\frac{\partial h^{a}[\Phi_{g(\zeta)}]}{\partial\zeta^{c}}
\right)\right|_{\zeta=0}\,\left.\frac{{\rm d}~}{{\rm d}s}\left(\frac{\partial\zeta^{c}(\theta,s\vartheta)}{\partial\theta^{b}}
\right)\right|_{\theta=0,\,s=0} }\\
{}&{}\\
{}&\D{=-\half\Delta^{a}{}_{c}[\Phi]\,C^{\,c}_{bd}\vartheta^{d}\,,}
\ea
\ee
while $h^{a}[\Phi]$ transforms as
\be
\delta h^{a}[\Phi]=\Delta^{a}{}_{b}[\Phi]\,\vartheta^{b}\,.
\ee
~\\

With a fermionic rigid (\textit{i.e.} independent of $x$) scalar parameter $\varepsilon$, the BRST transformation reads as a `gauge transformation' generated by $\vartheta^{a}=\varepsilon\w^{a}$ or, equivalently by
\be
\delta g=i\,\varepsilon\,\w^{a}T_{a}\,,
\ee
so that
\be
\ba{ll}
\D{\delta\Phi=\left.\frac{{\rm d}~}{{\rm d}s}\Phi_{g(s\varepsilon\w)}\right|_{s=0}\,,}~~~~~~&~~~~~~\D{\delta k_{a}=0\,,}\\
{}&{}\\
\D{\delta\w^{a}=-\half\varepsilon\,C^{\,a}_{bc}\w^{b}\w^{c}\,,}~~~~~~&~~~~~~
\D{\delta\bar{\w}_{a}=-\varepsilon\,k_{a}\,,}
\ea
\label{BRSTtr}
\ee
where we also defined the transformation of the ghosts $\w$ and $\bar\w\,$.\\

In Yang-Mills theories   the dynamical variables  $\{\Phi\}=\{A_\mu,\phi,\psi,\bar\psi\}$  consist  of a vector  field $A_\mu$,   matter fields $\psi$, $\bar\psi$ in the fundamental, anti-fundamental representations and matter fields $\phi$ in the adjoint representation,\footnote{In our analysis, the matter fields $\phi$, $\psi$ and $\bar\psi$ can be either  bosonic or fermionic.} such that  the standard gauge transformations are
\be
\ba{llll}
A_{g\mu}=gA_{\mu}g^{-1}-i\partial_{\mu}gg^{-1}\,,~~~&~~~
\phi_{g}=g\phi g^{-1}\,,~~~&~~~
\psi_{g}=g\psi\,,~~~&~~~\bar{\psi}_{g}=\bar{\psi}g^{-1}\,,~~~\mathit{etc.}
\ea
\label{gaugephysical}
\ee
Explicit expressions for the infinitesimal BRST transformations read\cite{Weinberg}
\be
\ba{llll}
\delta A_{\mu}=\varepsilon D_{\mu}\w\,,~~~&~~~
\delta\phi=i[\,\varepsilon\w\,,\,\phi]\,,~~~&~~~
\delta\psi=i\varepsilon\w\psi\,,~~~&~~~
\delta\bar{\psi}=-i\bar{\psi}\varepsilon\w\,,
\ea
\ee
where we set
\be
\w:=\w^{a}T_{a}\,.
\ee

We introduce the BRST charge $\QB$ which is the fermionic generator of the BRST transformations satisfying  $\delta\Phi=\varepsilon[\QB\,,\,\Phi\}$.  More explicitly,
\be
\ba{ll}
\D{[\QB\,,\,A_{\mu}]=+D_{\mu}\w\,,}~~~~&~~~~
\D{[\QB\,,\,\phi\}=+i[\w\,,\,\phi\}\,,}\\
{}&{}\\
{}[\QB\,,\,\psi\}=+i\w\psi\,,~~~~&~~~~
{}[\QB\,,\,\bar{\psi}\}=\mp i\bar{\psi}\w\,,\\
{}&{}\\
\D{\{\QB\,,\,\w^{a}\}=-\half C^{\,a}_{bc}\w^{b}\w^{c}\,,}~~~&~~~
\D{\{\QB\,,\,\bar{\w}_{a}\}=-k_{a}\,,}\\
{}&{}\\
\multicolumn{2}{l}{
\D{[\QB\,,\,k_{a}]=0\,,}}
\ea
\label{QBcom}
\ee
where the sign, $\mp$ depends on whether $\bar{\psi}$ is bosonic or fermionic. \\

In particular, the transformation was designed to satisfy  $\{\QB\,,\,\Delta^{a}{}_{b}[\Phi]\w^{b}\}=0$ in order for the Faddeev-Popov action (\ref{FPaction}) to be BRST closed, and such that
\be
\D{\{\QB\,,\,\w\}=+\,i\,\w^{2}\,.}
\ee
The BRST charge $\QB$ increases the ghost number by $+1$.\\
~\\

Because $\varepsilon^2=0$, the finite transformations are then given by
\be
\D{e^{\varepsilon\,{{\Ad}_\QB}}=1+{\varepsilon\,{\Ad}_\QB}\,,}
\ee
and, with
$\D{g:=e^{i\varepsilon\w}=1+i\varepsilon\w\,,}$  each field transforms to
\be
\ba{llll}
\Phi_{g}\,,~~~&~~~\w_{g}=
g\w=\w+i\varepsilon\w^{2}\,,~~~&~~~
\bar{\w}_{ga}=\bar{\w}_{a}-\varepsilon k_{a}\,,~~~&~~~k_{ga}=k_{a}\,.
\ea
\ee
Further under the successive transformations,
$\D{e^{\varepsilon_{1}{{\Ad}_\QB}}e^{\varepsilon_{2}{{\Ad}_\QB}}}\,,$
each field transforms as
\be
\ba{lllll}
\Phi~&\longrightarrow~&\Phi_{g_{2}}~&\longrightarrow~&\Phi_{g_{1}\circ g_{2}^{\prime}}=\Phi_{g_{3}}\,,\\
{}&{}^{}&{}&{}\\
\w~&\longrightarrow~&\w_{g_{2}}=\w+i\varepsilon_{2}\w^{2}~&\longrightarrow~&
\w_{g_{1}}+i\varepsilon_{2}(\w_{g_{1}})^{2}=\w+i(\varepsilon_{1}+\varepsilon_{2})\w^{2}=
\w_{g_{3}}\,,\\
{}&{}^{}&{}&{}\\
\bar{\w}_{a}~&\longrightarrow~&\bar{\w}_{g_{2}a}=\bar{\w}_{a}-\varepsilon_{2}k_{a}
~&\longrightarrow~&\bar{\w}_{g_{1}a}-\varepsilon_{2}k_{g_{1}a}=
\bar{\w}_{a}-(\varepsilon_{1}+\varepsilon_{2})k_{a}=\bar{\w}_{g_{3}a}\,,\\
{}&{}^{}&{}&{}\\
k_{a}~&\longrightarrow~&k_{g_{2}a}=k_{a}
~&\longrightarrow~&k_{g_{1}a}=k_{a}=k_{g_{3}a}\,,
\ea
\ee
where we set
\be
\ba{ll}
\D{g^{\prime}_{2}:=1+i\varepsilon_{2}\w_{g_{1}}
=1+i\varepsilon_{2}\w+\varepsilon_{1}\varepsilon_{2}\w^{2}\,,}~~~~&~~~~
\D{g_{3}:=e^{i(\varepsilon_{1}+\varepsilon_{2})\w}=1+i(\varepsilon_{1}+\varepsilon_{2})\w=
g_{1}\circ g_{2}^{\prime}\,.}
\ea
\ee
Thus, we note
\be
\D{e^{\varepsilon_{1}{{\Ad}_\QB}}e^{\varepsilon_{2}{{\Ad}_\QB}}=
e^{(\varepsilon_{1}+\varepsilon_{2}){{\Ad}_\QB}}}\,,
\ee
and hence, \textit{the BRST charge is nilpotent,}
\be
\ba{lll}
\left({\Ad}Q_{\BRST}\right)^{2}=0\,,~~~~&\mbox{or}&~~~~
[\QB\,,\,[\QB\,,\,\mbox{Field}\}\}=0\,.
\ea
\ee
The nilpotent property can be directly checked from (\ref{QBcom}),  using the Jacobi identity.\\

From the gauge invariance of the original action it follows that   $\calS[\Phi]$ is $\QB$-closed,
\be
\Big[\QB\,,\,\calS[\Phi]\Big]=0\,.
\ee
Hence, the Faddeev-Popov action reads as  a sum of   $\QB$-closed and  $\QB$-exact terms,
\be
\D{\calS_{\FP}[\Phi,\w,\bar{\w},k]=\calS[\Phi]
+\Big\{\QB\,,\,\bar{\w}_{a}V^{a}(k)-\bar{\w}_{a}h^{a}[\Phi]\Big\}\,,}
\ee
where, without loss of generality we have shifted the arbitrary function of the auxiliary field, $V(k)$, by a constant in order to satisfy  $V(0)=0$ and to write it as
\be
V(k):=\,k_{a}V^{a}(k)= -\,\Big\{\QB\,,\,\bar{\w}_{a}V^{a}(k)\Big\}\,.
\ee
~\\

In particular, for the Yang-Mills theory with the Lorentz gauge (\ref{LorentzG}),  one can rewrite the whole  action (\ref{FPYM}) as
\be
\ba{l}
~~~\displaystyle{\calS_{\scriptscriptstyle{YM}}[\Phi]+\int{\rm d}x^{D}\,\sqrt{g}\,\tr\Big(-V(k)+ k\nabla_{\mu}A^{\mu}-\nabla^{\mu}\bar{\w}D_{\mu}\w\Big)}\\
{}\\
\D{=\calS_{\scriptscriptstyle{YM}}[\Phi]+\int{\rm d}x^{D}\,\Big\{\QB\,,\,\sqrt{g}\,\tr\Big(\bar{\w}V(k)
-\bar{\w}\nabla_{\mu}A^{\mu}\Big)\Big\}\,.}
\ea
\ee

\subsection{Hodge charge and a number operator}

The Hodge charge is defined to be fermionic and acts only on the auxiliary fields $k_a$ as \be
\ba{ll}
{}\left[Q_{\Hodge}\,,\,k_{a}\right]=-\bar{\w}_{a}\,,~~~~&~~~~\left[Q_{\Hodge}\,,\,{\rm others}\right\}=0\,.
\ea
\ee
Hence, it is nilpotent,
\be
({\rm ad}_{Q_{\Hodge}})^{2}=0\,,
\ee
and satisfies
\be
\left\{Q_{\Hodge}\,,\,Q_{\BRST}\right\}=N_{\bar{\w},k}\,,
\ee
where $N_{\bar{\w},k}$ is the number operator counting the total number of $\bar{\w}_{a}$ and $k_{b}$ fields,
\be
\ba{lll}
{}\left[N_{\bar{\w},k}\,,\,\bar{\w}_{a}\right]=\bar{\w}_{a}\,,~~~~&~~~~
{}\left[N_{\bar{\w},k}\,,\,k_{a}\right]=k_{a}\,,~~~~&~~~~
\left[N_{\bar{\w},k}\,,\,{\rm others\,}\right]=0\,.
\ea
\ee
Both of $Q_{\BRST}$ and $Q_{\Hodge}$ do not change the total number of $\bar{\w}_{a}$ and $k_{b}$ fields  since they do not annihilate the quantity such as $\{Q_{\Hodge},\bar{\w}_{a}\}=0\,$. One can show straightforwardly that
\be
\ba{ll}
{}\left[Q_{\BRST}\,,\,N_{\bar{\w},k}\right]=0\,,~~~~&~~~~
{}\left[Q_{\Hodge}\,,\,N_{\bar{\w},k}\right]=0\,.
\ea
\ee
~\\

Consider a $Q_{\BRST}$-closed quantity $\Upsilon$,
\be
\left[Q_{\BRST}\,,\,\Upsilon\right\}=0\,.
\ee
The latter condition is extension of the gauge invariance condition for functionals depending on $\Phi$ only.
We can decompose it as a sum of $N_{\bar{\w},k}$ eigenstates,
\be
\ba{ll}
\displaystyle{\Upsilon=\sum_{N=0}^{\infty}\,\Upsilon_{N}\,,}~~~~&~~~~[N_{\bar{\w},k}\,,\,
\Upsilon_{N}]=N\Upsilon_{N}\,.
\ea
\ee
From  $\left[Q_{\BRST}\,,\,N_{\bar{\w},k}\right]=0$ one can deduce that
\be
\D{\left[Q_{\BRST}\,,\,\Upsilon_{N}\right\}=0\,.}
\ee
Therefore, any $Q_{\BRST}$-closed quantity can be written as
\be
\ba{lll}
\Big[Q_{\BRST}\,,\,\Upsilon\Big\}=0~~~~&\Longleftrightarrow&~~~~
\Upsilon[\Phi,\w,\bar{\w},k]=\Upsilon_{0}[\w,\Phi]
+\left[Q_{\BRST}\,,\,\widetilde{\Upsilon}[\Phi,\w,\bar{\w},k]\right\}\,,
\ea
\ee
where
\be
\ba{lll}
\Big[Q_{\BRST}\,,\,\Upsilon_{0}[\w,\Phi]\Big\}=0\,,~~~&~~~
\Big[N_{\bar{\w},k}\,,\,\Upsilon_{0}[\w,\Phi]\Big]=0\,,~~~&~~~
\D{\widetilde{\Upsilon}=\sum_{N=1}^{\infty}\,\frac{1}{N}
\Big[Q_{\Hodge}\,,\,\Upsilon_{N}\Big\}\,.}
\ea
\ee
In other words, the cohomology group of the BRST charge is trivial at non-vanishing grading $N_{\bar{\w},k}$. This reasoning is a particular example of a standard procedure for computing cohomology groups. In mathematical terms, the Hodge differential $Q_{\Hodge}$ is called a ``contracting homotopy'' for the number operator $N_{\bar{\w},k}$ with respect to the BRST differential $Q_{\BRST}\,$. More materials on general cohomology groups can be found in the introduction for physicists to graded differential algebras in the chapter 8 of \cite{Henneaux:1992ig}.\\

The importance of the BRST cohomology sits in the general theorem that the BRST cohomology group at ghost number zero is isomorphic\footnote{The reader may consult the section 11.1 of \cite{Henneaux:1992ig} for a proof and for more comments.} to the algebra of observables of the theory. In other words, for the quantum theory it is isomorphic to the physical spectrum. Apart from this, as was roughly shown in the above example, the importance of the BRST formalism in gauge theories is that it allows to write a gauge-fixed path integral and to make sure that the final results are independent of the choice of gauge (because the corresponding terms are BRST trivial), see \textit{e.g.} \cite{Henneaux:1985kr}. Nevertheless, the BRST cohomology is of high interest already at the classical level, {as explained in the report \cite{Barnich:2000zw} where some examples of applications are given.}
For further discussion on  the `Hamiltonian' BRST formalism presented here, we refer to the chapters 9 till 12 and the chapter 14 of \cite{Henneaux:1992ig}. The `Lagrangian' BRST formalism of Batalin and Vilkovisky \cite{Batalin:1983} has the advantage of being entirely general (it includes the case of gauge algebras that close only on-shell) and covariant (since it is Lagrangian). The Batalin-Vilkovisky (also called  ``antifield'') formalism is explained in the specific reviews \cite{Henneaux:1989} and in the chapters 17 \& 18 of \cite{Henneaux:1992ig}.\newpage

\section*{Acknowledgments}

We thank Nicolas Boulanger and Imtak Jeon for useful feedbacks and Glenn Barnich for discussions on some references.
We are grateful to the Institut des Hautes \'Etudes Scientifiques
(IH\'ES, Bures-sur-Yvette), to the Center for Quantum SpaceTime (CQUeST, Seoul)
and to the Laboratoire de Math\'ematiques et de Physique Th\'eorique (LMPT, Tours) for hospitality during visits, sorted by chronological order. The research of JHP is in part supported  by the Korea Foundation for International Cooperation of Science \& Technology with grant number K20821000003-08B1200-00310,  by   the Center for Quantum Spacetime of Sogang University with grant number R11-2005-021, and  by the Korea Science and Engineering Foundation with grant number R01-2007-000-20062-0.

\newpage
\appendix

\begin{center}
\Large{\textbf{Appendix}}
\end{center}

\section{Some proofs}\label{someProofs}

Here we present the proofs of some facts discussed in the main body of the text.
\begin{itemize}
\item Eq.(\ref{FdK}).
\begin{proposition} An arbitrary function on the jet space $F(q_{n},t)$ is a total derivative if and only if its Euler-Lagrange equations vanish identically,
\be
\D{\frac{\delta~F}{\delta q^{A}}
(q_{n},t)=0~~~\Longleftrightarrow~~~  F(q_{n},t)=\frac{{\rm d}K(q_{n},t)}{{\rm d}t}}\,\,.
\label{FdK2}
\ee
\end{proposition}

\proof{The proof of the necessity `$\Leftarrow$' is straightforward from (\ref{dtq}).
In order to show the sufficiency, `$\Rightarrow$',  we filter the set of all functions on the jet space by sets $\F_{N}$,
\be
\ba{ll}
\F_{N}:=\{F(q^{A}_{n},t)\,,\,0\leq n\leq N\}\,,~~~~~&~~~~N=0,1,2,3,\cdots\,.
\ea
\ee
We prove the sufficiency by mathematical induction on $N\,$:

\begin{itemize}
	\item When $N=0$, the left hand side of the claim (\ref{FdK2}) implies that the function depends, at most,  only on the explicit time, $t$,  being  independent of $q_{0}^{A}$, \textit{i.e.} $F(t)$. We can simply set $K(t)=\int^t_0{\rm d}t^{\prime}\,F(t^{\prime})$.
  \item Now we assume that the converse is true for any $0\leq N< M$, and consider the case, $N=M$. It is useful to note that if $F\in\F_{N}$, then $\D{\left(\frac{{\rm d}}{{\rm d}t}\right)^{\!n}F\in\F_{N+n}}$ and its only dependence on $q_{N+n}$ appears as
\be
\ba{ll}
\D{\left(\frac{{\rm d}}{{\rm d}t}\right)^{\!n}F=q^{A}_{N+n}\,\frac{\partial F}{\partial q^{A}_{N}}\,+\,\O_{N+n-1}\,,}~~~&~~~\O_{N+n-1}\in\F_{N+n-1}\,.
\ea
\ee
Hence for $F\in\F_{M}$, from
\be
\D{0=\frac{\delta F(q_{n},t)}{\delta q^{A}}=\sum_{n=0}^{M}\left(-\frac{{\rm d}}{{\rm d}t}\right)^{\!n}\frac{\partial F}{\partial q^{A}_{n}}=(-1)^{M}q^{B}_{2M}\frac{\partial^{2}F}{\partial q^{B}_{M}q^{A}_{M}}\,+\,\O_{2M-1}\,,}
\ee
we first note that $F\in\F_{M}$ is at most linear in $q_{M}$, \textit{i.e.}
\be
\ba{ll}
F(q_{n},t)=q_{M}^{A}F_{A}(q_{n},t)+\O_{M-1}\,,~~~&~~~F_{A}\in\F_{M-1}\,.
\ea
\ee
Consequently,
\be
\ba{ll}
0&\D{=\frac{\delta F(q_{n},t)}{\delta q^{A}}}\\
{}&\D{=\left(-\frac{{\rm d}}{{\rm d}t}\right)^{\!M}F_{A}\,+\,(-1)^{\#_{A}\#_{B}}
\left(-\frac{{\rm d}}{{\rm d}t}\right)^{\!M-1}\left(q^{B}_{M}\frac{\partial F_{B}~~}{\,\partial q^{A}_{M-1}}\right)\,+\,\O_{2M-2}}\\
{}&{}\\
{}&\D{=(-1)^{M}q^{B}_{2M-1}\frac{\partial F_{A}~~}{\partial q^{B}_{M-1}} +(-1)^{M-1}(-1)^{\#_{A}\#_{B}}q^{B}_{2M-1}\frac{\partial F_{B}~~}{\partial q^{A}_{M-1}}\,+\,\O_{2M-2}\,.    }
\ea
\ee
Thus,
\be
\D{\frac{\partial F_{A}~~}{\partial q^{B}_{M-1}} -(-1)^{\#_{A}\#_{B}}\frac{\partial F_{B}~~}{\partial q^{A}_{M-1}}=0\,,}
\ee
so, by the usual Poincar\'e lemma in the space of one-forms $F_A(q^B_{M-1})$, there exists a function $K(q_{n},t)$, such that
\be
\ba{ll}
\D{F_{A}=\frac{\partial K(q_{n},t)}{\partial q^{A}_{M-1}}\,,}~~~&~~~K(q_{n},t)\in\F_{M-1}\,.
\ea
\ee
Finally, if we define $\D{F^{\prime}:=F-\frac{{\rm d}K}{{\rm d}t}}$, then
\be
\ba{ll}
\D{\frac{\delta F^{\prime}}{\delta q^{A}}=0}\,,~~~~&~~~~F^{\prime}\in\F_{M-1}\,.
\ea
\ee
Thus, from the induction hypothesis, $F^{\prime}$ is a total derivative, and hence so is $F$ itself.
\end{itemize}
This completes our proof.}

\item Eq.(\ref{usefulQq}).\\\\
Using Eq.(\ref{totaltrivial}), direct manipulation shows the following chain of identities, for an arbitrary function $F(q_n,t)$ on the jet space and for $m\geq 1\,$,
\be
\ba{ll}
\D{\sum_{n=0}^{\infty}\,\left(-\frac{{\rm d}}{{\rm d}t}\right)^{\! n}\left(
\frac{\partial \q_{m}^{B}}{\partial q^{A}_{n}}F\right)}&=
\D{\sum_{n=0}^{\infty}\,\left(-\frac{{\rm d}}{{\rm d}t}\right)^{\! n}\left[\left\{
\frac{{\rm d}}{{\rm d}t}\left(\frac{\partial \q_{m-1}^{B}}{\partial q^{A}_{n}}\right)+
\frac{\partial \q_{m-1}^{B}}{\partial q^{A}_{n-1}}\right\}F\right]}\\
{}&{}\\
{}&=\D{\sum_{n=0}^{\infty}\,\left(-\frac{{\rm d}}{{\rm d}t}\right)^{\! n}\left[\frac{{\rm d}}{{\rm d}t}
\left(\frac{\partial \q_{m-1}^{B}}{\partial q^{A}_{n}}F\right)+
\frac{\partial \q_{m-1}^{B}}{\partial q^{A}_{n-1}}F-\frac{\partial \q_{m-1}^{B}}{\partial q^{A}_{n}}\frac{{\rm d}F}{{\rm d}t}\right]}\\
{}&{}\\
{}&=\D{\sum_{n=0}^{\infty}\,\left(-\frac{{\rm d}}{{\rm d}t}\right)^{\! n}\left[\frac{\partial \q_{m-1}^{B}}{\partial q^{A}_{n}}\left(-\frac{{\rm d}}{{\rm d}t}\right)F\right]}\\
{}&{}\\
{}&=\D{\sum_{n=0}^{\infty}\,\left(-\frac{{\rm d}}{{\rm d}t}\right)^{\! n}\left[\frac{\partial \q_{0}^{B}}{\partial q^{A}_{n}}\left(-\frac{{\rm d}}{{\rm d}t}\right)^{\! m}F\right]\,.}
\ea
\label{usefulQqproof}
\ee

\item Eq.(\ref{mainc1}).\\\\
We show the relation  (\ref{mainc1})  by induction on the power of $s$.  We assume that the  following relation is true up to the power $N-1$ in $s$,
\be
\D{f^{B}_{l}(q_{m},t)\frac{\partial~}{\partial q_{l}^{B}}\q_{n}^{A}= f_{n}^{A}(\q_{m},t)+{\cal O}(s^{N}) \,,}
\ee
which is clearly true for $N=1$, as $\q_{n}^{A}=q_{n}^{A}$ if $s=0$.  Now differentiating the left hand side with respect to $s$, we get, up to the power $N$ in $s$,
\be
\ba{ll}
\D{\frac{{\rm{d}}\,}{{\rm{d}s}}\left(
f^{B}_{l}(q_{m},t)\frac{\partial~}{\partial q_{l}^{B}}\q_{n}^{A}\right)}&=
\D{f^{B}_{l}(q_{m},t)\frac{\partial~}{\partial q_{l}^{B}}\left(
f^{C}_{p}(q_{k},t)\frac{\partial~}{\partial q_{p}^{C}}\q_{n}^{A}\right)}\\
{}&{}\\
{}&=\D{f^{B}_{l}(q_{m},t)\frac{\partial~}{\partial q_{l}^{B}}f_{n}^{A}(\q_{k},t)}+{\cal O}(s^{N})\\
{}&{}\\
{}&=\D{f^{B}_{l}(q_{p},t)\frac{\partial\q_{k}^{C}}{\partial q_{l}^{B}}\frac{\partial~}{\partial \q_{k}^{C}}f_{n}^{A}(\q_{m},t)}+{\cal O}(s^{N})\\
{}&{}\\
{}&=\D{\frac{{\rm{d}}\q^{C}_{k}}{{\rm{d}s}}\frac{\partial~}{\partial \q_{k}^{C}}f_{n}^{A}(\q_{m},t)}+{\cal O}(s^{N})\\
{}&{}\\
{}&=\D{\frac{{\rm{d}}\,}{{\rm{d}s}}f_{n}^{A}(\q_{m},t)+{\cal O}(s^{N})\,.}
\ea\label{usefulQqproof2}
\ee
Thus, the relation holds up to power $N$, and this completes our proof.

\item Eq.(\ref{VddV}).
\begin{proposition}For an arbitrary quantity, $F(p,q,t)$,  the two actions, namely taking  the time derivative and taking the restriction on $\V$, commute each order.\end{proposition}
\proof{Using the coordinates, $(x,\phi)$ for the whole phase space, (\ref{newco}), we first define $\widetilde{F}(x,\phi,t)=F(p,q,t)$. Then
\be
\ba{ll}
\D{\left.\left(\frac{{\rm d}~}{{\rm d}t}F(p,q,t)\right)\right|_{\V}}&=
\D{\left.\left(\frac{{\rm d}~}{{\rm d}t}\widetilde{F}(x,\phi,t)\right)\right|_{\V}}\\
{}&{}\\
{}&=\D{\left.\left(\dot{x}^{\hat{\imath}}\frac{\partial \widetilde{F}}{\partial x^{\hat{\imath}}}+\dot{\phi}_{h}
\frac{\partial\widetilde{F}}{\partial\phi_{h}}+\frac{\partial \widetilde{F}}{\partial t}\right)\right|_{\V}}\\
{}&{}\\
{}&=\D{\left.\left(\dot{x}^{\hat{\imath}}\frac{\partial \widetilde{F}(x,0,t)}{\partial x^{\hat{\imath}}}+\frac{\partial \widetilde{F}(x,0,t)}{\partial t}\right)\right|_{\V}}\\
{}&{}\\
{}&=\D{\frac{{\rm d}~}{{\rm d}t}\widetilde{F}(x,0,t)}\\
{}&{}\\
{}&=\D{\frac{{\rm d}~}{{\rm d}t}\Big(\left.F(p,q,t)\right|_{\V}\Big)}\,.
\ea
\label{VddVproof}
\ee
}

\item Eq.(\ref{tiP}).

\begin{proposition}The Poisson bracket is independent of time,
\be
\D{\left[~~~,~~~\right\}_{P.B.}=\left.\left[~~~,~~~\right\}_{P.B.}\right|_{t_{0}}}\,,
\ee
or
\be
\D{(-1)^{\#_{A}}\frac{\overleftarrow{\partial} ~~~}{\partial q^{A}}\,\frac{\overrightarrow{\partial}~~~ }{\partial p_{A}}~-~\frac{\overleftarrow{\partial} ~~~}{\partial p_{A}}\,\frac{\overrightarrow{\partial}~~~ }{\partial q^{A}}=
(-1)^{\#_{A}}\frac{\overleftarrow{\partial} ~~~}{\partial q_{0}^{A}}\,\frac{\overrightarrow{\partial}~~~ }{\partial p_{0A}}~-~\frac{\overleftarrow{\partial} ~~~}{\partial p_{0A}}\,\frac{\overrightarrow{\partial}~~~ }{\partial q_{0}^{A}}\,.}
\label{tiPproof}
\ee
\end{proposition}
Roughly speaking, the Poisson bracket is independent of time, \textit{i.e.} preserved, because time evolution is a symplectic transformation generated by the Hamiltonian.

\proof{
We show the proposition for a set $\{\free^{i}(t)\}$ of local functions which are piecewise infinitely differentiable. Then the time independence holds globally,  since $(p,q)$ are globally continuous.  A direct manipulation gives
\be
\ba{ll}
\D{\left.\left[~~~,~~~\right\}_{P.B.}\right|_{t_{0}}}
&\D{=
(-1)^{\#_{A}}\frac{\overleftarrow{\partial} ~~~}{\partial q_{0}^{A}}\,\frac{\overrightarrow{\partial}~~~ }{\partial p_{0A}}~-~\frac{\overleftarrow{\partial} ~~~}{\partial p_{0A}}\,\frac{\overrightarrow{\partial}~~~ }{\partial q_{0}^{A}}}\\
{}&{}\\
{}&=\D{\frac{\overleftarrow{\partial} ~~~}{\partial q^{A}}\left.\left[q^{A},q^{B}\right\}_{P.B.}\right|_{t_{0}}\frac{\overrightarrow{\partial} ~~~}{\partial q^{B}}+
\frac{\overleftarrow{\partial} ~~~}{\partial q^{A}}\left.\left[q^{A},p_{B}\right\}_{P.B.}\right|_{t_{0}}\frac{\overrightarrow{\partial} ~~~}{\partial p_{B}}}\\
{}&{}\\
{}&~~~\D{\,+\frac{\overleftarrow{\partial} ~~~}{\partial p_{A}}\left.\left[p_{A},q^{B}\right\}_{P.B.}\right|_{t_{0}}\frac{\overrightarrow{\partial} ~~~}{\partial q^{B}}+
\frac{\overleftarrow{\partial} ~~~}{\partial p_{A}}\left.\left[p_{A},p_{B}\right\}_{P.B.}\right|_{t_{0}}\frac{\overrightarrow{\partial} ~~~}{\partial p_{B}}}\,.
\ea
\ee
Obviously, the equality we want to show holds for the zeroth order in $t-t_{0}$. Now we suppose that it holds up to the order $(t-t_{0})^{k}$, so that up to the power $k$,
\be
\ba{ll}
\D{\left.\left[q^{A},q^{B}\right\}_{P.B.}\right|_{t_{0}}\simeq 0}\,,~~~~&~~~~
\D{\left.\left[q^{A},p_{B}\right\}_{P.B.}\right|_{t_{0}}\simeq (-1)^{\#_{A}}\delta^{A}{}_{B}}\,,\\
{}&{}\\
\D{\left.\left[p_{A},q^{B}\right\}_{P.B.}\right|_{t_{0}}\simeq -\delta_{A}{}^{B}}\,,~~~~&~~~~
\D{\left.\left[p_{A},p_{B}\right\}_{P.B.}\right|_{t_{0}}\simeq 0}\,.
\ea
\label{korder}
\ee
Also for two generic functions, $F(p,q)$ and $G(p,q)$, which do not have explicit time dependence, we get up to the power, $k$,
\be
\ba{ll}
\D{\frac{{\rm d}~}{{\rm d}t}\Big(\left.[F,G\}_{P.B.}\right|_{t_{0}}\Big)}&=\D{
\left.\left[\frac{{\rm d}F}{{\rm d}t},G\right\}_{P.B.}\right|_{t_{0}}+
\left.\left[F,\frac{{\rm d}G}{{\rm d}t}\right\}_{P.B.}\right|_{t_{0}}}\\
{}&{}\\
{}&\simeq\D{\left[\frac{{\rm d}F}{{\rm d}t},G\right\}_{P.B.}+\left[F,\frac{{\rm d}G}{{\rm d}t}\right\}_{P.B.}}\\
{}&{}\\
{}&=\D{\Big[\left[F,H_{T}\right\}_{P.B.},G\Big\}_{P.B.}+\Big[F,\left[G,H_{T}\right\}_{P.B.}\Big\}_{P.B.}}\\
{}&{}\\
{}&=\D{-\Big[H_{T},\left[F,G\right\}_{P.B.}\Big\}_{P.B.}}\,.
\ea
\ee
This shows that the relations, (\ref{korder}), actually hold up to the power, $k+1$, completing the proof.}

\item Eq.(\ref{dpqm})

{The first equality in (\ref{dpqm}) follows from the algebraic identity (\ref{KL1}) on the tangent space variables $(q,\dq,t)\,$. Firstly, one considers (\ref{KKL1}) for $A=\hat{m}\,$,
\be
\D{\frac{\partial (\delta q^{B})}{\partial \dq^{\hat{m}}}\frac{\partial L}{\partial \dq^{B}}=\frac{\partial (\delta K)}{\partial \dq^{\hat{m}}}}\,,
\label{KKKL1}
\ee
Secondly, one takes the partial derivative of each side of (\ref{KKKL1}) with respect to $q^A\,$,
\be
\D{\frac{\partial^2 (\delta q^{B})}{\partial q^{A}\,\partial \dq^{\hat{m}}}\,\frac{\partial L}{\partial \dq^{B}}+(-1)^{\#_{A}(\#_{B}+\#_{\hat{m}})}\,\frac{\partial (\delta q^{B})}{\partial \dq^{\hat{m}}}\,\frac{\partial^2 L}{\partial q^A\,\partial \dq^{B}}=\frac{\partial^2 (\delta K)}{\partial q^{A}\,\partial \dq^{\hat{m}}}}\,.
\label{COMP}
\ee
Thirdly, the partial derivative of each side of (\ref{Deltap}) reads explicitly as
\be
\frac{\partial\,\Deltap_{A}}{\partial\dq^{\hat{m}}}\,=\,
\frac{\partial^2\,(\delta K)}{\partial\dq^{\hat{m}}\,\partial q^{A}}\,-\,\frac{\partial^2\,(\delta q^{B})}{\partial\dq^{\hat{m}}\,\partial q^{A}}\,\widehat{p}_B\,,
\label{Deltapp}
\ee
where we made use of (\ref{widehatp}). Fourthly, making use of (\ref{COMP}) in (\ref{Deltapp}) leads to (\ref{dpqm}).}\\

The second equality in (\ref{dpqm}) holds {because of the integrability condition (\ref{ddqL}). Proceeding step by step, one may start by using the chain rule in order to show the identity
\be
(-1)^{\#_{A}\#_{B}}
\frac{\partial(\Deltaq^{B})}{\partial\dq^{\hat m}}\frac{\partial h^{\hat a}}{\partial q^{A}}
\frac{\partial^2 L}{\partial \dq^{\hat a}\partial\dq^{B}}
=(-1)^{(\#_{A}+\#_{\hat a})\#_{\hat m}}\frac{\partial h^{\hat a}}{\partial q^{A}}
\left(\frac{\partial(\delta q^{B})}{\partial\dq^{\hat m}}+
\frac{\partial h^{\hat b}}{\partial\dq^{\hat m}}\frac{\partial(\delta q^{B})}{\partial \dq^{\hat b}}\right)
\frac{\partial^{2}L}{\partial\dq^{B}\partial \dq^{\hat a}} \,.
\label{maexch}
\ee
Then, the relation (\ref{ddqL}) is used to exchange some indices in Eq.(\ref{maexch}) as follows
\begin{eqnarray}
&&(-1)^{(\#_{A}+\#_{\hat a})\#_{\hat m}}\left(\frac{\partial(\delta q^{B})}{\partial\dq^{\hat m}}+
\frac{\partial h^{\hat b}}{\partial\dq^{\hat m}}\frac{\partial(\delta q^{B})}{\partial \dq^{\hat b}}\right)
\frac{\partial^{2}L}{\partial\dq^{B}\partial \dq^{\hat a}}\nonumber\\
&&=
(-1)^{(\#_{A}+\#_{\hat B})\#_{\hat m}}\frac{\partial(\delta q^{B})}{\partial \dq^{\hat a}}\left(
\frac{\partial^{2}L}{\partial\dq^{\hat m}\partial\dq^{B}}
+\frac{\partial h^{\hat b}}{\partial\dq^{\hat m}}\frac{\partial^{2}L}{\partial \dq^{\hat b}\partial\dq^{B}}\right)\,.
\label{exchind}
\end{eqnarray}
Finally, one observes that the sum of terms in the parenthesis of Eq.(\ref{exchind}) vanishes since
\be
\frac{\partial^{2}L}{\partial\dq^{\hat m}\partial\dq^{B}}
+\frac{\partial h^{\hat b}}{\partial\dq^{\hat m}}\frac{\partial^{2}L}{\partial \dq^{\hat b}\partial\dq^{B}}=
\frac{\partial \widehat{p}_{B}(q,p_{b},t)}{\partial\dq^{\hat m}}=0\,,
\label{ouf}
\ee
due to (\ref{widehatp}).
The set of Eqs.(\ref{maexch})-(\ref{ouf}) implies that
\be
\left(
\frac{\partial^{2}L(q,\dq,t)}{\partial q^{A}\,\partial\dq^{B}}\right)_{\dq^{\hat a}\,=\, h^{\hat{a}}(q,\,p_{a},\,\dq^{\hat m},\,t)}
\,=\,\frac{\partial \widehat{p}_B(q,\,p_{a},\,\dq^{\hat m},\,t)}{\partial q^{A}}
\ee
which ends the proof of Eq.(\ref{dpqm}) due to (\ref{primary1}).
}

\end{itemize}

\newpage
\section{Grassmann algebra}\label{subsectionGrassmann}

In principle, in order to be able to discuss  rigorously a generic dynamical system (\textit{i.e.} which contains both bosons and fermions), one needs to introduce the ``Grassmann algebra'' $\Lambda_{\hat{N}}$ which is generated by the $\hat{N}$ anti-commuting Grassmann variables~\cite{Buchbinder:1995uq},
\be
\ba{ll}
\zeta^{\alpha}\zeta^{\beta}+\zeta^{\beta}\zeta^{\alpha}=0\,,~~~~&~~~~\alpha,\beta=1,2,\cdots, \hat{N}\,.
\ea
\ee
They generate the  following basis for the Grassmann algebra
\be
\ba{l}
{}~~~~~~\left.~~~~~~1~~~~~~~~~~~~~~~~~~~~~~~\right\}~\mbox{``body''}\\
{}\\
{}\\
~~~~~~\left.
\ba{l}
~~~~~~\zeta^{\alpha}\\
{}\\
~~~~~~\zeta^{\alpha}\zeta^{\beta},~~~~~\alpha<\beta\\
~~~~~~~\cdot\\
~~~~~~~\cdot\\
~~~~~~~\cdot\\
~~~~~~\zeta^{1}\zeta^{2}\cdots\zeta^{\hat{N}}~~~~~~~~~\ea
\right\}~\mbox{``soul''}
\ea
\label{basisL}
\ee
which has the dimension, $\mbox{dim}(\Lambda_{\hat{N}})=2^{\hat{N}}$ while $\hat{N}$ can be infinity .\\

Any quantity in the Grassmann  algebra, $\Lambda_{\hat{N}}$, can be expressed as an expansion in terms of the above basis over the field $\mathbb{R}$ of real numbers field (or the field $\mathbb{C}$ of complex numbers),
\be
\D{x=x_{0}+\sum_{n=1}^{\hat{N}}\,\frac{1}{n!}\,x_{\alpha_{1}\alpha_{2}\cdots \alpha_{n}}\zeta^{\alpha_{1}}\zeta^{\alpha_{2}}\cdots\zeta^{\alpha_{n}}\,,}
\ee
where $x_{0}$ and $x_{\alpha_{1}\alpha_{2}\cdots \alpha_{n}}$ are real (or complex numbers) carrying totally anti-symmetric indices. Naturally, the bosons allow the expansion of even $n$'s only, while fermions allow only odd $n$'s. \\

\textit{It is crucial to note that if and only if $x_{0}\neq 0$, the inverse, $x^{-1}$, exists.}\\

It is convenient to  rename the elements in the basis with a given ordering as
\be
\D{\left\{\,\Z^{J},~1\leq J\leq 2^{\hat{N}}\,\right\}=\left\{\,1,~\zeta^{\alpha},~\cdots\,,~{\zeta^{1}\zeta^{2}\cdots\zeta^{\hat{N}}} \,\right\}\,,}
\ee
and to write
\be
\D{x=\sum_{J=1}^{2^{\hat{N}}}\,x_{J}\Z^{J}\,.}
\ee
We also introduce the following notation to pick up the real or complex number coefficient,
\be
\D{\left[x\right]_{J}=x_{J}\,.}
\ee
~\\

In terms of the Grassmann algebra, the Lagrangian is a bosonic variable, but not necessarily a pure `body'. Furthermore  the actual dynamical variables are the real numbers, $\left[ q^{A}\right]_{J}$, leading to a much bigger phase space.\\

In deriving the equations of motion for a Lagrangian,  what we  actually encounter is the expression,
\be
\int {\rm d}t~\delta\L(q_{n},t)=\int {\rm d}t~\delta\!\left[ q^{A}\right]_{\!J}\Z^{J}
\,\frac{\delta}{\delta q^A}\L(q_{m},t)\,.
\label{deltaL2}
\ee
This implies that for the bosons, by considering especially the variations of the pure body, the equations of motion can be indeed collectively expressed as usual (\ref{EOML}), but the equations of motion for the fermions should be refined to hold in a weaker form,
\be
\ba{ll}
\D{\left[\frac{\delta\L}{\delta q^A}(q_{m},t)\right]_{J}\equiv 0}\,,~~~~~&~~~~~J\neq 2^{\hat{N}}\,,
\ea
\ee
when $\hat{N}$ is odd. Namely for the fermions,  the usual equation of motion is true except the highest order in `soul'.  However this subtle issue can be neglected either by imposing the missing equation for  $J= 2^{\hat{N}}$ by hand, or by letting $\hat{N}\rightarrow\infty$.\\

The purpose of the present subsection was to provide a rigorous way to analyze the dynamical systems containing both bosons and fermions. Nevertheless,  in practice we will favor  the  Lagrangian  systems which do not require any explicit use of the basis for the Grassmann algebra, especially when they are transformed into Hamiltonian form for constrained systems.


\newpage
\section{Basics on supermatrices}\label{supermatrix}

A generic   $(n_{1}+n_{2})\times (m_{1}+m_{2})$  supermatrix, $M$,  over a Grassmann algebra, $\Lambda_{\hat{N}}$,\\ (see Section \ref{subsectionGrassmann}),   is of the form,
\be
M=\left(\ba{cc}
\,A_{n_{1}\times m_{1}}\,&\,\Psi_{n_{1}\times m_{2}}\,\\
\,\Theta_{n_{2}\times m_{1}}\,&\,B_{n_{2}\times m_{2}}\,\ea\right)\,,
\ee
where $A$, $B$ are bosonic and $\Psi$, $\Theta$ are fermionic.  \\

The complex conjugation, transpose, and the Hermitian conjugation read respectively \cite{Buchbinder:1995uq},
\be
\ba{ll}
M^{\ast}=\left(\ba{cc}\,A\,&\,\Psi\,\\
\,\Theta\,&\,B\,\ea\right)^{\ast}=\left(\ba{cc}\,A^{\ast}\,&\,-\Psi^{\ast}\,\\
\,\Theta^{\ast}\,&\,B^{\ast}\,\ea\right)~~&\mbox{or}~~~~
(M^{\ast})_{ab}=(-1)^{\#{b}(\#a\,+\,\#{b})}(M_{ab})^{\ast}\,,\\
{}&{}\\
M^{T}=\left(\ba{cc}\,A\,&\,\Psi\,\\
\,\Theta\,&\,B\,\ea\right)^{T}=\left(\ba{cc}\,A^{T}\,&\,\Theta^{T}\,\\
\,-\Psi^{T}\,&\,B^{T}\,\ea\right)~~&\mbox{or}~~~~
(M^{T})_{ab}=(-1)^{\#{a}(\#a\,+\,\#{b})}M_{ba}\,,\\
{}&{}\\
M^{\dagger}=(M^{\ast})^{T}=\left(\ba{cc}\,A\,&\,\Psi\,\\
\,\Theta\,&\,B\,\ea\right)^{\dagger}=
\left(\ba{cc}\,A^{\dagger}\,&\,\Theta^{\dagger}\,\\
\,\Psi^{\dagger}\,&\,B^{\dagger}\,\ea\right)~~&\mbox{or}~~~~
(M^{\dagger})_{ab}=(M_{ba})^{\ast}\,.
\ea
\label{Transpose}
\ee
Note that
\be
\ba{lll}
(M^{\ast})^{\ast}=M\,,~~~&~~~(M^{\dagger})^{\dagger}=M\,,~~~&~~~
(M^{T})^{\dagger}=M^{\ast}\,,\\
{}&{}&{}\\
(M_{1}M_{2})^{\ast}=M_{1}^{\ast}M_{2}^{\ast}\,,~~~~&~~~
(M_{1}M_{2})^{T}=M_{2}^{T}M_{1}^{T}\,,~~~~&~~~
(M_{1}M_{2})^{\dagger}=M_{2}^{\dagger}M_{1}^{\dagger}\,.
\ea
\ee
However,
\be
\ba{lll}
(M^{T})^{T}\neq M\,,~~~~&~~~(M^{\ast})^{T}\neq (M^{T})^{\ast}\,,~~~~&~~~(M^{\dagger})^{T}\neq M^{\ast}\,,~~~~~~~etc.
\ea
\ee
~\\
In particular, a {real supermatrix} is of the generic form,
\be
\ba{ll}
M=M^{\ast}=\left(\ba{cc}
\,A\,&\,i\Psi\,\\
\,\Theta\,&\,B\,\ea\right)\,,~~~~&~~~~M^{\dagger}=M^{T}\,,
\ea
\label{Reals}
\ee
where every variable is real, $A=A^{\ast}$, $B=B^{\ast}$, $\Psi=\Psi^{\ast}$, $\Theta=\Theta^{\ast}$.\newpage

For the  $(n_{1}+n_{2})\times (n_{1}+n_{2})$ square supermatrix, $M$,
\be
M=\left(\ba{cc}
\,A_{n_{1}\times n_{1}}\,&\,\Psi_{n_{1}\times n_{2}}\,\\
\,\Theta_{n_{2}\times n_{1}}\,&\,B_{n_{2}\times n_{2}}\,\ea\right)\,,
\ee
the inverse  can be expressed as
\begin{equation}
M^{-1}=\left(\begin{array}{cc}
(A-\Psi B^{-1}\Theta)^{-1}&-A^{-1}\Psi(B-\Theta A^{-1}\Psi)^{-1}\\
-B^{-1}\Theta(A-\Psi B^{-1}\Theta)^{-1}&(B-\Theta A^{-1}\Psi)^{-1}
\end{array}\right)\,,\label{M-1}
\end{equation}
where we may write
\begin{equation}
(A-\Psi B^{-1}\Theta)^{-1}=A^{-1}+\sum_{p=1}^{\infty}\,(A^{-1}\Psi B^{-1}\Theta)^{p}A^{-1}\,.
\end{equation}
Note that due to the fermionic property of $\Psi,\Theta$, the power series
terminates at $p\leq n_{1}n_{2}+1$. \newline

The supertrace and the superdeterminant of $M$ are defined as \cite{Buchbinder:1995uq}\footnote{The last equality comes from $
\det(1-A^{-1}\Psi B^{-1}\Theta)=\mbox{det}{}^{-1} (1-B^{-1}\Theta A^{-1}\Psi),$
~which can be shown\\ using
$\det (1-X)=\mbox{exp}\left(
-\sum_{p=1}^{\infty}\,
\frac{1}{p}\mbox{tr}\,X^{p}\right),$
and observing
$
\mbox{tr}\,(A^{-1}\Psi B^{-1}\Theta)^{p}=-\mbox{tr}\,(B^{-1}\Theta A^{-1}\Psi)^{p}\,.
$}
\begin{eqnarray}
&&\mbox{str}\,M=\mbox{tr}\,A-\mbox{tr}\,B\,,\\
{}\nonumber\\
&&\mbox{sdet}\,M=\det(A-\Psi B^{-1}\Theta)/\det B=\det A/\det (B-\Theta A^{-1}\Psi)\,.
\label{sdetdef}
\end{eqnarray}
From Eq.(\ref{M-1}),
$$\mbox{sdet}\,M\neq 0\Longleftrightarrow\det A\det B\neq 0$$ is the necessary and sufficient condition for  the existence of $M^{-1}$.   \\
~\\

The supertrace and the superdeterminant have the properties,
\be
\ba{ll}
\mbox{str}\,(M_{1}M_{2})=\mbox{str}\,(M_{2}M_{1})\,,~~~~&~~~~
\mbox{sdet}\,(M_{1}M_{2})=\mbox{sdet}\,M_{1}\,\mbox{sdet}\,M_{2}\,.
\ea
\ee
~\\

For a generic $n\times n$ bosonic  matrix, $A$, in a similar fashion to above, decomposing it
into the `body' and `soul' (see Section \ref{subsectionGrassmann}),
\be
A=A_{\rm{\scriptstyle{body}}}+A_{\rm{\scriptstyle{soul}}}\,,
\ee
we have
\begin{equation}
A^{-1}=A_{\rm{\scriptstyle{body}}}^{-1}+\sum_{p=1}^{n^{2}+1}\,
(-A_{\rm{\scriptstyle{body}}}^{-1}A_{\rm{\scriptstyle{soul}}})^{p}
A_{\rm{\scriptstyle{body}}}^{-1}\,.
\label{inverseA}
\end{equation}
\textit{ Thus, $A^{-1}$ exists if and only if $A_{\rm{\scriptstyle{body}}}^{-1}$  exists.}

\newpage
\section{Lemmas on the canonical transformations of supermatrices}\label{lemmasupermatrix}

In this appendix, we do not explicitly state which entries in the supermatrices which are Grassmann even or odd. The way the supermatrices have been written is supposed to be self-explanatory.\\\\

\noindent{\bf Fact 1.}\\
For any $n\times m$ bosonic matrix $M$ over a Grassmann algebra $\Lambda_{\hat{N}}$
(see Section \ref{subsectionGrassmann}),
\be
\ba{ll}
M=\Big(v_{1}\,v_{2}\,\cdots\,v_{m}\Big)\,,~~~~&~~~~
v_{j}=(v_{1j}\,\,v_{2j}\,\,\cdots\,\,v_{nj})^{T}\,,
\ea
\ee
there exists an $m\times m$ nondegenerate  matrix $Q$ satisfying
\be
MQ=\Big(w_{1}\,\,w_{2}\,\,\cdots\,\,w_{k}\,\,s_{1}\,\,s_{2}\,\,\cdots\,\,s_{m-k}\Big)\,,
\ee
where all vector (\textit{i.e.} columns) $w_{i}$ are orthogonal to each other and each body of them is nonzero, while the $s_{j}$'s are pure souls or zero.

\proof{To show this, one needs to separate the vectors $v_{i}$ into two groups: pure soul ones and other ones with nontrivial bodies. Then one only needs to orthogonalize\footnote{Note that, since the body is nonzero, the inverse of the scalar product $(w^{\dagger}_{j}w_{j})^{-1}$ exists, and the orthogonalization can be done by a generalization of the Gram-Schmidt procedure for the graded case.} the latter.}

~\\
~\\
\noindent{\bf Fact 2.}\\
For any $n\times m$  bosonic matrix $M$ over  a Grassmann algebra $\Lambda_{\hat{N}}$, there exists two nondegenerate  matrices, $P$ and  $P^{\prime}$, which transform $M$ into the canonical form,
\be
\ba{ll}
P^{\prime}MP=\left(\ba{ll}\,~\Lambda_{k\times k}\,&\,0_{k\times (m-k)}\,\\
\,0_{(n-k)\times k}\,&\,s_{(n-k)\times(m-k)}\,\ea\right)\,,
\ea\label{fact2}
\ee
where $\Lambda_{k\times k}$ is a $k\times k$ nondegenerate  diagonal matrix,
\be
\ba{ll}
\Lambda_{k\times k}=\mbox{diag}(\lambda_{1},\lambda_{2},\cdots,\lambda_{k})\,,
~~~&~~~(\lambda_{i})_{\rm{\scriptstyle{body}}}\neq 0\,,~~~1\leq i\leq k\,,
\ea
\ee
of rank $k$ so that $k\leq m$ and $k\leq n\,$, and
 the $(n-k)\times(m-k)$ matrix $s_{(n-k)\times(m-k)}$ is a pure soul.
\proof{From \textit{Fact} 1, we construct $P^{\prime}$ out of the orthogonal vectors
$\{w_{1}^{\dagger},w_{2}^{\dagger},\cdots,w_{k}^{\dagger}\}$ and their complementary vectors to get
\be
P^{\prime}MP=\left(\ba{ll}\,\Lambda\,&\,s^{\prime}\,\\
\,0\,&\,s\,\ea\right)\,.
\ee
Now we only need to take one more step for the final result,
\be
\left(\ba{ll}\Lambda\,&\,s^{\prime}\,\\
0\,&\,s\,\ea\right)
\left(\ba{cc}\,1\,&\,-\Lambda^{-1}s^{\prime}\,\\
0\,&\,1\,\ea\right)=\left(\ba{ll}\Lambda\,&\,0\,\\
0\,&\,s\,\ea\right)\,.
\ee
}

~\\
~\\
\noindent{\bf Fact 3. - \textit{Corollary}} \\
For any nondegenerate $n\times n$  bosonic matrix $M$ over a Grassmann algebra $\Lambda_{\hat{N}}$, meaning
\be
\det M_{\rm{\scriptstyle{body}}}\neq 0\,,
\ee
there exist two nondegenerate matrices $P$ and  $P^{\prime}$ which transforms $M$ into the identity,
\be
P^{\prime}MP=1\,.
\ee
\proof{The proof is straightforward from \textit{Fact} 2 and its \textit{Proof}, since the matrix is non-degenerate and $n=m=k$.}

~\\
~\\
\noindent{\bf Fact 4.}\\
For any nondegenerate $(n_{1}+n_{2})\times (n_{1}+n_{2})$  square supermatrix $M$ over a Grassmann algebra, $\Lambda_{\hat{N}}$,
\be
\ba{ll}
M=\left(\ba{cc}A\,&\,\psi\\
\chi\,&\,B\ea\right)\,,~~~~&~~~~\det A\det B\neq 0\,,
\ea
\ee
there exists two nondegenerate supermatrices $P$ and  $P^{\prime}$ satisfying
\be
P^{\prime}MP=1\,.
\ee
\proof{After the transformation,
\be
\left(\ba{cc}1\,&\,0\\
-\chi A^{-1}\,&\,1\ea\right)
\left(\ba{cc}A\,&\,\psi\\
\chi\,&\,B\ea\right)
\left(\ba{cc}1\,&\,-A^{-1}\psi\\
0\,&\,1\ea\right)=\left(\ba{cc}A\,&\,0\\
0\,&\,B-\chi A^{-1}\psi\ea\right)\,,
\ee
we only need to apply \textit{Fact} 3.}

~\\
~\\
\noindent{\bf Fact 5.} \\
For any nondegenerate, bosonic, real, symmetric or anti-symmetric matrix $A_{\pm}$ over a Grassmann algebra $\Lambda_{\hat{N}}$,
\be
\ba{ll}
A_{\pm}=A_{0}+A_{\rm{\scriptstyle{soul}}}\,,~~~&~~~~
A^{\ast}_{\pm}=A_{\pm}\,,\\
{}&{}\\
(\det A_{\pm})_{\rm{\scriptstyle{body}}}=(\det A_{0})_{\rm{\scriptstyle{body}}}\neq 0\,,~~~&~~~~A_{\pm}^{T}=\pm A_{\pm}\,,
\ea
\ee
there exists a  nondegenerate real matrix $P$, the transformation induced by which, removes  the  pure soul $A_{\rm{\scriptstyle{soul}}}$  completely,
\be
PA_{\pm}P^{T}=A_{0}\,.
\label{transfPP}
\ee
{\bf Remark:} The point of \textit{Fact} 5 is the removal or addition of any soul to the original matrix, $A_{\pm}$, via the insertion between two appropriately chosen real matrices (\ref{transfPP}).\\

\proof{We present explicitly the real transformation,
\be
P=P^{\ast}=1+\D{\sum_{n=1}^{\infty}\,a_{n}
\big(A_{\rm{\scriptstyle{soul}}}A_{0}^{-1}\big)^{n}}\,,
\label{Pexp}
\ee
where the coefficients are  given by a recurrence relation~\cite{Park:2003ku}, with  $a_{0}=0$, $a_{1}=-\frac{1}{2}$,
\begin{equation}
\ba{ll}
a_{n+1}=
-a_{n}-\textstyle{\frac{1}{2}}\,\D{\sum_{j=1}^{n}\,a_{j}(a_{n+1-j}+a_{n-j})\,,}~~&~~\mbox{for~\,} n\geq 1\,.\ea
\end{equation}

Due to the Grassmannian property, the sum in (\ref{Pexp}) terminates at a finite order.}
~~\\
~~\\
\noindent{\bf Fact 6.}\\
For any bosonic, real, symmetric or anti-symmetric matrix, $A_{\pm}$, over a Grassmann algebra, $\Lambda_{\hat{N}}$,
\be
\ba{ll}
A_{\pm}=A_{\pm}^{\ast}\,,~~~&~~~~A_{\pm}^{T}=\pm A_{\pm}\,,
\ea
\ee
there exists a  nondegenerate real matrix, $P$,  which transforms $A$ into the canonical form,
\be
\ba{ll}
PA_{\pm}P^{T}=\left(\ba{ll}\,b_{\pm}\,&\,0\,\\
\,0\,&\,s_{\pm}\,\ea\right)\,,~~~~&~~~~P=P^{\ast}\,,
\ea
\ee
where the bosonic matrix, $b_{\pm}=\pm b_{\pm}^{T}$, is nondegenerate,
$(\det b_{\pm})_{\rm{\scriptstyle{body}}}\neq 0$,  while $s=\pm s^{T}$ is a pure soul or zero.

\proof{First, using a real orthonormal matrix $O$, one can  transform the `body' of $A$ into the canonical form $\left(\ba{ll}\,b\,&\,0\,\\
\,0\,&\,0\,\ea\right)$, which gives
\be
OA_{\pm}O^{T}=\left(\ba{cc}
b+s_{1}&s_{2}\\
\pm s_{2}^{T}&s_{3}\ea\right)\,,
\ee
where $b=\pm b^{T}$ is nondegenerate and real, while $s_{1}=\pm s_{1}^{T}, s_{2}, s_{3}=\pm s_{3}^{T}$ are all real and pure souls. We further transform it as
\begin{eqnarray}
&&\left(\ba{cc}1&0\\\mp s_{2}^{T}(b+s_{1})^{-1}&1\ea\right)
\left(\ba{cc}
b+s_{1}&s_{2}\\
\pm s_{2}^{T}&s_{3}\ea\right)
\left(\ba{cc}1&-(b+s_{1})^{-1}s_{2}\\0&1\ea\right)\nonumber\\
&&=
\left(\ba{cc}
b+s_{1}&0\\
0&s_{3}\mp s_{2}^{T}(b+s_{1})^{-1}s_{2}\ea\right)\,.
\end{eqnarray}
To complete the proof, we only need to apply \textit{Fact} 5 to $b+s_{1}$.}

~\\
~\\
\noindent{\bf Fact 7.}\\
For a generic  $(n_{1}+n_{2})\times (m_{1}+m_{2})$ supermatrix $M$ over a Grassmann algebra $\Lambda_{\hat{N}}$,
\be
M=\left(\ba{cc}A_{n_{1}\times m_{1}}\,&\,\Psi_{n_{1}\times m_{2}}\\
\Theta_{n_{2}\times m_{1}}\,&\,B_{n_{2}\times m_{2}}\ea\right)\,,
\ee
there exists two nondegenerate supermatrices $P$ and  $P^{\prime}$ which transforms it into the canonical form,
\be
P^{\prime}MP=
\left(
\ba{cccc}
\,1\,&\,0\,&\,0\,&\,0\,\\
\,0\,&\,s_{1}\,&\,0\,&\,\psi\,\\
\,0\,&\,0\,&\,1\,&\,0\,\\
\,0\,&\,\chi\,&\,0\,&s_{2}\,
\ea
\right)\,,
\label{Lemma3}
\ee
where $s_{1}$, $s_{2}$ are bosonic pure soul, and $\psi$, $\chi$ are fermionic.  The partition of the canonical form reads,
\be
{\left[k_{1}+(n_{1}-k_{1})+k_{2}+(n_{2}-k_{2})\right]\times
\left[k_{1}+(m_{1}-k_{1})+k_{2}+(m_{2}-k_{2})\right]}\,,
\ee
where $k_{1}$, $k_{2}$ are respectively the ranks  of the bosonic matrices $A$, $B$.
\proof{
From \textit{Fact} 2, we can first transform the $A_{n_{1}\times m_{1}}$  into the canonical form, in order to put $M$ into the form
\be
\left(
\ba{lll}
\,1\,&\,0\,&\,\psi_{1}\,\\
\,0\,&\,s_{1}\,&\,\psi_{2}\,\\
\,\chi_{1}\,&\,\chi_{2}\,&\,B\,
\ea
\right)\,,
\ee
and to further have
\be
\left(
\ba{ccc}
\,1\,&\,0\,&\,0\,\\
\,0\,&\,1\,&\,0\,\\
\,-\chi_{1}\,&\,0\,&\,1\,
\ea
\right)\left(
\ba{lll}
\,1\,&\,0\,&\,\psi_{1}\,\\
\,0\,&\,s_{1}\,&\,\psi_{2}\,\\
\,\chi_{1}\,&\,\chi_{2}\,&\,B\,
\ea
\right)\left(
\ba{ccc}
\,1\,&\,0\,&\,-\psi_{1}\,\\
\,0\,&\,1\,&\,0\,\\
\,0\,&\,0\,&\,1\,
\ea
\right)=
\left(
\ba{llc}
\,1\,&\,0\,&\,0\,\\
\,0\,&\,s_{1}\,&\,\psi_{2}\,\\
\,0\,&\,\chi_{2}\,&\,~B-\chi_{1}\psi_{1}\,
\ea
\right)\,.
\ee
Now we apply \textit{Fact} 3 to   $(B-\chi_{1}\psi_{1})$ to get
\be
\left(
\ba{cccc}
\,1\,&\,0\,&\,0\,&\,0\,\\
\,0\,&\,s_{1}\,&\,\psi_{3}\,&\,\psi_{4}\,\\
\,0\,&\,\chi_{3}\,&\,1\,&\,0\,\\
\,0\,&\,\chi_{4}\,&\,0\,&s_{2}\,
\ea
\right)\,.
\ee
Finally, one completes the proof by the equality,
\be
\ba{l}
\left(
\ba{cccc}
\,1\,&\,0\,&\,0\,&\,0\,\\
\,0\,&\,1\,&\,-\psi_{3}\,&\,0\,\\
\,0\,&\,0\,&\,1\,&\,0\,\\
\,0\,&\,0\,&\,0\,&1\,
\ea
\right)
\left(
\ba{cccc}
\,1\,&\,0\,&\,0\,&\,0\,\\
\,0\,&\,s_{1}\,&\,\psi_{3}\,&\,\psi_{4}\,\\
\,0\,&\,\chi_{3}\,&\,1\,&\,0\,\\
\,0\,&\,\chi_{4}\,&\,0\,&s_{2}\,
\ea
\right)\left(
\ba{cccc}
\,1\,&\,0\,&\,0\,&\,0\,\\
\,0\,&\,1\,&\,0\,&\,0\,\\
\,0\,&\,-\chi_{3}\,&\,1\,&\,0\,\\
\,0\,&\,0\,&\,0\,&1\,
\ea
\right)\\
{}\\
{}\\
=
\left(
\ba{cccc}
\,1\,&\,0\,&\,0\,&\,0\,\\
\,0\,&\,s_{1}-\psi_{3}\chi_{3}\,&\,0\,&\,\psi_{4}\,\\
\,0\,&\,0\,&\,1\,&\,0\,\\
\,0\,&\,\chi_{4}\,&\,0\,&s_{2}\,
\ea
\right)\,.
\ea
\ee
}
{\bf Remark:} Note also that \textit{Fact} 3 follows as a corollary too.
{}\\
{}\\
\noindent{\bf Fact 8.} \\
Consider a $(n_{-}+n_{+})\times(n_{-}+n_{+})$ anti-Hermitian supermatrix,
over a Grassmann algebra $\Lambda_{\hat{N}}$, which has the symmetry property, $\Omega_{ab}=-(-1)^{\#_{a}\#_{b}}\Omega_{ba}$, or equivalently
\be
\ba{ll}
\Omega=-\Omega^{\dagger}\,,~~~~&~~~~\Omega^{T}=\left(\ba{cc}-1&0\\0&1\ea\right)\Omega\,.
\label{propO}
\ea
\ee
It is of the general form,
\be
\Omega=\left(\ba{cc}\,A_{-}\,&\,\Psi\,\\-\Psi^{T}\,&\,iA_{+}\,\ea\right)\,,
\ee
where every variable is real, $A_{\pm}=A_{\pm}^{\ast}$, $\Psi=\Psi^{\ast}$, and $A_{\pm}=\pm A_{\pm}^{T}$.\\

We note that the anti-Hermiticity and symmetry properties (\ref{propO}) are preserved under the transformations by a real supermatrix, (\ref{Reals}),
\be
\ba{ll}
\Omega\Longrightarrow L\Omega L^{T}\,,~~~~&~~~~L=L^{\ast}\,,
\ea
\ee
since
\be L^{\dagger}=(L^{\ast})^{T}\,,\quad (L^{T})^{\dagger}=L^{\ast}\,, \quad(L^{T})^{T}=
{\left(\ba{cc}-1&0\\0&1\ea\right)}L{\left(\ba{cc}-1&0\\0&1\ea\right)}\,.
\ee

The claim is that there exists a nondegenerate real supermatrix, $L=L^{\ast}$, which transforms $\Omega$ into the following canonical form,
\be
L\Omega L^{T}=\left(
\ba{cccc}
\,b_{-}\,&\,0\,&\,0\,&\,0\,\\
\,0\,&\,s_{-}\,&\,0\,&\,\psi\,\\
\,0\,&\,0\,&\,ib_{+}\,&\,0\,\\
\,0\,&\,-\psi^{T}\,&\,0\,&is_{+}\,
\ea
\right)\,,
\label{Lemma4}
\ee
where all the variables are real, $b_{\pm}=b_{\pm}^{\ast}$,\, $s_{\pm}=s_{\pm}^{\ast}$,\,  $\psi=\psi^{\ast}$~; ~$b_{\pm}$ are nondegenerate  bosonic matrices, $(\det b_{\pm})_{\rm{\scriptstyle{body}}}\neq 0$~; ~$s_{\pm}$ are pure souls~; ~and $b_{\pm}=\pm b_{\pm}^{T}$, $s_{\pm}=\pm s_{\pm}^{T}$\,.\\

The partition reads
\be
{\Big[k_{-}+(n_{-}-k_{-})\,+\,k_{+}+(n_{+}-k_{+})\Big]^{2}\,,}
\ee
where $k_{\mp}$ are respectively the ranks  of the bosonic matrices,  $A_{\mp}$, so that $(b_{\pm})_{k_{\pm}\times k_{\pm}}$.
\proof{From \textit{Fact} 6, we transform $A_{-}$ into the canonical form, in such a way that
\be
R\Omega R^T=\left(\ba{ccc}
b_{-}&0&\psi_{1}\\
0&s^{\prime}_{-}&\psi_{2}\\
-\psi_{1}^{T}&-\psi_{2}^{T}&iA_{+}\ea\right)\,,
\ee
where $R=\left(\ba{cc}P&0\\0&1\ea\right)$.
We take it further to
\be
\ba{l}
\left(\ba{ccc}
1&0&0\\
0&1&0\\
\psi_{1}^{T}b_{-}^{-1}&0&1\ea\right)
\left(\ba{ccc}
b_{-}&0&\psi_{1}\\
0&s^{\prime}_{-}&\psi_{2}\\
-\psi_{1}^{T}&-\psi_{2}^{T}&iA_{+}\ea\right)
\left(\ba{ccc}
1&0&-b_{-}^{-1}\psi_{1}\\
0&1&0\\
0&0&1\ea\right)\\
{}\\
=
\left(\ba{ccc}
b_{-}&0&0\\
0&\,s^{\prime}_{-}&\psi_{2}\\
0&-\psi_{2}^{T}&~iA_{+}+\psi_{1}^{T}b_{-}^{-1}\psi_{1}\ea\right)\,.
\ea
\ee
Now apply \textit{Fact} 6 to $(iA_{+}+\psi_{1}^{T}b_{-}^{-1}\psi_{1})$ to get
\be
\left(\ba{cccc}
b_{-}&0&0&0\\
0&\,s^{\prime}_{-}&\chi&\psi\\
0&-\chi^{T}&~ib_{+}&0\\
0&-\psi^{T}&0&is_{+}\ea\right)\,.
\ee
Finally,  to complete the proof, we only need to take the following transformation,
\be
\ba{l}
\left(\ba{cccc}
1&0&0&0\\
0&1&i\chi b_{+}^{-1}&0\\
0&0&1&0\\
0&0&0&1\ea\right)
\left(\ba{cccc}
b_{-}&0&0&0\\
0&\,s^{\prime}_{-}&\chi&\psi\\
0&-\chi^{T}&~ib_{+}&0\\
0&-\psi^{T}&0&is_{+}\ea\right)
\left(\ba{cccc}
1&0&0&0\\
0&1&0&0\\
0&-ib_{+}^{-1}\chi^{T}&1&0\\
0&0&0&1\ea\right)\\
{}\\
=
\left(\ba{cccc}
b_{-}&0&0&0\\
0&\,s^{\prime}_{-}-i\chi b_{+}^{-1}\chi^{T}&0&\psi\\
0&0&~ib_{+}&0\\
0&-\psi^{T}&0&is_{+}\ea\right)\,.
\ea
\ee
}

\newpage
\section{A paradigmatic example}\label{parexample}

As an illustration of the general case, discussed in the core of the text, a most simple case is presented here: a Lagrangian
\begin{equation}
L(q^A,\dq^B)=T(\dq^B)-V(q^A)\,,
\label{Lexa}
\end{equation}
which
\begin{enumerate}
	\item is a function on a bosonic tangent space of finite dimension ${\cal N}\,$,
	\item is of homogeneity degree equal to two (\textit{i.e.} the system is free),
	\item leads to a positive definite energy, and
	\item does not lead to tertiary constraints.
\end{enumerate}
Hopefully, this example combines three virtues: (i) its simplicity should allow to displace the focus from the technical onto the conceptual, (ii) it includes both cases of first and second class constraints, and (iii) it provides the starting point of usual perturbative expansion, so it is not merely academical.\\

The Lagrangian is assumed to be quadratic, therefore
the kinetic energy $T(\dq)=\frac12 T_{AB}\dq^A\dq^B$ and the potential energy $V(q)=\frac12 V_{AB}q^A q^B$ are both quadratic forms.
The Lagrangian (\ref{Lexa}) leads to a conserved energy equal to
\begin{equation}
E(q,\dq)=T(\dq)+V(q)\,,
\label{LexaE}
\end{equation}
which is positive definite, $E(q,\dq)\geq 0$, if and only if the kinetic and potential energy are separately positive definite: $T(\dq)\geq 0\,$, $V(q)\geq 0\,$. Without loss of generality, one may assume that the variables $q^A$ have been `rotated' so that the symmetric matrix $T_{AB}=\partial^2 L/\partial q^A\partial q^B$ is diagonal:
\begin{equation}
T(\dq^a,\dq^m)=\frac12\,\sum\limits_a M_a\,(\dq^a)^2\,,
\end{equation}
where the index $a$ corresponds to the $\N-\M$ strictly positive eigenvalues $M_a>0\,$, while the index $m$ corresponds to the remaining $\M$ zero eigenvalues.
The corresponding momentas are respectively given by $p_a=M_a\dq^a$ (no sum on the index $a$!) and $p_m=0\,$.\footnote{In this example, notice that the distinction between hatted and unhatted indices is not necessary.} Therefore, one gets $\M$ primary constraints $\phi_m(q,p,t)=p_m$ and the primary constraint surface $V$ is the hyperplane $p_m=0$ of codimension $\M$ embedded in the $2\N$-dimensional phase space.
The canonical Hamiltonian (\ref{Hamiltonian}) reads
\be
\left.H(q^A,p_b,t)\right|_V\,=\,\sum\limits_a \frac{(p_a)^2}{2m_a}\,-\,\frac12 \sum\limits_{A,B}V_{AB}q^A q^B\,.
\ee
The total Hamiltonian (\ref{totH}) is thus given by
\be
H_T(q^A,p_B,u^m,t)\,=\,H\,+\,u^m\,p_m \,.
\ee
The time evolution of $q^m$ through the Poisson bracket with the total Hamiltonian leads to the equality $\dq^m=\{q^m,H_T\}_{P.B.}=u^m\,$. The one-to-one maps (\ref{11maps}) can be seen explicitly in the present case since $p_a=M_a\dq^a$ (no summation) and $u^m=\dq^m\,$.

The preservation (\ref{cons}) of the primary constraints under the time evolution leads to the secondary constraints
\begin{equation}
\phi^\prime_{m}(q,p,t)=V_{mm}q^m\,+\,\frac12\sum\limits_{n\neq m}V_{mn}q^n\,+\,\frac12\sum\limits_{a}V_{ma}q^a\,.
\label{preserv}
\end{equation}
Now the point is that the preservation (\ref{dotphi}) of the secondary constraints under the time evolution would leads to new, \textit{i.e.} tertiary constraints if some entries $V_{ma}$ were non-vanishing. Therefore, in the particular example we are considering, one assumes that the potential does not include mixed terms:
\be
V(q^a,q^m)\,=\,\frac12 \sum\limits_{a,b}V_{ab}q^a q^b\,+\,\frac12\sum\limits_{m,n}V_{mn}q^m q^n\,.
\ee
For that reason, one may perform a rotation in the plane of the variables $q^a/\sqrt{M_a}$ in order to make the symmetric matrix $V_{ab}$ diagonal without modifying the kinetic energy. Without loss of generality, the symmetric matrix $V_{mn}$ may also be assumed to be diagonal. Therefore,
\be
V(q^{\overline{a}},q^\alpha,q^{\overline{m}},q^\mu)\,=\,\frac12 \sum\limits_{\overline{a}}(\omega_{\overline{a}})^2(q^{\overline{a}})^2\,
+\,\frac12\sum\limits_{\overline{m}}(q^{\overline{m}})^2\,.
\ee
where the `barred' indices correspond to the strictly positive eigenvalues while the `Greek' indices correspond to the vanishing eigenvalues of the matrices $V_{ab}$ and $V_{mn}\,$.
The secondary constraints (\ref{preserv}) become simply $\phi^{\prime}_{\overline{m}}(q,p,t)=q^{\overline{m}}\,$. There are no tertiary constraints because the preservation (\ref{inheq}) of the secondary constraints under the time evolution only leads to the fact that the Lagrange multipliers $u^{\overline{m}}=0\,$, while the $u^\mu\,$'s can be arbitrary functions of time, which signals the presence of some gauge freedom. The constraint surface $\V$ is thus the hyperplane defined by the system $p_\mu=p_{\overline{m}}=q^{\overline{n}}=0\,$. It is straightforward to check that the constraints $p_\mu$ are (primary) first class constraints and that the set $\{p_{\overline{m}},q^{\overline{n}}\}$ contains all the second class constraints. Notice that, in this example, the total and extended Hamiltonians are identical since there are no secondary first class constraints.

Under the sole hypotheses stated above, the Lagrangian and the Hamiltonian have been decomposed into a sum of four pieces\footnote{The present ``paradigmatic'' example has been inspired from the two examples given in the section 1.6.2 of \cite{Henneaux:1992ig} which correspond to $L^{\mbox{1st}}$ and $L^{\mbox{2nd}}$. The straightforward quantization procedure for these two cases is respectively carried on in the sections 13.1.1 and 13.1.2.}
\begin{eqnarray}
L&=&L^{\mbox{free}}+L^{\mbox{harmonic}}+L^{\mbox{1st}}+L^{\mbox{2nd}}\,,\nonumber\\
H&=&H^{\mbox{free}}+H^{\mbox{harmonic}}+H^{\mbox{1st}}+H^{\mbox{2nd}}\,,
\end{eqnarray}
where each piece corresponds to one of the following four distinct physical cases:
\begin{itemize}
	\item {\bf Free particles:}
\begin{eqnarray}
L^{\mbox{free}}(\dq^\alpha)&=&\frac12\sum\limits_\alpha M_\alpha(\dq^\alpha)^2\,,\nonumber\\
H^{\mbox{free}}(p_\alpha)&=&\sum\limits_\alpha \frac{(p_\alpha)^2}{2\,M_\alpha}\,.\nonumber
\end{eqnarray}
	\item {\bf Harmonic oscillators:}
\begin{eqnarray}
L^{\mbox{harmonic}}(q^{\overline{a}},\dq^{\overline{b}})&=&\frac12\sum\limits_{\overline{a}}	 \Big[M_{\overline{a}}(\dq^{\overline{a}})^2-(\omega_{\overline{a}})^2(q^{\overline{a}})^2\Big]\,.\nonumber\\
H^{\mbox{harmonic}}(q^{\overline{a}},p_{\overline{b}})&=&\frac12\sum\limits_\alpha
\Big[\frac{(p_{\overline{a}})^2}{M_{\overline{a}}}+(\omega_{\overline{a}})^2(q^{\overline{a}})^2\Big]\,.\nonumber
\end{eqnarray}
  \item {\bf First class variables:}
  \begin{eqnarray}
L^{\mbox{1st}}&=&0\,,\nonumber\\
H_T^{\mbox{1st}}(p_\mu,u^\nu)&=&\sum\limits_\mu u^\mu p_\mu=
H_E^{\mbox{1st}}(p_\mu,u^\nu)\,.\nonumber
\end{eqnarray}
  \item {\bf Second class variables:}
  \begin{eqnarray}
L^{\mbox{2nd}}(q^{\overline{m}})&=& -\frac12\sum\limits_{\overline{m}}(q^{\overline{m}})^2,\nonumber\\
H_T^{\mbox{2nd}}(q^{\overline{m}},p_{\overline{n}},u^{\overline{r}})&=& \frac12\sum\limits_{\overline{m}}(q^{\overline{m}})^2=
H_E^{\mbox{2nd}}(q^{\overline{m}},p_{\overline{n}})\,.\nonumber
\end{eqnarray}

\end{itemize}

\newpage



\newpage

\begin{thebibliography}{99}

\bibitem{Chandrasekhar}
H.~Weyl, \textit{Symmetry} (Princeton University Press,
1952);\\
S.~Chandrasekhar, \textit{Truth and Beauty: Aesthetics and motivations in science} (University of Chicago Press, 1990).

\bibitem{Raifeartaigh}
L.~O'Raifeartaigh, \textit{The dawning of gauge theory} (Princeton University Press,
1997).

\bibitem{Dirac:1950pj}
P.~A.~M.~Dirac,
``Generalized Hamiltonian dynamics,''
Can.\ J.\ Math.\  {\bf 2} (1950) 129;
``The Hamiltonian form of field dynamics''
Can.\ J.\ Math.\  {\bf 3} (1951) 1;
``Generalized Hamiltonian dynamics'' \& ``The Theory of gravitation in Hamiltonian form,''
Proc.\ Roy.\ Soc.\ Lond.\  A {\bf 246} (1958) 326 \& 333.

\bibitem{Dirac}
P.~A.~M.~Dirac,
\textit{Lectures on quantum mechanics} (Yeshiva University, 1964).

\bibitem{Hanson}
A.~Hanson, T.~Regge, and C.~Teitelboim,
\textit{Constrained Hamiltonian Systems} (Accademia Nazionale dei Lincei, 1976).

\bibitem{Tyutin:1990}
D.M. Gitman and I.V. Tyutin, \textit{Quantization of fields with constraints} (Springer-Verlag, 1990).

\bibitem{Henneaux:1992ig}
M.~Henneaux and C.~Teitelboim, \textit{Quantization of gauge systems}
(Princeton University Press, 1992).

\bibitem{Blago}
K.~B.~Marathe, \textit{Constrained Hamiltonian systems}, Lecture Notes in Physics {\bf 180} (Springer, 1983);\\ 
J.~Govaerts, \textit{Hamiltonian Quantisation and Constrained Dynamics} (Leuven University, 1991);\\
M.~Blagojevic, \textit{Gravitation and Gauge Symmetries} (Institute of Physics Publishing, 2001);\\
P.~Spindel, \textit{Mécanique analytique} (Scientifiques GB, 2002).

\bibitem{Sardanachvily}
G.~Sardanashvily, \textit{Generalized Hamiltonian formalism for field theory} (World Scientific, 1995).

\bibitem{Buchbinder:1995uq}
I.~L.~Buchbinder and S.~M.~Kuzenko,
\textit{Ideas and methods of supersymmetry and supergravity: A walk through Superspace} (Institute of Physics Publishing, 1998).

\bibitem{Barnich:2000zw}
G.~Barnich, F.~Brandt and M.~Henneaux,
``Local BRST cohomology in gauge theories,''
Phys.\ Rept.\  {\bf 338} (2000) 439
[{\tt hep-th/0002245}].

\bibitem{Henneaux:1990au}
M.~Henneaux, C.~Teitelboim and J.~Zanelli,
``Gauge invariance and degree of freedom count,''
Nucl.\ Phys.\  B {\bf 332} (1990) 169.

\bibitem{Souriau}
J.~M.~Souriau, \textit{Structure des syst\`emes dynamiques} (Dunod, 1970);\\
J.~Butterfield, ``On symplectic reduction in classical mechanics'' in J.~Butterfield and J.~Earman Eds, \textit{Philosophy of Physics} (North Holland, 2006) 1 [{\tt physics/0507194}].

\bibitem{Dresse:1990dj}
A.~Dresse, P.~Gregoire and M.~Henneaux,
``Path integral equivalence between the extended and nonextended Hamiltonian formalisms,''
Phys.\ Lett.\  B {\bf 245} (1990) 192.

\bibitem{Park:2003ku}
J.-H.~Park,
 ``Superfield theory and supermatrix model,''\\
JHEP {\bf 0309} (2003) 046
[{\tt hep-th/0307060}].

\bibitem{Faddeev:1967fc}
L.~D.~Faddeev and V.~N.~Popov,
``Feynman diagrams for the Yang-Mills field,''
Phys.\ Lett.\  B {\bf 25} (1967) 29.

\bibitem{Klebanov:1991qa}
I.~R.~Klebanov,
``String theory in two-Dimensions,''
{\tt hep-th/9108019}.

\bibitem{Plyushchay:1993hs}
M.~S.~Plyushchay and A.~V.~Razumov,
``Dirac versus reduced phase space quantization for systems admitting no
gauge conditions,''
Int.\ J.\ Mod.\ Phys.\  A {\bf 11} (1996) 1427
[{\tt hep-th/9306017}].

\bibitem{Haar}
J.~Conway, \textit{A Course in Functional Analysis} (Springer-Verlag, 1990).

\bibitem{Becchi:1975nq}
C.~Becchi, A.~Rouet and R.~Stora,
``Renormalization of the Abelian Higgs-Kibble model,''
Commun.\ Math.\ Phys.\  {\bf 42} (1975) 127;
``Renormalization of gauge theories,''
Annals Phys.\  {\bf 98} (1976) 287.

\bibitem{Tyutin:1975qk}
I.~V.~Tyutin,
``Gauge invariance in field theory and statistical physics in operator formalism,''
preprint LEBEDEV-75-39.

\bibitem{Weinberg}
S.~Weinberg, \textit{The quantum theory of fields, Vol. 2: Modern applications}
(Cambridge University Press, 1996).

\bibitem{Henneaux:1985kr}
M.~Henneaux,
``Hamiltonian form of the path integral for theories with a gauge freedom,''
Phys.\ Rept.\  {\bf 126} (1985) 1;\\
L.~Baulieu,
``Perturbative Gauge Theories,''
Phys.\ Rept.\  {\bf 129} (1985) 1.

\bibitem{Batalin:1983}
I.~A. Batalin and G.~A. Vilkovisky, ``Quantization of gauge theories with
linearly dependent generators,'' {\em Phys. Rev.} {\bf D28} (1983)
2567 [Erratum-ibid.\  D {\bf 30} (1984) 508];
``Closure of the gauge algebra, generalized Lie equations and Feynman
rules,'' Nucl.\ Phys.\  B {\bf 234} (1984) 106.

\bibitem{Henneaux:1989}
M.~Henneaux, ``Lectures on the antifield-BRST formalism for gauge theories,''
{\em Nucl. Phys. Proc. Suppl.} {\bf 18A} (1990) 47;\\
J.~Gomis, J.~Paris, and S.~Samuel, ``Antibracket, antifields and gauge theory
  quantization,'' {\em Phys. Rept.} {\bf 259} (1995) 1
[{\tt hep-th/9412228}].

\end{thebibliography}
\end{document}